\documentclass[pra,twocolumn,showpacs,superscriptaddress,floatfix]{revtex4}

\usepackage{amsmath}
\usepackage{times}
\usepackage{txfonts}
\usepackage{graphicx}

\mathchardef\ogon="012C%
\newcommand{\as}{a\kern-0.22em\lower.40ex\hbox{$_{\ogon}$}}

\begin{document}

  \title{Adiabatic association of ultracold molecules via magnetic field 
    tunable interactions}
  \author{Krzysztof G{\'o}ral}
    \affiliation{Clarendon Laboratory, Department of Physics,  
    University of Oxford, Parks Road, Oxford, OX1 3PU, United Kingdom}
  \affiliation{Center for Theoretical Physics, Polish Academy of 
    Sciences, Al.\ Lotnik\'ow 32/46, 02-668 Warsaw, Poland}
  \author{Thorsten K\"{o}hler}
  \affiliation{Clarendon Laboratory, Department of Physics,  
    University of Oxford, Parks Road, Oxford, OX1 3PU, United Kingdom}
  \author{Simon A.~Gardiner} 
  \affiliation{JILA, University of Colorado and National Institute of 
    Standards and Technology, Boulder, Colorado 80309-0440}
  \author{Eite Tiesinga}
    \affiliation{Atomic Physics Division, National Institute of Standards and 
    Technology, 100 Bureau Drive Stop 8423, Gaithersburg, Maryland 20899-8423}
  \author{Paul S.\ Julienne}
  \affiliation{Atomic Physics Division, National Institute of Standards and 
    Technology, 100 Bureau Drive Stop 8423, Gaithersburg, Maryland 20899-8423}

\begin{abstract}
We consider in detail the situation of applying a time dependent external 
magnetic field to a $^{87}$Rb atomic Bose-Einstein condensate held in a 
harmonic trap, in order to adiabatically sweep the interatomic interactions 
across a Feshbach resonance to produce diatomic molecules. To this end, we 
introduce a minimal two-body Hamiltonian depending on just five 
measurable parameters of a Feshbach resonance, which accurately determines all 
low energy binary scattering observables, in particular, the molecular 
conversion efficiency of just two atoms. Based on this description of the 
microscopic collision phenomena, we use the many-body theory of T.~K\"ohler 
and K.~Burnett [Phys.~Rev.~A \textbf{65}, 033601 (2002)] to study the 
efficiency of the association of molecules in a $^{87}$Rb Bose-Einstein 
condensate during a linear passage of the magnetic field strength across the 
100 mT Feshbach resonance. We explore different, experimentally accessible, 
parameter regimes, and compare the predictions of Landau-Zener, configuration 
interaction, and two level mean field calculations with those of the 
microscopic many-body approach. Our comparative studies reveal a remarkable 
insensitivity of the molecular conversion efficiency with respect to both the 
details of the microscopic binary collision physics and the coherent nature 
of the Bose-Einstein condensed gas, provided that the magnetic field strength 
is varied linearly. We provide the reasons for this universality of the 
molecular production achieved by {\em linear} ramps of the magnetic field 
strength, and identify the Landau-Zener coefficient determined by
F.H. Mies {\em et al.} [Phys.~Rev.~A \textbf{61}, 022721 (2000)] as the 
main parameter that controls the efficiency.
\end{abstract}
\date{\today}
\maketitle

\section{Introduction}
The prospect of achieving quantum degenerate molecular gases has attracted 
considerable attention for some time now \cite{Levi00}. Such an accomplishment 
may open new avenues for research, for instance, bright sources of molecules 
for cold collision studies \cite{Weiner99}, precise molecular spectroscopy, 
elucidating the nature of a possible BCS-BEC crossover in Fermi gases 
\cite{Randeria95} and, possibly, 
the exploitation of dipole-dipole interactions \cite{Baranov02}. It has been 
clear from the outset, however, that laser cooling techniques, essential for 
the production of Bose-Einstein condensates and degenerate Fermi gases of 
atoms, are difficult to apply in the case of molecules due to their typically 
complicated rovibrational energy spectrum. The association of ultracold atoms 
into diatomic molecules, which may also be quantum degenerate, therefore seems 
a very promising route. The molecular conversion can be achieved by 
photoassociation \cite{Wynar00} or by application of time dependent magnetic 
fields in the vicinity of Feshbach resonances \cite{Inouye98}. The Feshbach 
resonance technique has recently been exploited with great success to 
produce large, ultracold assemblies of diatomic molecules, using as a source 
both atomic Bose-Einstein condensates 
\cite{Donley02,Claussen03,Herbig03,Duerr03,Xu03} and 
quantum degenerate two component gases of Fermionic atoms 
\cite{Regal03,Strecker03,Cubizolles03,JochimPRL03,Regal04,Greiner03,JochimScience03,Zwierlein03,Bartenstein04,Regal04-2,Zwierlein04,Bartenstein04-2}.

The experiments reported in Refs.~\cite{Herbig03,Duerr03,Xu03} achieve 
conversion of alkali atoms in dilute Bose-Einstein condensates to diatomic 
molecules by an adiabatic sweep of the strength of a homogeneous magnetic 
field across a Feshbach resonance from negative to positive scattering 
lengths. The observations indicate molecular fractions much smaller than the 
ideal limit of half the number of initial condensate atoms, even in the limit 
of perfect adiabaticity. This may be due to limitations in the initial 
production of molecules, or their long term stability in the presence of the 
surrounding gas.  

The long term stability of the highest excited vibrational molecular bound
state, produced by the adiabatic association technique, may be limited due to 
collisional deexcitation. Both  
theoretically and experimentally, very little is known about the 
associated rate constants for highly excited diatomic molecules composed of
alkali atoms. To our knowledge, the only available exact calculations have 
been performed for the deexcitation of tightly bound Na$_2$ molecules upon 
collision with a Na atom \cite{Soldan02}. From an experimental viewpoint, 
the comparatively high relative momenta of the products of collisional 
deexcitation have so far prevented their detection. Conclusive experimental 
studies of the loss rates and the underlying microscopic processes are 
therefore difficult to achieve.

In such an uncertain situation it is important to first understand the 
production of the diatomic molecules. In this paper we consider the situation
of adiabatically sweeping the magnetic field strength across a Feshbach 
resonance to form molecules from an initially Bose-Einstein condensed dilute 
gas of atoms. The calculations presented in this paper consider the 
100 mT Feshbach resonance of $^{87}$Rb \cite{Marte02}. The underlying concepts 
can be extended to arbitrary species of Bosonic atoms, while the two-body 
physics is also applicable to any pair of atoms interacting via $s$ waves, 
including Fermionic species.

The structure of the paper is as follows: Section \ref{sec:twobody}
describes the binary physics of the association of ultracold atoms to a 
diatomic molecule, and Section \ref{sec:manybody} analyses the 
many-body aspects of this process in a Bose-Einstein condensate. 
Section \ref{sec:conclusions} summarises and presents our conclusions, and 
there is an appendix, which expands upon the explicit description of the 
binary collision physics specific to this work.

Section \ref{sec:twobody} presents a two channel description of the magnetic 
field tunable resonance 
enhanced binary scattering and its relationship to the properties of the 
highest excited vibrational bound state. We develop the concept of a two-body 
Hamiltonian that accurately describes the relevant physics in 
terms of a minimal set of five parameters, which can usually be deduced from 
measurable properties of a Feshbach resonance. We then focus on the near 
resonance universal properties of the highest excited vibrational bound state, 
and identify the smooth transition between free and bound atoms in an ideally 
adiabatic passage across a Feshbach resonance. We then consider the dynamics
of the association of just two atoms during a linear ramp of the magnetic field
strength across a Feshbach resonance, and reveal the dependence of molecular
conversion on the physical parameters. The last subsection is concerned with  
the determination of dissociation energy spectra, as dissociation of molecules 
by ramps of the magnetic field strength is frequently a necessary precursor to 
their detection. We show that both the molecular conversion efficiency and the 
dissociation spectra are determined by the same physical parameters of a 
Feshbach resonance, and are subject to a remarkable insensitivity with respect 
to the details of the binary collision dynamics, provided that the magnetic 
field strength is varied linearly.

Section \ref{sec:manybody} introduces the relevant elements of the microscopic 
quantum dynamics approach \cite{KB02}, which properly accounts for both the 
microscopic binary collision physics and the macroscopic coherent nature of an 
inhomogeneous atomic Bose-Einstein condensate. Using this approach, we study 
the many-body aspects of the molecular production during a linear adiabatic 
passage of the magnetic field strength across a Feshbach resonance. From the 
first order microscopic quantum dynamics approach, by applying the Markov 
approximation, we derive the commonly used two level mean field model 
\cite{Drummond98,Timmermans99,Yurovsky00,Goral01,Salgueiro01,Adhikari01,Cusack02,Abdullaev03,Naidon}. 
We discuss the deficits of two level models with respect to the description of 
the intermediate dynamics during passage of the magnetic field strength 
across a Feshbach resonance. Our comparative studies show, however, that the 
many-body approaches predict virtually the same final molecular production at
all levels of approximation considered in this paper, provided that the ramp 
of the magnetic field strength across the Feshbach resonance is linear. 
We provide the reasons for this universality of the molecular conversion in 
linear ramps of the magnetic field strength with respect to the details of 
the underlying microscopic binary collision dynamics and to the 
coherent nature of the Bose-Einstein condensed gas. We then show that the 
molecular production efficiency is determined by the same physical parameters 
that we have previously identified in the associated two-body problem. Our 
findings strongly indicate that measurements of molecular production 
efficiencies as well as dissociation spectra obtained from {\em linear} ramps 
of the magnetic field strength are largely inconclusive with respect to the 
details of the underlying binary collision physics. 

\section{Adiabatic association of  two atoms}
\label{sec:twobody}
We consider a configuration of two atoms exposed to a homogeneous magnetic
field whose strength can be varied.
The concept underlying the adiabatic association of diatomic molecules in 
ultracold gases can be understood solely on the basis of binary collision 
physics. The key feature of this experimental technique is the adiabatic 
transfer of the zero energy binary scattering state into the highest excited 
diatomic vibrational bound state. In this section we shall study the 
binding energies of the diatomic molecules that determine the positions of 
the Feshbach resonances. We shall then show how the resonance enhanced 
interatomic collisions can be accurately described in terms of a minimal set 
of five quantities that can be determined from current experiments. 
Specifically, we consider the association of two 
asymptotically free ultracold $^{87}$Rb atoms with a total angular momentum 
quantum number of $F=1$ and an orientation quantum number of $m_F=+1$, at 
magnetic field strengths close to the broadest Feshbach resonance at about 
100 mT. We shall also study the dissociation of the molecules, which plays an 
important role in the direct detection of ultracold molecular gases.

\subsection{Feshbach resonances and vibrational bound states in $^{87}$Rb}
\label{subsec:Feshbachandboundstates}
\begin{figure}[htb] \includegraphics[width=\columnwidth,clip]{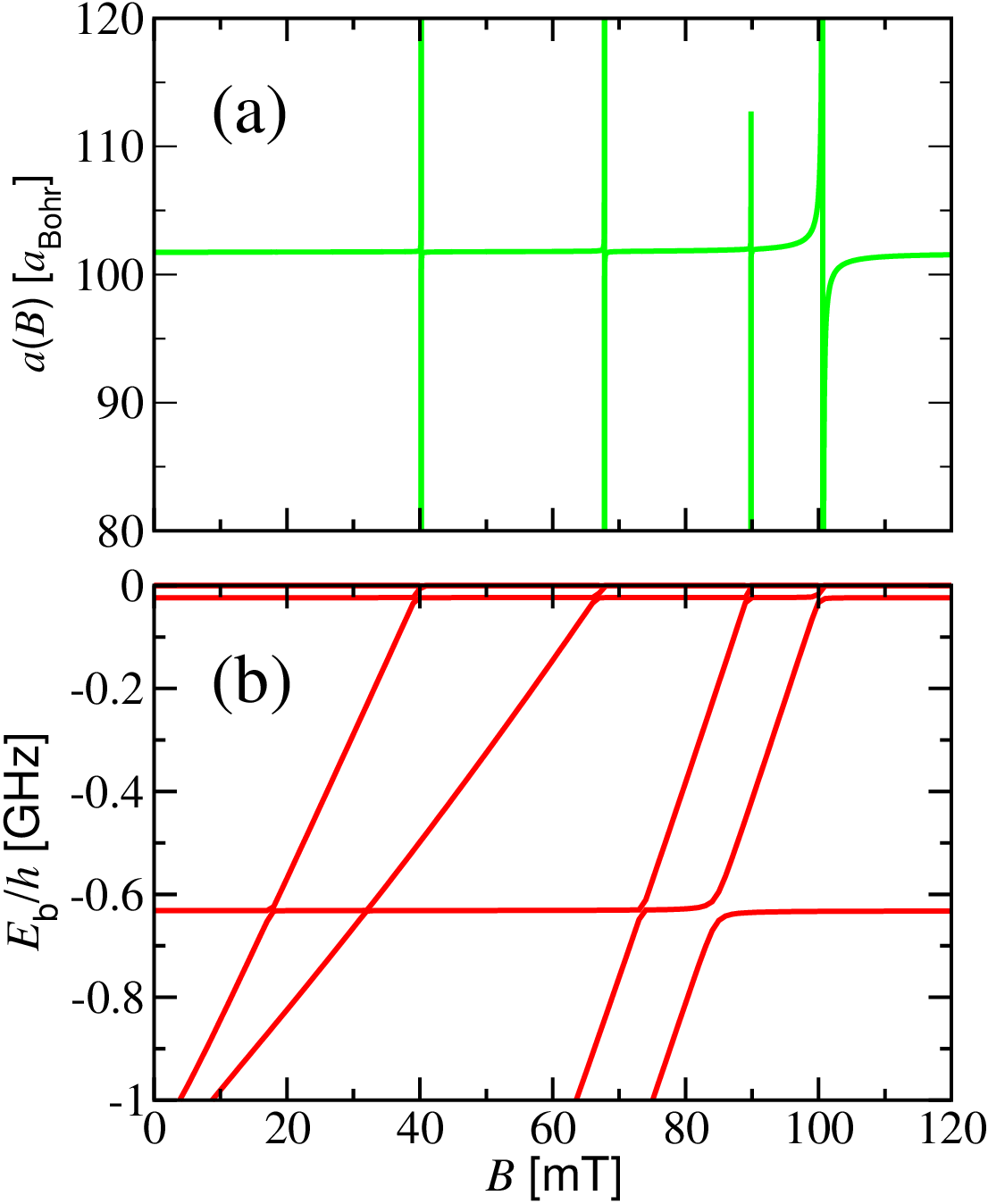}
    \caption{Magnetic field strength dependence of the $s$ wave scattering
    length (a) and the energies of the highest excited $s$-wave vibrational
    molecular bound states (b) of two $^{87}$Rb atoms.  At each magnetic
    field strength $B$ the zero of energy is set at the threshold for
    dissociation into two asymptotically free atoms in the $(F=1,m_F=+1)$
    hyperfine state.  The lines in (b) at about -0.02 and -0.63 GHz parallel 
    to the $E=0$ axis represent the energies of the last $s$-wave bound states
    of the open background channel. These states have the same magnetic
    moment as the separated atoms. The slanted lines in (b) represent
    closed channel molecular states that have different magnetic moments
    from the separated atoms.  The weak avoided crossings at the
    intersection of parallel and slanted lines are due to interactions
    between the background and closed channels, as noted in 
    Ref.~\cite{Duerr03}.
    When the closed channel molecular states cross threshold at $E=0$, the
    zeros of the binding energy in (b) correspond to the positions of the
    Feshbach resonances in (a). Our calculated resonance positions are
    within one per cent of the measured positions~\cite{Marte02}.}
\label{fig:Eboverview}
\end{figure}

Throughout this paper, we will denote the open scattering channel of
two asymptotically free atoms in the $(F=1,m_F=+1)$ electronic ground
state as the open channel with an associated reference potential (the
background scattering potential) $V_\mathrm{bg}(r)$.  The dissociation
threshold of $V_\mathrm{bg}(r)$ is determined by the internal energy of
the noninteracting atoms, i.e.~twice the energy corresponding to the
$(F=1,m_F=+1)$ hyperfine state.  When the atoms are exposed to an
external homogeneous magnetic field the $m_F$ degeneracy of the atomic
hyperfine levels is removed by the Zeeman effect. As a consequence,
the potentials associated with the different asymptotic binary
scattering channels are shifted with respect to each other.  Although
in general the interchannel coupling is weak, in the vicinity of
certain magnetic field strengths the open channel can be strongly
coupled to closed channels.  This strong coupling leads to
singularities of the $s$ wave scattering length known as Feshbach
resonances.  Figure \ref{fig:Eboverview} (a) shows the theoretically
predicted $s$ wave scattering length of two colliding $^{87}$Rb atoms
in the electronic ground state that are exposed to a homogeneous
magnetic field of strength $B$.  The theoretical calculations use five
coupled equations for one open and four closed channels to describe the
$s$-wave collision of two $(F=1,m_F=+1)$ atoms~\cite{Mies00}.  We use
standard methods to solve these equations for either scattering or
bound states.

The Feshbach resonances are related to the binding energy of the
highest excited vibrational state in a simple way; the singularities of
the scattering length in Fig.~\ref{fig:Eboverview} (a) exactly match
those magnetic field strengths that correspond to the zeros of the
binding energy with respect to the threshold energy for dissociation
into two asymptotically free atoms. Figure \ref{fig:Eboverview} (b)
gives an overview of the magnetic field dependence of the binding
energies of $s$-wave symmetry molecular states of two $^{87}$Rb atoms
below the dissociation threshold energy of the open channel.

\subsection{Two channel energy states}
\label{subsec:energystates}
The scattering lengths and binding energies in Fig.~\ref{fig:Eboverview} have 
been obtained from exact solutions of the multichannel two-body Schr\"odinger 
equation as described in Subsection \ref{subsec:Feshbachandboundstates}. These 
calculations were performed with a realistic potential matrix 
that accurately describes the bound and free molecular states over a wide 
range of energies and magnetic field strengths. Based on these exact 
considerations, Fig.~\ref{fig:Eboverview} reveals that the binding energy of 
the highest excited vibrational state determines the singularities of the 
scattering length and, in turn, all resonance enhanced low energy scattering 
properties of two atoms. Collisions in ultracold gases involve only a quite 
limited range of energies, and the adiabatic association of molecules takes 
place at magnetic field strengths in the close vicinity of a particular 
Feshbach resonance. We shall therefore restrict our analysis to an 
appropriately smaller range of energies and magnetic field strengths around 
the 100 mT Feshbach resonance. 

\subsubsection{Background scattering}
\label{subsubsec:background}
We shall first show how the low energy background scattering can be accurately 
described in terms of experimentally known physical quantities. At 
magnetic field strengths asymptotically far from the resonance the 
interchannel coupling is weak and the highest excited multichannel vibrational 
bound state can be determined directly from a single channel description with 
the background scattering potential $V_\mathrm{bg}(r)$. Considerations beyond 
the scope of this paper 
\cite{Gao98}
show that the corresponding binding energy is 
determined, to an excellent approximation, by the long range asymptotic 
behaviour of $V_\mathrm{bg}(r)$ and the background scattering length 
$a_\mathrm{bg}$, i.e.~the scattering length associated with the background 
scattering potential $V_\mathrm{bg}(r)$. Neglecting retardation phenomena, at 
large interatomic separations $V_\mathrm{bg}(r)$ has the universal form 
$V_\mathrm{bg}(r)\underset{r\to\infty}{\sim}-C_6/r^6$, where $C_6$ is the 
van der Waals dispersion coefficient. The low energy background scattering is  
determined by the same parameters $a_\mathrm{bg}$ and $C_6$
\cite{Gao98}. This universality is due to the fact that at typical 
ultracold collision energies the de Broglie wavelengths
are much larger than the van der Waals length
\begin{equation}
  l_\mathrm{vdW}=\frac{1}{2}\left(\frac{mC_6}{\hbar^2}\right)^{1/4}.
  \label{lvdW}
\end{equation} 
Here $m$ is the atomic mass. As the van der Waals length is the characteristic 
length scale set by the long range tail of the background scattering potential,
the details of the potential $V_\mathrm{bg}(r)$ are not resolved 
in the collisions. At zero collision energy the background scattering length 
$a_\mathrm{bg}$ incorporates all the unresolved details of $V_\mathrm{bg}(r)$ 
into a single length scale. The van der Waals length determines the 
first correction at finite collision energies, which accounts 
for the long range asymptotic behaviour of $V_\mathrm{bg}(r)$. We note that 
any model of the background scattering potential will recover the binding 
energy of the highest excited vibrational state and all low energy scattering 
properties of the exact potential $V_\mathrm{bg}(r)$ to an excellent 
approximation, if it properly accounts for the parameters $C_6$ and 
$a_\mathrm{bg}$. We provide an appropriate minimal background scattering 
potential in the appendix. 

\subsubsection{Two channel Schr\"odinger equation}
We assume in the following that all the bound and free energy states 
associated with the background scattering potential have been determined, and 
on this basis derive the resonance enhanced collision properties of two atoms. 
In the vicinity of a Feshbach resonance, the strong coupling between the 
open channel and other asymptotic scattering channels originates 
from the near degeneracy of the magnetic field dependent energy 
$E_\mathrm{res}(B)$ of a closed channel vibrational state 
(a Feshbach resonance level) $\phi_\mathrm{res}(r)$ with the dissociation 
threshold energy of the open channel. Consequently, the resonance 
enhanced collision physics of two $^{87}$Rb atoms can be accurately described 
by the general form of a two-channel Hamiltonian matrix of the relative motion 
of the atoms: 
\begin{align}
  H_\mathrm{2B}=
  \left(
  \begin{array}{cc}
    -\frac{\hbar^2}{m}\nabla^2+V_\mathrm{bg}(r) & W(r)\\
    W(r) & -\frac{\hbar^2}{m}\nabla^2+V_\mathrm{cl}(B,r)
  \end{array}
  \right).
  \label{H2B2channel}
\end{align}
Here $m$ is the atomic mass, $r$ is the distance between the atoms, $W(r)$ 
determines the strength of the coupling between the channels, and the closed 
channel potential $V_\mathrm{cl}(B,r)$ supports the resonance state:
\begin{equation}
  \left[-\frac{\hbar^2}{m}
    \nabla^2+V_\mathrm{cl}(B,r)\right]\phi_\mathrm{res}(r)=
  E_\mathrm{res}(B)\phi_\mathrm{res}(r). 
  \label{SEphires}
\end{equation}
In the following the resonance state $\phi_\mathrm{res}(r)$ is normalised to 
unity. In Eq.~(\ref{H2B2channel}), as elsewhere in this paper, we have chosen 
the zero of energy as the dissociation threshold of the open channel, 
i.e.~$V_\mathrm{bg}(r)\underset{r\to\infty}{\to}0$. The dissociation 
threshold of $V_\mathrm{cl}(B,r)$ is determined accordingly by the energy of 
two noninteracting atoms in the closed channel that is strongly coupled to 
the open channel. The relative Zeeman energy shift between the channels can 
be tuned by varying the magnetic field strength. 

The bound and free energy states of the general Hamiltonian matrix in 
Eq.~(\ref{H2B2channel}) relate the remaining potentials $W(r)$ and 
$V_\mathrm{cl}(B,r)$ to a minimal set of measurable properties of a Feshbach 
resonance. The two channel states that we shall consider in the following 
are of the general form $|\mathrm{bg}\rangle\phi^\mathrm{bg}(\mathbf{r})
+|\mathrm{cl}\rangle\phi^\mathrm{cl}(\mathbf{r})$,
where $|\mathrm{bg}\rangle$ and $|\mathrm{cl}\rangle$ denote the internal
states of an atom pair in the open channel and the closed channel strongly 
coupled to it, respectively. The two components of the energy states 
are solutions of the stationary coupled Schr\"odinger equations
\begin{align}
  \label{SEphibg}
  \left[-\frac{\hbar^2}{m}\nabla^2+V_\mathrm{bg}(r)\right]
  \phi^\mathrm{bg}(\mathbf{r})
  +W(r)\phi^\mathrm{cl}(\mathbf{r})&=E\phi^\mathrm{bg}(\mathbf{r}),\\
  \label{SEphicl}
  W(r)\phi^\mathrm{bg}(\mathbf{r})+
  \left[-\frac{\hbar^2}m\nabla^2+V_\mathrm{cl}(B,r)\right]
  \phi^\mathrm{cl}(\mathbf{r})&=E\phi^\mathrm{cl}(\mathbf{r}).
\end{align}

\subsubsection{Continuum states}
Bound and continuum energy states are distinguished by their energies and 
by their asymptotic behaviour at large interatomic distances. In the 
scattering  continuum above the dissociation threshold all energies are in the 
spectrum of the Hamiltonian in Eq.~(\ref{H2B2channel}) and can be associated 
with the momentum $\mathbf{p}$ of the relative motion of two asymptotically 
noninteracting atoms in the open channel through $E=p^2/m$. Due to the 
continuous scattering angles between the atoms at a 
definite collision energy, the scattering energy states are infinitely 
degenerate. In the following, we will choose their wave functions to behave 
at large interatomic distances like: 
\begin{equation}
  \phi_\mathbf{p}^\mathrm{bg}(\mathbf{r})
  \underset{r\to \infty}{\sim}\frac{1}{(2\pi\hbar)^{3/2}}
  \left[e^{i\mathbf{p}\cdot\mathbf{r}/\hbar}+f(\vartheta,p)
    \frac{e^{ipr/\hbar}}{r}\right].
  \label{BCphipbg}
\end{equation}
This long range asymptotic behaviour corresponds to an incoming plane wave 
and an outgoing spherical wave in the  open channel. Here and 
throughout this paper we will assume the plane wave momentum states to be 
normalised as 
$\exp(i\mathbf{p}\cdot\mathbf{r}/\hbar)/(2\pi\hbar)^{3/2}$. 
The function $f(\vartheta,p)$ in Eq.~(\ref{BCphipbg}) is the scattering 
amplitude, which depends on $p=\sqrt{mE}$ and on the scattering angle 
$\vartheta$ between the momentum $\mathbf{p}$ of the relative motion of the 
asymptotically noninteracting incoming atoms and their final relative 
position $\mathbf{r}$. The closed channel component 
$\phi_\mathbf{p}^\mathrm{cl}(\mathbf{r})$ of the wave function vanishes at 
asymptotically large distances between the colliding atoms. 
We shall also introduce the energy states $\phi_\mathbf{p}^{(+)}(\mathbf{r})$ 
of the background scattering that satisfy the Schr\"odinger equation
\begin{equation}
  \left[-\frac{\hbar^2}{m}\nabla^2+V_\mathrm{bg}(r)\right]
  \phi_\mathbf{p}^{(+)}(\mathbf{r})=
  \frac{p^2}{m}\phi_\mathbf{p}^{(+)}(\mathbf{r}),
\end{equation}
with the long range asymptotic behaviour:
\begin{equation}
  \phi_\mathbf{p}^{(+)}(\mathbf{r})\underset{r\to \infty}{\sim}
  \frac{1}{(2\pi\hbar)^{3/2}}
  \left[e^{i\mathbf{p}\cdot\mathbf{r}/\hbar}+f_\mathrm{bg}(\vartheta,p)
    \frac{e^{ipr/\hbar}}{r}\right].
  \label{BCphipplus}
\end{equation}

The coupled Schr\"odinger equations (\ref{SEphibg}) and (\ref{SEphicl}) can be 
expressed in terms of the energy dependent Green's functions:
\begin{align}
  \label{Gbg}
  G_\mathrm{bg}(z)&=
  \left[z-\left(-\frac{\hbar^2}{m}\nabla^2+V_\mathrm{bg}\right)\right]^{-1},\\
  G_\mathrm{cl}(B,z)&=
  \left[z-\left(-\frac{\hbar^2}{m}\nabla^2+V_\mathrm{cl}(B)\right)\right]^{-1}.
  \label{Gcl}
\end{align} 
Here $z$ is a complex parameter with the dimension of an energy. The coupled 
Schr\"odinger equations then read: 
\begin{align}
  \label{LSphipbg}
  \phi_\mathbf{p}^\mathrm{bg}&=\phi_\mathbf{p}^{(+)}+
  G_\mathrm{bg}(E+i0)W\phi_\mathbf{p}^\mathrm{cl},\\
  \phi_\mathbf{p}^\mathrm{cl}&=G_\mathrm{cl}(B,E)W\phi_\mathbf{p}^\mathrm{bg}.
  \label{LSphipcl}
\end{align}
The argument ``$z=E+i0$'' of the Green's function $G_\mathrm{bg}(z)$ indicates 
that the physical energy $E=p^2/m$ is approached from the upper half of the 
complex plane. This choice of the energy argument ensures that the scattering 
wave function $\phi_\mathbf{p}^\mathrm{bg}(\mathbf{r})$ has the long range 
asymptotic form of Eq.~(\ref{BCphipbg}), in accordance with the asymptotic 
behaviour of the Green's function at large interatomic distances:
\begin{equation}
  G_\mathrm{bg}(E+i0,\mathbf{r},\mathbf{r}')\underset{r\to\infty}{\sim}-
  (2\pi\hbar)^{3/2}\frac{m}{4\pi\hbar^2}\frac{e^{ipr/\hbar}}{r}
  \left[\phi_\mathbf{p}^{(-)}(\mathbf{r}')\right]^*.
  \label{asympGbg}
\end{equation}
Here 
%\begin{equation}
  $\phi_\mathbf{p}^{(-)}(\mathbf{r}')=
  \left[\phi_{-\mathbf{p}}^{(+)}(\mathbf{r}')\right]^*$
  %\label{phipminus}
%\end{equation}
is the incoming continuum energy state associated with the background 
scattering \cite{Newton82}, 
and $\mathbf{p}=\left(\sqrt{mE}\right)\mathbf{r}/r$ can be 
interpreted as the asymptotic momentum associated with the relative motion of 
the scattered atoms.

As the resonance state $\phi_\mathrm{res}(r)$ fulfils the Schr\"odinger 
equation (\ref{SEphires}), according to Eq.~(\ref{Gcl}) the Green's function 
$G_\mathrm{cl}(B,z)$ has a singularity at $z=E_\mathrm{res}(B)$, i.e.
\begin{equation} 
  \langle\phi_\mathrm{res}|G_\mathrm{cl}(B,z)|\phi_\mathrm{res}\rangle=
  \frac{1}{z-E_\mathrm{res}(B)}. 
  \label{poleapproximation1}
\end{equation}
At magnetic field strengths in the vicinity of a Feshbach resonance 
$E_\mathrm{res}(B)$ is nearly degenerate with the dissociation threshold 
energy of the  open channel. Furthermore, the kinetic energies 
$E=p^2/m$ in ultracold collisions are small in comparison with the typical 
spacing between molecular vibrational bound states. As a consequence, the
denominator in Eq.~(\ref{poleapproximation1}) becomes sufficiently small for  
the Green's function $G_\mathrm{cl}(B,E)$ in Eq.~(\ref{LSphipcl}) to be 
excellently approximated by its resonance state component \cite{Child74}: 
\begin{equation}
  G_\mathrm{cl}(B,E)\approx |\phi_\mathrm{res}\rangle
  \frac{1}{E-E_\mathrm{res}(B)}
  \langle\phi_\mathrm{res}|.
  \label{poleapproximation}
\end{equation} 
Inserting this pole approximation of the Green's function into 
Eq.~(\ref{LSphipcl}) determines the functional form of the closed channel 
component of the scattering wave function to be
\begin{equation}
  |\phi_\mathbf{p}^\mathrm{cl}\rangle=|\phi_\mathrm{res}\rangle A(B,E).
  \label{phipcl}
\end{equation}
The wave function $\phi_\mathbf{p}^\mathrm{bg}(\mathbf{r})$ is then determined 
by eliminating $\phi_\mathbf{p}^\mathrm{cl}$ on the right hand side of 
Eq.~(\ref{LSphipbg}) in terms of Eq.~(\ref{phipcl}) which gives
\begin{equation}
  |\phi_\mathbf{p}^\mathrm{bg}\rangle=|\phi_\mathbf{p}^{(+)}\rangle+
  G_\mathrm{bg}(E+i0)W|\phi_\mathrm{res}\rangle A(B,E).
  \label{phipbg}
\end{equation}
The as yet unknown amplitude 
\begin{equation}
  A(B,E)=\frac{\langle\phi_\mathrm{res}|W|\phi_\mathbf{p}^\mathrm{bg}\rangle}
  {E-E_\mathrm{res}(B)}
\end{equation}
can be determined straightforwardly by multiplying Eq.~(\ref{phipbg}) 
by $\langle\phi_\mathrm{res}|W$ from the left. This yields, after a short 
calculation:
%\begin{widetext}
\begin{equation}
  A(B,E)=\frac{\langle\phi_\mathrm{res}|W|\phi_\mathbf{p}^{(+)}\rangle}
  {E-E_\mathrm{res}(B)-
    \langle\phi_\mathrm{res}|WG_\mathrm{bg}(E+i0)W|\phi_\mathrm{res}\rangle}.
  \label{amplitude}
\end{equation} 
%\end{widetext}
Once all the energy states associated with the single channel potential 
$V_\mathrm{bg}(r)$ and the resonance state $\phi_\mathrm{res}(r)$ are known, 
Eqs.~(\ref{phipcl}), (\ref{phipbg}) and (\ref{amplitude}) establish the 
complete solution of the coupled Schr\"odinger equations (\ref{LSphipbg}) 
and (\ref{LSphipcl}) in the pole approximation. Under the 
assumption that the configuration of two atoms in the closed channels is 
restricted to the resonance state $\phi_\mathrm{res}$, the pole 
approximation becomes exact. This assumption implies the replacement
\begin{equation}
  -\frac{\hbar^2}{m}\nabla^2+V_\mathrm{cl}(B)\to 
  |\phi_\mathrm{res}\rangle E_\mathrm{res}(B)\langle\phi_\mathrm{res}|
  \label{replacementHcl}
\end{equation}
in the two channel Hamiltonian in Eq.~(\ref{H2B2channel}). 
Equations (\ref{phipcl}), (\ref{phipbg}) and (\ref{amplitude}) then determine 
the exact scattering energy states of the resulting restricted two channel 
Hamiltonian. 

The scattering length $a$ is determined by the long range asymptotic form of 
the scattering wave function $\phi_\mathbf{p}^\mathrm{bg}(\mathbf{r})$ in 
Eq.~(\ref{BCphipbg}) through 
\begin{equation}
  f(\vartheta,p)\underset{p\to 0}{\sim}-a.
\end{equation}
Because the incoming plane wave is isotropic at zero energy,
this limit is obviously independent of the scattering angle. 
In an analogous way, the 
background scattering length $a_\mathrm{bg}$ is related to the asymptotic form 
of the scattering wave function $\phi_\mathbf{p}^\mathrm{(+)}(\mathbf{r})$ in 
Eq.~(\ref{BCphipplus}), in the limit of zero energy.
Inserting the known asymptotic form of the Green's function 
$G_\mathrm{bg}(E+i0)$ at large interatomic distances in Eq.~(\ref{asympGbg}) 
and the amplitude in Eq.~(\ref{amplitude}) into Eq.~(\ref{phipbg}), a short 
calculation determines the scattering length to be: 
\begin{equation}
  a=a_\mathrm{bg}-
  \frac{\frac{m}{4\pi\hbar^2}(2\pi\hbar)^3
    \left|\langle\phi_\mathrm{res}|W|\phi_0^{(+)}\rangle\right|^2}
       {E_\mathrm{res}(B)+
	 \langle\phi_\mathrm{res}|WG_\mathrm{bg}(0)W|\phi_\mathrm{res}\rangle}.
       \label{aofEres}
\end{equation}
The energy $E_\mathrm{res}(B)$ of the closed channel state is determined by 
the Zeeman energy shift between the asymptotic open channel and 
the closed channel strongly coupled to it. Within the limited range of 
magnetic field strengths that we shall consider in this paper,   
the Zeeman effect is approximately linear in the magnetic field strength $B$. 
We shall denote by $B_\mathrm{res}$ the magnetic field strength at which 
$E_\mathrm{res}(B)$ crosses the dissociation threshold of the  open 
channel, i.e.~$E_\mathrm{res}(B_\mathrm{res})=0$. An expansion of 
$E_\mathrm{res}(B)$ about $B=B_\mathrm{res}$ yields:
\begin{equation}
  E_\mathrm{res}(B)=\left[\frac{dE_\mathrm{res}}{dB}(B_\mathrm{res})\right]
  (B-B_\mathrm{res}).
  \label{slope}
\end{equation}
Equations (\ref{aofEres}) and (\ref{slope}) then determine the magnetic field 
dependence of the scattering length by the well known formula
\begin{equation}
a(B)=a_\mathrm{bg}\left[1-\frac{(\Delta B)}{B-B_0}\right],
\label{aofB}
\end{equation}  
where
\begin{equation}
  (\Delta B)=\frac{m}{4\pi\hbar^2 a_\mathrm{bg}}
  \frac{(2\pi\hbar)^3
    \left|\langle\phi_\mathrm{res}|W|\phi_0^{(+)}\rangle\right|^2}
       {\left[\frac{dE_\mathrm{res}}{dB}(B_\mathrm{res})\right]}
       \label{resonancewidth}
\end{equation}
is termed the resonance width and 
\begin{equation}
  B_0=B_\mathrm{res}-
  \frac{\langle\phi_\mathrm{res}|WG_\mathrm{bg}(0)W|\phi_\mathrm{res}\rangle}
       {\left[\frac{dE_\mathrm{res}}{dB}(B_\mathrm{res})\right]}
       \label{resonanceshift}
\end{equation}
is the measurable resonance position, i.e.~the magnetic field strength at 
which the scattering length has a singularity. We note that according to 
Eq.~(\ref{resonanceshift}) the resonance position $B_0$ is shifted with 
respect to the magnetic field strength $B_\mathrm{res}$ at which the energy 
of the resonance state becomes degenerate with the dissociation threshold 
energy of the open channel. 

\subsubsection{Bound states}
The molecular bound states vanish at asymptotically large interatomic 
distances, and their energies are below the dissociation threshold of the 
open channel. Similarly to Eqs.~(\ref{LSphipbg}) and 
(\ref{LSphipcl}), the coupled Schr\"odinger equations (\ref{SEphibg}) and 
(\ref{SEphicl}) can be expressed in terms of the coupled integral equations
\begin{align}
  \label{LSphibbg}
  \phi_\mathrm{b}^\mathrm{bg}&=
  G_\mathrm{bg}(E_\mathrm{b})W\phi_\mathrm{b}^\mathrm{cl},\\
  \phi_\mathrm{b}^\mathrm{cl}&=
  G_\mathrm{cl}(B,E_\mathrm{b})W\phi_\mathrm{b}^\mathrm{bg},
  \label{LSphibcl}
\end{align}
which incorporate the long range asymptotic behaviour of the components, 
$\phi_\mathrm{b}^\mathrm{bg}$ and $\phi_\mathrm{b}^\mathrm{cl}$, of the bound 
state. Here $E_\mathrm{b}$ is the binding energy. A short calculation verifies 
that the pole approximation in Eq.~(\ref{poleapproximation}) leads to the 
normalised solutions
\begin{equation}
  \left(
  \begin{array}{c}
    \phi_\mathrm{b}^\mathrm{bg}\\
    \phi_\mathrm{b}^\mathrm{cl}
  \end{array}
  \right)=\frac{1}{\mathcal{N}_\mathrm{b}}
  \left(
  \begin{array}{c}
    G_\mathrm{bg}(E_\mathrm{b})W\phi_\mathrm{res}\\
    \phi_\mathrm{res}
  \end{array}
  \right)
  \label{phib}
\end{equation}
with the normalisation constant
\begin{equation}
  \mathcal{N}_\mathrm{b}=\sqrt{1+
    \langle\phi_\mathrm{res}|W
    \left[G_\mathrm{bg}(E_\mathrm{b})\right]^2W|\phi_\mathrm{res}\rangle},
  \label{twochannelnormalisation}
\end{equation} 
whenever the binding energy $E_\mathrm{b}$ fulfils the condition: 
\begin{equation}
  E_\mathrm{b}=E_\mathrm{res}(B)+
  \langle\phi_\mathrm{res}|WG_\mathrm{bg}(E_\mathrm{b})W|
  \phi_\mathrm{res}\rangle.
  \label{determinationEb}
\end{equation}
We note that Eq.~(\ref{determinationEb}) recovers Eq.~(\ref{resonanceshift}) 
when the binding energy and the magnetic field strength are inserted as 
$E_\mathrm{b}=0$ and $B=B_0$, respectively, i.e.~the binding energy, indeed, 
vanishes at the position of the resonance.

\subsubsection{Minimal two channel Hamiltonian}
Both the resonance width in Eq.~(\ref{resonancewidth}) and 
the shift in Eq.~(\ref{resonanceshift}) depend only on the product 
$W(r)\phi_\mathrm{res}(r)$. Consequently, for a minimal description of the 
resonance enhanced scattering we only need to specify 
$W(r)\phi_\mathrm{res}(r)$ in terms of two parameters to recover the magnetic 
field dependence of the scattering length in Eq.~(\ref{aofB}) and its 
relationship with the binding energy of the highest excited vibrational 
state.  A derivation beyond the scope of this paper shows that, once the width 
of the resonance is known experimentally, a determination of the resonance 
shift does not require a full solution of the coupled channel two-body 
Schr\"odinger equation with a realistic potential matrix. It turns out that 
Eq.~(\ref{resonanceshift}) is excellently approximated by \cite{Julienne89} 
\begin{equation}
  B_0-B_\mathrm{res}=(\Delta B)\frac{\frac{a_\mathrm{bg}}{\overline{a}}
    \left(1-\frac{a_\mathrm{bg}}{\overline{a}}\right)}
  {1+\left(1-\frac{a_\mathrm{bg}}{\overline{a}}\right)^2}.
  \label{magicformula}
\end{equation}
Here $\overline{a}$ is the mean scattering length of the background scattering 
potential $V_\mathrm{bg}(r)$, which is related to the van der Waals length in 
Eq.~(\ref{lvdW}) and Euler's $\Gamma$ function by \cite{GribakinFlambaum93}
\begin{equation}
  \overline{a}=\frac{1}{\sqrt{2}}
  \frac{\Gamma(3/4)}{\Gamma(5/4)}l_\mathrm{vdW}.
\end{equation}
This approximation is consistent with the treatment of the background 
scattering in terms of just the parameters $a_\mathrm{bg}$ and $l_\mathrm{vdW}$
in \ref{subsubsec:background}. 

\begin{figure}[htb]
  \includegraphics[width=\columnwidth,clip]{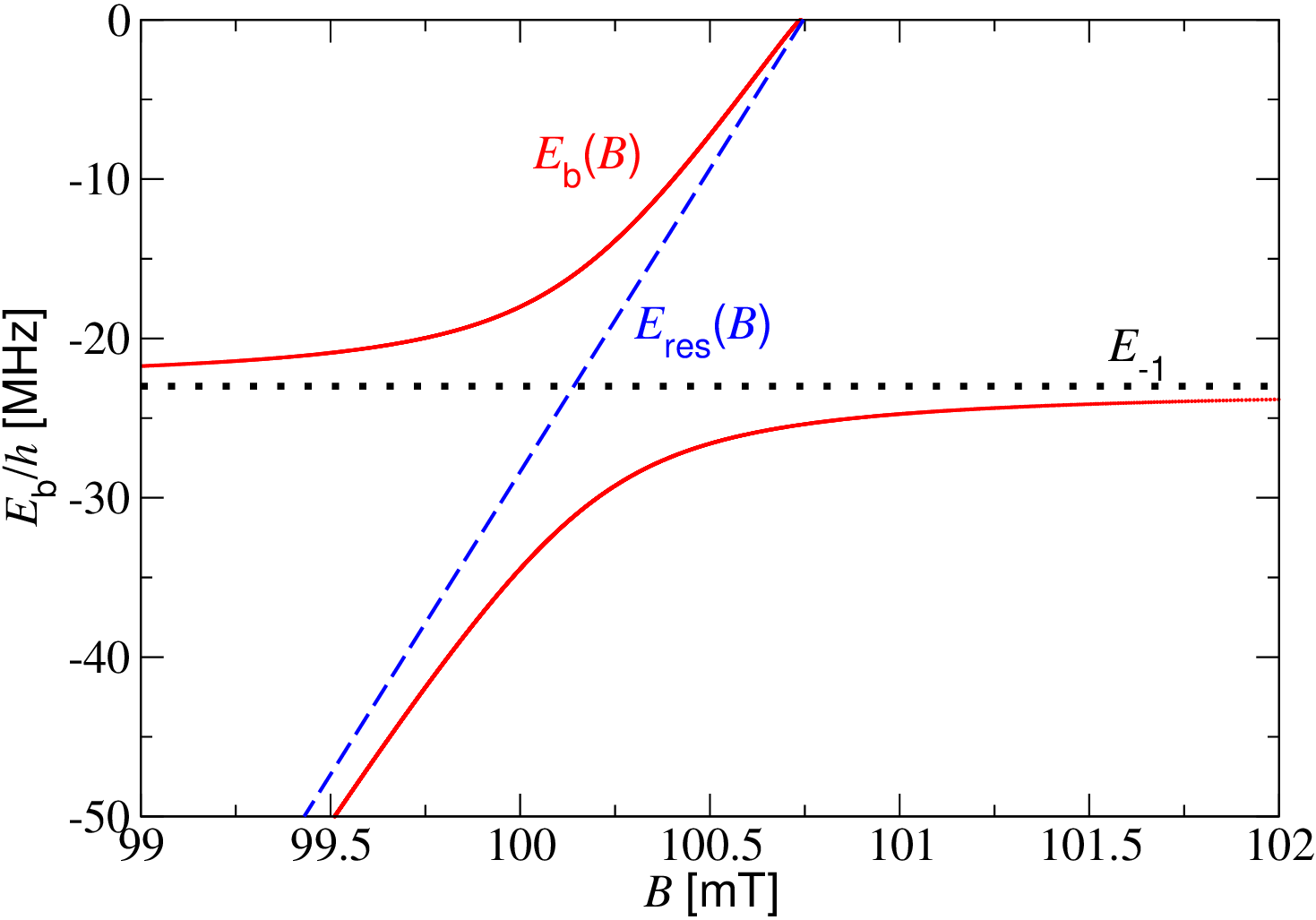}
  \caption{The magnetic field dependence of the binding energy of the 
    vibrational bound states of the two channel Hamiltonian in the appendix 
    (solid curves). On the low field side of the 
    resonance a new bound state $\phi_\mathrm{b}$ emerges at the resonance 
    position $B_0=100.74$ mT \cite{Volz03}
    whose binding energy we have denoted by 
    $E_\mathrm{b}(B)$. At magnetic field strengths asymptotically far from the 
    resonance, the highest excited two channel vibrational bound state becomes 
    identical with the highest excited single channel vibrational bound state 
    $\phi_{-1}(r)$ that is associated with $V_\mathrm{bg}(r)$ and whose 
    binding energy is denoted by $E_{-1}$ (dotted horizontal line). The 
    dashed line indicates the energy $E_\mathrm{res}(B)$ of the resonance 
    state.}
  \label{fig:EbofBtwochannel}
\end{figure}

In addition to the parameters $C_6$ and $a_\mathrm{bg}$ of the background 
scattering potential $V_\mathrm{bg}(r)$, a minimal description of the 
resonance enhanced scattering in a two channel Hamiltonian thus requires 
us to account for the width $(\Delta B)$ in Eq.~(\ref{resonancewidth}) and the 
shift $B_0-B_\mathrm{res}$ in Eqs.~(\ref{resonanceshift}) and 
(\ref{magicformula}), in terms of the interchannel coupling 
$W(r)\phi_\mathrm{res}(r)$, while the slope 
$\left[\frac{dE_\mathrm{res}}{dB}(B_\mathrm{res})\right]$
of the resonance in Eq.~(\ref{slope}) determines the component of the 
Hamiltonian in the closed channel that is strongly coupled to the open 
channel. For the 100 mT Feshbach resonance of $^{87}$Rb, four of the five 
parameters of the two channel Hamiltonian, 
i.e.~$C_6=4660$ a.u.~\cite{Roberts01} (1 a.u. = 0.095734 yJ nm$^6$), 
$a_\mathrm{bg}=100 \ a_\mathrm{Bohr}$ \cite{vanKempen02}
($a_\mathrm{Bohr}=0.052918\ $nm), 
$(\Delta B)=0.02$ mT and $B_0-B_\mathrm{res}=-0.006371$ mT, can be either
directly deduced from experiments \cite{Volz03} or from 
Eq.~(\ref{magicformula}). The only parameter that is not easily accessible is 
the slope of the resonance. We have obtained 
$\frac{1}{h}\left[\frac{dE_\mathrm{res}}{dB}(B_\mathrm{res})\right]=38$ MHz/mT 
from the binding energies in Fig.~\ref{fig:Eboverview}. 
In the appendix we provide the explicit form of the 
minimal two channel Hamiltonian in the pole approximation that we apply 
in the following to determine the dynamics of the adiabatic association of 
molecules. Figure \ref{fig:EbofBtwochannel} shows the magnetic field 
dependence of the binding energies as obtained from this minimal Hamiltonian. 
The low energy scattering properties 
of two asymptotically free $^{87}$Rb atoms in the open channel 
and the properties of the highest excited vibrational bound state are 
insensitive with respect to the details of the implementation of the five 
parameter two channel Hamiltonian. 

\subsection{Universal properties of near resonant bound states}
\label{subsec:universal}
\begin{figure}[htb]
  \hspace{0.1cm}
  \includegraphics[width=\columnwidth,clip]{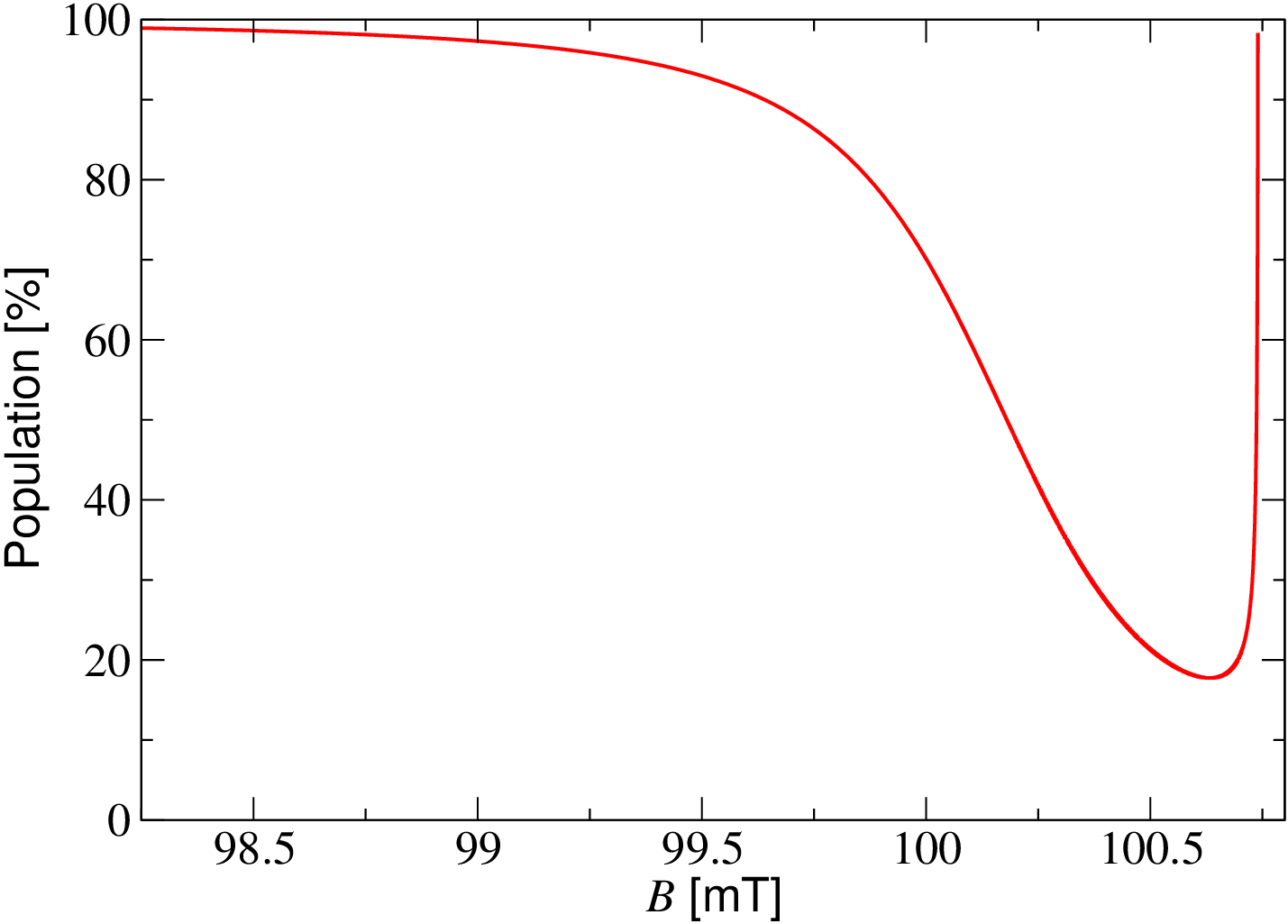}
  \caption{The population of the  open channel component of the 
    highest excited vibrational bound state $\phi_\mathrm{b}$ as a function 
    of the magnetic field strength on the low field side of the Feshbach 
    resonance. The population was determined from the minimal two channel 
    Hamiltonian in the appendix.}
  \label{fig:mixingcoefficient}
\end{figure}

We shall now focus on the properties of the highest excited vibrational bound 
state $\phi_\mathrm{b}$ on the low field side of the Feshbach resonance whose 
emergence at the resonance position $B_0=100.74$ mT causes the singularity of 
the scattering length (see Fig.~\ref{fig:EbofBtwochannel}). The vibrational 
state $\phi_\mathrm{b}$ is determined by its components in the  
open channel and in the closed channel strongly coupled to it, as given by 
Eq.~(\ref{phib}), while the binding energy is determined by 
Eq.~(\ref{determinationEb}). We note that the closed channel component 
$\phi_\mathrm{b}^\mathrm{cl}(r)$ in Eq.~(\ref{phib}) has the functional form 
of the resonance state $\phi_\mathrm{res}(r)$, which we have
normalised to unity. Figure \ref{fig:mixingcoefficient} shows the population
\begin{equation}
  4\pi \int_0^\infty r^2 dr \  
  \left|\phi_\mathrm{b}^\mathrm{bg}(r)\right|^2=
  \frac{\mathcal{N}_\mathrm{b}^2-1}{\mathcal{N}_\mathrm{b}^2}
\end{equation}    
of the  open channel component of the highest excited vibrational 
bound state $\phi_\mathrm{b}$ as a function of the magnetic field strength. 

At magnetic field strengths asymptotically far from the resonance, 
Fig.~\ref{fig:mixingcoefficient} shows that $\phi_\mathrm{b}$ is transferred
entirely into the bound state $\phi_{-1}(r)$ of 
$V_\mathrm{bg}(r)$, i.e.~the highest 
excited vibrational bound state in the absence of interchannel coupling. 
Figure \ref{fig:mixingcoefficient} also reveals that $\phi_\mathrm{b}$ is 
dominated by its component in the open channel also in a small region 
of magnetic field strengths in the close vicinity of the Feshbach resonance. 

\subsubsection{Universal binding energy}
We shall study the binding energy $E_\mathrm{b}$ that is determined by 
Eq.~(\ref{determinationEb}) in this small region of magnetic field strengths 
on the low field side of the Feshbach resonance. As the binding energy 
vanishes at the resonance position, these studies will involve the asymptotic 
behaviour of Eq.~(\ref{determinationEb}) in the limit $E_\mathrm{b}\to 0$. 
Inserting the resolvent identity \cite{Newton82}
\begin{equation}
  G_\mathrm{bg}(E_\mathrm{b})=G_\mathrm{bg}(0)-
  E_\mathrm{b}G_\mathrm{bg}(0)G_\mathrm{bg}(E_\mathrm{b})
\end{equation}
as well as Eqs.~(\ref{slope}) and (\ref{resonanceshift}) into 
Eq.~(\ref{determinationEb}) yields:
%\begin{widetext}
\begin{align}
  \nonumber
  E_\mathrm{b}=&\left[\frac{dE_\mathrm{res}}{dB}(B_\mathrm{res})\right]
  (B-B_0)\\
  &-E_\mathrm{b}
  \langle\phi_\mathrm{res}|WG_\mathrm{bg}(0)G_\mathrm{bg}(E_\mathrm{b})  
  W|\phi_\mathrm{res}\rangle.
  \label{Ebuniversal1}
\end{align}
%\end{widetext}
In accordance with Eq.~(\ref{Gbg}), the Green's functions $G_\mathrm{bg}(0)$ 
and $G_\mathrm{bg}(E_\mathrm{b})$ in Eq.~(\ref{Ebuniversal1}) can be 
decomposed into the complete orthogonal set of bound and continuum energy 
states associated with the background scattering potential $V_\mathrm{bg}(r)$. 
These decompositions yield:
%\begin{widetext}
\begin{align}
  \nonumber
  E_\mathrm{b}=&
  \left[\frac{dE_\mathrm{res}}{dB}(B_\mathrm{res})\right](B-B_0)\\
  &-E_\mathrm{b}
  \left[
    \int d\mathbf{p} \
    \frac{\left|\langle\phi_\mathrm{res}|W|\phi_\mathbf{p}^{(+)}\rangle
      \right|^2}
	 {-\frac{p^2}{m}
	   \left(E_\mathrm{b}-\frac{p^2}{m}\right)}+
	 \sum_v
	 \frac{\left|\langle\phi_\mathrm{res}|W|\phi_v\rangle\right|^2}
	      {-E_v\left(E_\mathrm{b}-E_v\right)}
	      \right].   
  \label{spectraldecompositionEb}
\end{align}
%\end{widetext}
Here the sum includes the indices $v=-1,-2,-3,\ldots$ of all vibrational bound 
states $\phi_v(r)$ associated with the potential $V_\mathrm{bg}(r)$. In the 
limit of vanishing binding energy $E_\mathrm{b}$, the momentum integral on the 
right hand side of Eq.~(\ref{spectraldecompositionEb}) is singular at $p=0$. 
As a consequence, the slowly varying matrix element 
$\langle\phi_\mathrm{res}|W|\phi_\mathbf{p}^{(+)}\rangle$ can be evaluated at 
$p=0$, while the sum over the vibrational bound states can be neglected. Using 
the general  relationship between the matrix element 
$\langle\phi_\mathrm{res}|W|\phi_0^{(+)}\rangle$ and the resonance width in 
Eq.~(\ref{resonancewidth}), the asymptotic form of the remaining momentum 
integral is then given by:
%\begin{widetext}
  \begin{align}
    \nonumber
    \int d\mathbf{p} \ \frac{\left|
      \langle\phi_\mathrm{res}|W|\phi_\mathbf{p}^{(+)}\rangle\right|^2}
	 {-\frac{p^2}{m}\left(E_\mathrm{b}-\frac{p^2}{m}\right)}
	 &\underset{E_\mathrm{b}\to 0}{\sim}
	 4\pi\int_0^\infty p^2dp \ 
	 \frac{\left|\langle\phi_\mathrm{res}|W|\phi_0^{(+)}\rangle\right|^2}
	      {-\frac{p^2}{m}\left(E_\mathrm{b}-\frac{p^2}{m}\right)}\\
	      &=\left[\frac{dE_\mathrm{res}}{dB}(B_\mathrm{res})\right]
	      \left(\Delta B\right)
	      a_\mathrm{bg}\sqrt{\frac{m}{\hbar^2\left|E_\mathrm{b}\right|}}.
  \end{align} 
%\end{widetext}
With this evaluation of the integral, Eq.~(\ref{spectraldecompositionEb}) 
can be solved for $E_\mathrm{b}$ and recovers the universal form
\begin{equation}
  E_\mathrm{b}(B)=-\frac{\hbar^2}{m[a(B)]^2}
  \label{Ebuniversal}
\end{equation}   
of the binding energy. 

\subsubsection{Universal bound states}
\begin{figure}[htb]
  \includegraphics[width=\columnwidth,clip]{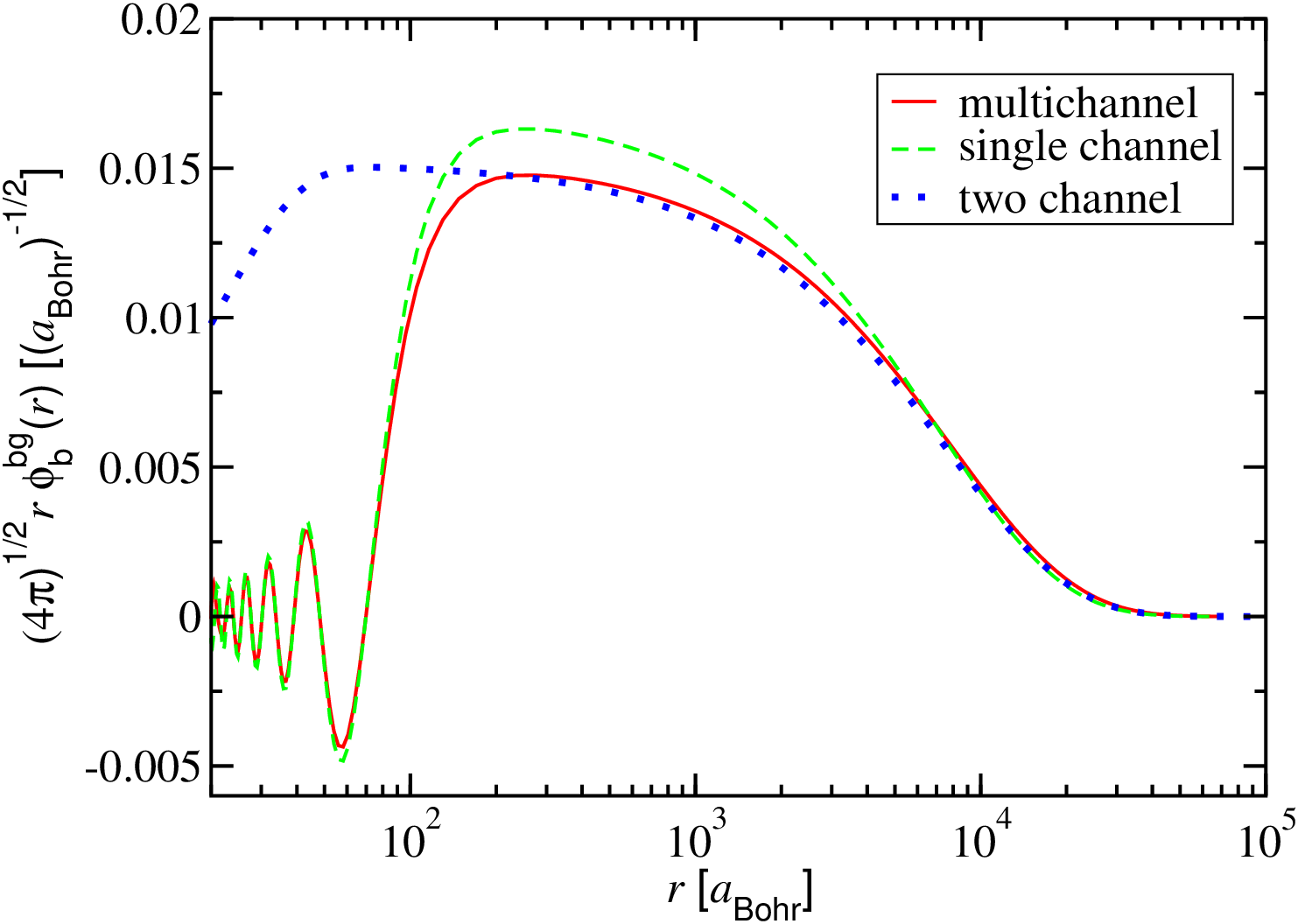}
  \caption{Radial wave functions associated with the component 
    $\phi_\mathrm{b}^\mathrm{bg}$ of the highest excited 
    vibrational bound state as determined from a full multichannel calculation 
    (solid curve) and from a single channel Hamiltonian (dashed curve) with 
    the background scattering potential modified in such a way that it 
    recovers the same scattering length $a(B)=7200 \ a_\mathrm{Bohr}$ as the 
    full multichannel Hamiltonian. The dotted curve shows the result of an 
    analogous calculation with the minimal two channel Hamiltonian in the 
    appendix. The components $\phi_\mathrm{b}^\mathrm{bg}$ of the two- and 
    multichannel wave functions agree at interatomic distances $r$ that are 
    large in comparison with the van der Waals length of about 
    $80 \ a_\mathrm{Bohr}$. The minimal two channel description does not 
    account for all the bound states of the exact background scattering 
    potential. The nodes of the multichannel wave function are therefore not 
    recovered. The differences between the single channel wave function and 
    the solid and dotted curves result from their different normalisation due 
    to the small but still relevant 10 \% admixture of the resonance state in 
    the multichannel wave functions. We note that the radial coordinate is 
    given on a logarithmic scale.}
  \label{fig:haloboundstates}  
\end{figure}

In accordance with Eq.~(\ref{phib}), near resonance the highest 
excited bound state is strongly dominated by its component in the asymptotic 
open channel, which is given by
\begin{equation}
  \phi_\mathrm{b}(r)=
  \phi_\mathrm{b}^\mathrm{bg}(r)=\frac{1}{\sqrt{2\pi a(B)}}
  \frac{e^{-r/a(B)}}{r}
  \label{universalwavefunction}
\end{equation}
at interatomic distances that are large in comparison with the van der Waals 
length. This wave function is extended far beyond the outer classical turning 
point 
$r_\mathrm{classical}=
\left[a(B)\left(2 l_\mathrm{vdW}\right)^2\right]^{1/3}$
of the background scattering potential $V_\mathrm{bg}(r)$, with a mean 
interatomic distance on the order of the scattering length: 
\begin{equation}
\langle r \rangle=
4\pi\int_0^\infty r^2 dr \ \left|\phi_\mathrm{b}(r)\right|^2 r=a(B)/2.
\label{bondlength}
\end{equation}
As a consequence, the coupling between the channels in the close vicinity of a 
Feshbach resonance can be treated, to an excellent approximation, as a 
perturbation of the background scattering potential $V_\mathrm{bg}(r)$. In 
fact, to describe the universal low energy scattering properties as well as 
the highest excited vibrational bound state, the whole potential matrix can 
be replaced by 
a single channel potential $V(B,r)$ that recovers the correct magnetic field 
dependence of the scattering length and the length scale associated with the 
long range van der Waals interaction between the atoms. This involves the 
replacement of the two-body Hamiltonian matrix in Eq.~(\ref{H2B2channel}) by 
the single channel Hamiltonian 
\begin{equation}
  H_\mathrm{2B}=-\frac{\hbar^2}{m}\nabla^2+V(B,r).
  \label{H2B1channel}
\end{equation}
Figure \ref{fig:haloboundstates} shows a comparison between the  
component $\phi_\mathrm{b}^\mathrm{bg}$ of the highest excited vibrational 
bound state as determined from a full multichannel calculation, and from a 
single channel Hamiltonian with the background scattering potential modified 
in such a way that it recovers the scattering length of the full multichannel 
Hamiltonian.

When the magnetic field strength approaches the resonance position from below, 
the spatial extent of $\phi_\mathrm{b}(r)$ becomes infinite and the bound 
state wave function becomes degenerate with the zero energy scattering state 
of two asymptotically free atoms in the  open channel. 
Consequently, by sweeping the magnetic field strength adiabatically across the 
resonance from negative to positive scattering lengths the zero energy 
scattering state is transferred smoothly into the bound state 
$\phi_\mathrm{b}(r)$. This is the key feature of the adiabatic association of 
molecules in ultracold atomic gases. We note that applying the sweep of the 
magnetic field strength in the direction from positive to negative scattering 
lengths is not suited to associate molecules because there is no energetically 
accessible vibrational bound state on the high field side of the Feshbach 
resonance. Both the association mechanism and its asymmetry in the direction 
of the sweep can be understood purely in terms of two-body considerations.

\subsection{Transition probability}
\label{subsec:2Bapproach}
We shall show in the following how the dynamics of the adiabatic association 
technique can be described in terms of the previous considerations. We assume 
that the magnetic field strength is swept linearly across the resonance 
position starting asymptotically far from the resonance on the high field side.
The two-body Hamiltonian in Eq.~(\ref{H2B2channel}) becomes explicitly 
time dependent through its dependence on the magnetic field strength, 
i.e.~$H_\mathrm{2B}=H_\mathrm{2B}(t)$. The probability for the adiabatic 
association of a pair of atoms in the initial state $|\Psi_i\rangle$ is 
determined in terms of the two-body time evolution operator 
$U_\mathrm{2B}(t_f,t_i)$ by:
\begin{equation}
  P_{fi}=\left|\langle\phi_\mathrm{b}(B_f)| 
  U_\mathrm{2B}(t_f,t_i)|\Psi_i\rangle\right|^2.
  \label{transitionprobability}
\end{equation}
Here $t_i$ and $t_f$ are the initial and final times before and after the 
linear ramp of the magnetic field strength, respectively, while 
$\phi_\mathrm{b}(B_f)$ is the highest excited vibrational bound state of the 
two-body Hamiltonian $H_\mathrm{2B}(t_f)$ at the final magnetic field strength 
$B(t_f)$. The time evolution operator is determined by the Schr\"odinger 
equation:
\begin{equation}
  i\hbar\frac{\partial}{\partial t} 
  U_\mathrm{2B}(t,t_i)=H_\mathrm{2B}(t)U_\mathrm{2B}(t,t_i).
  \label{SEU2B}
\end{equation}
The transition amplitude in Eq.~(\ref{transitionprobability}) for the minimal 
two channel Hamiltonian in the appendix can be 
obtained with methods similar to those applied to the determination of the 
single channel two-body time evolution operator in Ref.~\cite{KGB03}. 
A transition amplitude similar to that in Eq.~(\ref{transitionprobability}) 
serves as an input to the microscopic quantum dynamics approach to the 
association of molecules in a trapped dilute Bose-Einstein condensate that 
is presented in Subsection \ref{subsec:microscopicquantumdynamics}. 

We shall use Eq.~(\ref{transitionprobability}) to determine the probability 
for the association of two atoms in the ground state of a large box of 
volume $\mathcal{V}$ that is later taken to infinity. In the homogeneous limit 
the appropriate initial state in Eq.~(\ref{transitionprobability}) is given by 
the product state
\begin{equation}
  |\Psi_i\rangle=|0,\mathrm{bg}\rangle\sqrt{\frac{(2\pi\hbar)^3}{\mathcal{V}}},
  \label{Psiinitialbox}
\end{equation}  
where $|0\rangle$ is the isotropic zero momentum plane wave of the relative 
motion of two atoms in free space and $|\mathrm{bg}\rangle$ denotes their 
internal state in the asymptotic  open channel. The factor 
$\sqrt{(2\pi\hbar)^3/\mathcal{V}}$ in Eq.~(\ref{Psiinitialbox}) provides the 
appropriate normalisation. In the homogeneous limit the product 
$P_{fi}\mathcal{V}$ is independent of the volume $\mathcal{V}$.

\begin{figure}[htb]
  \includegraphics[width=\columnwidth,clip]{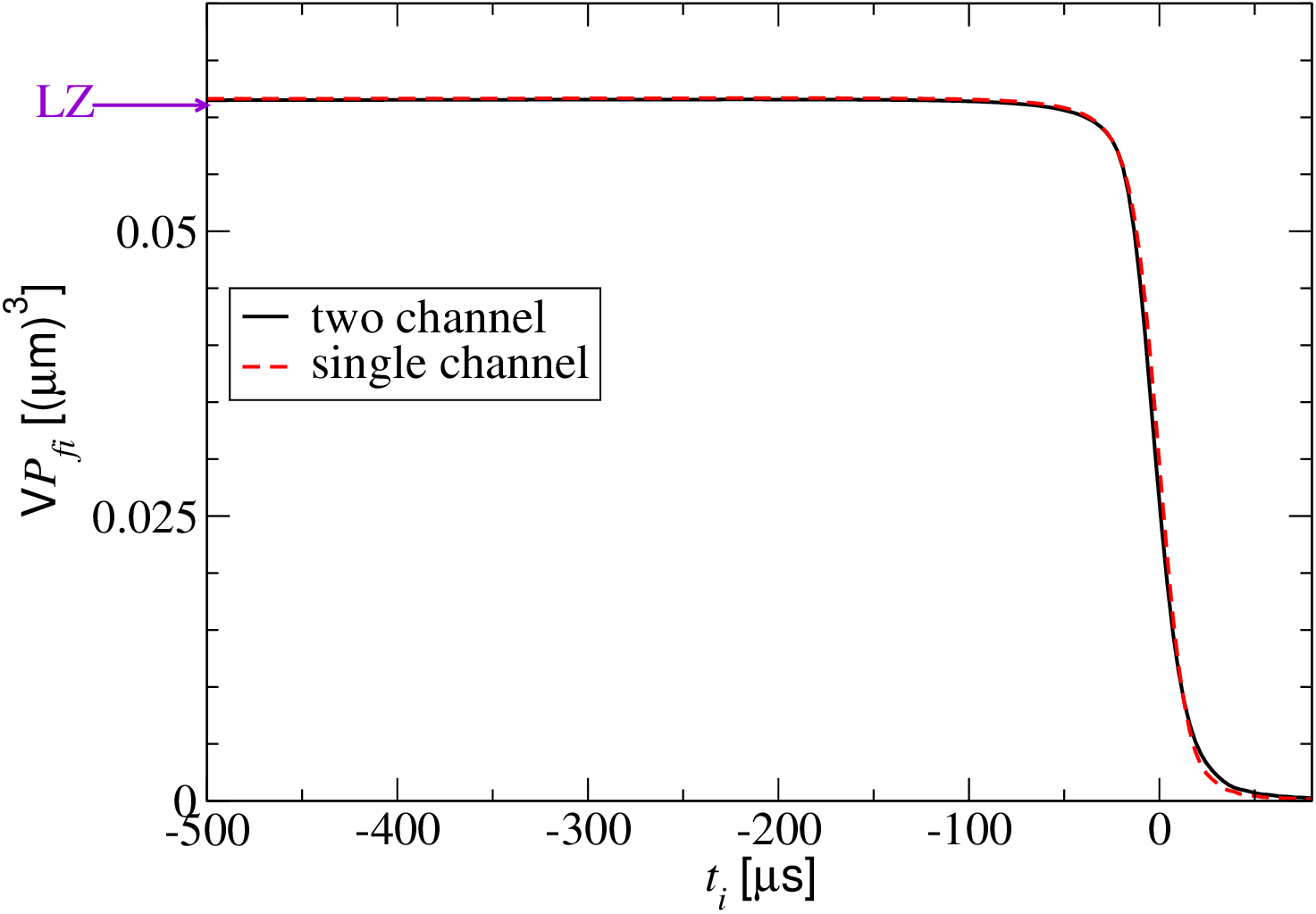}
  \caption{The product $P_{fi}\mathcal{V}$ as a function of the initial time 
    $t_i$ for two $^{87}$Rb atoms in the ground state of a large box with the 
    volume $\mathcal{V}$. The solid curve shows results based on the two 
    channel approach of the appendix, while the 
    dashed curve corresponds to an analogous calculation with a single 
    channel Hamiltonian that properly accounts for the width of the resonance. 
    The final time in these calculations is $t_f=80\ \mu$s; $t_i=0$ 
    corresponds to the initial time at which the linear ramp of the magnetic 
    field strength starts at the position of the Feshbach resonance.
    In the case of $t_i<0$ the linear ramp crosses the resonance position 
    from above. The horizontal arrow indicates the asymptotic Landau-Zener 
    prediction for $P_{fi}\mathcal{V}$ as obtained from Eq.~(\ref{deltaLZ}) in 
    Subsection \ref{subsec:LandauZener}.}
  \label{fig:transprob}
\end{figure}

Figure \ref{fig:transprob} shows the product $P_{fi}\mathcal{V}$ as a function 
of the initial time $t_i$ for a linear sweep of the magnetic field strength 
across the 100.74 mT Feshbach resonance of $^{87}$Rb with a ramp speed of 
0.1 mT/ms. The calculations were performed on the basis of 
Eq.~(\ref{transitionprobability}) with the two channel approach of the
appendix (solid curve) and with a single channel 
Hamiltonian (dashed curve) that properly accounts for the width of the 
resonance. Although the detailed time evolution is slightly different in the 
single and two channel case the transition probabilities are virtually equal 
once the ramp starts sufficiently far outside the width of the resonance. 
We shall show in Subsection \ref{subsec:LandauZener} how this independence of 
the final molecular population from the details of the two-body description 
can be derived in a systematic way.  This derivation will reveal that in the 
limits $t_i\to-\infty$ and $t_f\to\infty$ the product $P_{fi}\mathcal{V}$ 
depends only on the atomic mass $m$, the background scattering length 
$a_\mathrm{bg}$, the width of the resonance $(\Delta B)$, and the ramp speed. 

\subsection{Landau-Zener approach}
\label{subsec:LandauZener}
An intuitive extension of the two-body dynamics to the adiabatic association 
of molecules in a dilute Bose-Einstein condensate has been 
developed by Mies {\em et al.}
\cite{Mies00}. This approach is based on the time dependent 
two-body Schr\"odinger equation 
\begin{equation}
  i\hbar\frac{\partial}{\partial t}\Psi(t)=H_\mathrm{2B}(t)\Psi(t)
  \label{SEtimedependent}
\end{equation}
in the spherically symmetric harmonic potential of an optical atom trap
\cite{opticaltrap}. 
The confining atom trap modifies the potentials in Eq.~(\ref{H2B2channel}) to 
$V_\mathrm{bg}(r)\to V_\mathrm{bg}(r)+\frac{1}{2}\frac{m}{2}
\omega_\mathrm{ho}^2 r^2$ and $V_\mathrm{cl}(B,r)\to V_\mathrm{cl}(B,r)+
\frac{1}{2}\frac{m}{2}\omega_\mathrm{ho}^2 r^2$, where $\omega_\mathrm{ho}$ is 
the angular trap frequency. To extend the binary dynamics to the association 
of molecules in a dilute Bose-Einstein condensate, Mies {\em et al.}~have 
formulated Eq.~(\ref{SEtimedependent}) in terms of a basis set expansion with 
respect to the single channel energy states in the open channel 
and the closed channel strongly coupled to it. 
In this subsection, we shall develop an improved version of this approach. 

\subsubsection{Two-body configuration interaction approach}
We label the spherically symmetric energy states associated 
with the background scattering by the vibrational quantum numbers 
$v=0,1,2,\ldots$. These energy states fulfil the stationary Schr\"odinger 
equation
\begin{equation}
  \left[-\frac{\hbar^2}{m}\nabla^2+V_\mathrm{bg}(r)\right]\phi_v(r)=
  E_v\phi_v(r).
\end{equation}
In contrast to the free space continuum energy states 
$\phi_\mathbf{p}^{(+)}(\mathbf{r})$ in Eq.~(\ref{BCphipplus}), the vibrational 
trap states $\phi_v(r)$ are confined in space and we shall assume them to be 
unit normalised. In the atom trap the dissociation threshold energy in the 
open channel is given by $E_0$. In analogy to 
Subsection \ref{subsec:universal} we will label the vibrational bound states 
$\phi_v(r)$ below this threshold by negative quantum numbers 
$v=-1,-2,-3,\dots$. For realistic atom traps and background scattering 
potentials, we can presuppose the condition 
$|E_{-1}|\gg \hbar\omega_\mathrm{ho}$ to be fulfilled. 
As a consequence, the vibrational bound states below the dissociation 
threshold are hardly modified by the trapping potential. We shall assume 
furthermore that the configuration of the atoms in the 
closed channel that is strongly coupled to the  open channel,
is restricted to the resonance state $\phi_\mathrm{res}(r)$. 
This assumption is analogous to the pole approximation in 
Eq.~(\ref{poleapproximation}). The basis set expansions for the  
open channel and the closed channel components of $\Psi(t)$ are then given by 
\begin{align}
  \Psi_\mathrm{bg}(r,t)&=\sum_v\phi_v(r)C_v(t)\\
  \Psi_\mathrm{cl}(r,t)&=\phi_\mathrm{res}(r)C_\mathrm{res}(t), 
\end{align}
respectively. The expansion coefficients $C_v(t)$ and $C_\mathrm{res}(t)$ 
are determined by Eq.~(\ref{SEtimedependent}) in terms of the equivalent 
infinite set of coupled configuration interaction (CI) equations:
\begin{align}
  \label{CIbg}
  i\hbar\dot{C}_v(t)&=E_vC_v(t)+\langle\phi_v|W|\phi_\mathrm{res}\rangle 
  C_\mathrm{res}(t),\\
  i\hbar\dot{C}_\mathrm{res}(t)&=E_\mathrm{res}(B(t))C_\mathrm{res}(t)+
  \sum_v\langle\phi_\mathrm{res}|W|\phi_v\rangle C_v(t).
  \label{CIcl}
\end{align} 
These equations determine the exact dynamics of the adiabatic association of 
two atoms in the confining potential of an atom trap.  

\subsubsection{Many-body configuration interaction approach}
There are two essential phenomena that should be taken into account to extend 
the two-body CI approach to the many-body physics of the adiabatic association 
of molecules in a dilute Bose-Einstein condensate. 
First, the many-body mean field 
interactions can cause the size of the atom cloud to occupy a much larger 
volume than a single atom in the harmonic trapping potential. This size is 
determined by the nonlinearity parameter 
$k_\mathrm{bg}=Na_\mathrm{bg}/l_\mathrm{ho}$ of the Gross-Pitaevskii 
equation \cite{Dalfovo99}. Here $N$ is the number of atoms and 
$l_\mathrm{ho}=\sqrt{\hbar/(m\omega_\mathrm{ho})}$ is the oscillator length 
of the atom trap. We shall assume in the following that the Thomas Fermi 
condition $k_\mathrm{bg}\gg 1$ is fulfilled. This directly implies that the 
extension of the atom cloud is characterised by the Thomas Fermi radius:
\begin{equation}
  l_\mathrm{TF}=l_\mathrm{ho}\left(15 \ k_\mathrm{bg}\right)^{1/5}.
\end{equation}
This radius determines the mean kinetic energy per atom \cite{Fetter98}
\begin{equation}
  \frac{\langle E_\mathrm{kin}\rangle}{N}=\hbar\omega_\mathrm{ho}\frac{5}{2}
  \left(\frac{l_\mathrm{ho}}{l_\mathrm{TF}}\right)^2
  \mathrm{ln}\left(\frac{l_\mathrm{TF}}{1.2683 \ l_\mathrm{ho}}\right).
\end{equation}
The physical intuition underlying the Thomas Fermi limit relies upon the 
observation that the mean field potential cancels the trap potential at 
distances smaller than the Thomas Fermi radius. Under these assumptions 
the single atoms experience an effective flat potential rather than the 
harmonic potential of the atom trap. Motivated by this physical intuition, 
Mies {\em et al.} replaced the vibrational trap states in the matrix elements 
in Eqs.~(\ref{CIbg}) and (\ref{CIcl}) by those that correspond to a spherical 
box with a zero point energy of $\langle E_\mathrm{kin}\rangle/N$. 
The radius of this box is then determined by:  
\begin{equation}
  l_\mathrm{box}=l_\mathrm{TF}\left(\frac{2}{5}\right)^{1/2}
  \frac{\pi}{\sqrt{\mathrm{ln}
      \left(\frac{l_\mathrm{TF}}{1.2683 \ l_\mathrm{ho}}\right)}}.
\end{equation}
The second many-body phenomenon taken into account by Mies 
{\em et al.} is the macroscopic occupation of the lowest energy 
mode corresponding to the Bose-Einstein condensate. 
Neglecting the curve crossing between $E_\mathrm{res}(B)$ and the energy 
$E_{-1}$ of the highest excited bound state of the background scattering 
potential in Fig.~\ref{fig:EbofBtwochannel}, in a downward ramp 
of the magnetic field strength across the Feshbach resonance the prevailing 
asymptotic transition involves only the condensate mode and the resonance 
state. As each atom has $N-1$ atoms to interact with, the element of the 
potential matrix that is associated with this prevailing transition is 
enhanced by a factor of $\sqrt{N-1}$. 

Combining these results leads to a two level Rabi flopping model for the 
adiabatic association of molecules in a Bose-Einstein condensate:
\begin{align}
  \label{Rabibg}
  i\hbar\dot{C}_0(t)&=E_0C_0(t)+\frac{1}{2}\hbar\Omega^* C_\mathrm{res}(t),\\
  i\hbar\dot{C}_\mathrm{res}(t)&=E_\mathrm{res}(B(t))C_\mathrm{res}(t)+
  \frac{1}{2}\hbar\Omega C_0(t).
  \label{Rabicl}
\end{align}
Here the Rabi frequency is given by
\begin{equation}
  \Omega=2\sqrt{N-1}\frac{1}{\hbar}
  \sqrt{\frac{(2\pi\hbar)^3}{\frac{4\pi}{3}l_\mathrm{box}^3}}
  \langle\phi_\mathrm{res}|W|\phi_0^{(+)}\rangle.
  \label{Rabifrequency}
\end{equation}
We note that we have quoted the Rabi frequency in terms of the free space 
zero energy background scattering state $\phi_0^{(+)}(\mathbf{r})$ as 
introduced in Eq.~(\ref{BCphipplus}). The factor 
$\sqrt{(2\pi\hbar)^3/\left(\frac{4\pi}{3}l_\mathrm{box}^3\right)}$ 
provides the proper normalisation that accounts for the finite volume 
$\mathcal{V}=\frac{4\pi}{3}l_\mathrm{box}^3$ of the box. 
In accordance with Eq.~(\ref{resonancewidth}), the matrix element
$\langle\phi_\mathrm{res}|W|\phi_0^{(+)}\rangle$ in Eq.~(\ref{Rabifrequency})
can be expressed in terms of the resonance width $(\Delta B)$, the slope 
$\left[\frac{d E_\mathrm{res}}{d B}(B_\mathrm{res})\right]$
of the resonance and the background scattering length $a_\mathrm{bg}$.

When the magnetic field strength is swept linearly across the Feshbach 
resonance the energy of the resonance state changes linearly in time:
\begin{equation}
  E_\mathrm{res}(B(t))=E_0+
  \left[\frac{d E_\mathrm{res}}{d B}(B_\mathrm{res})\right]
  \left[\frac{d B}{d t}(t_\mathrm{res})\right](t-t_\mathrm{res}).
  \label{Eresoft}
\end{equation}
Here $t_\mathrm{res}$ is the time at which the energy of the resonance state 
crosses the dissociation threshold energy of the  open channel, 
i.e.~$B(t_\mathrm{res})=B_\mathrm{res}$ and 
$E_\mathrm{res}(B_\mathrm{res})=E_0$.  
Under the assumption of a linear Feshbach resonance crossing with the initial 
populations $|C_0(t_i)|^2=1$ and $|C_\mathrm{res}(t_i)|^2=0$, the final 
populations $|C_0(t_f)|^2$ and $|C_\mathrm{res}(t_f)|^2$ can be determined by
the Landau-Zener formulae:
\begin{align}
  \label{LZbg}
  |C_0(t_f)|^2&=e^{-2\pi\delta_\mathrm{LZ}},\\
  |C_\mathrm{res}(t_f)|^2&=1-e^{-2\pi\delta_\mathrm{LZ}},
  \label{LZcl}
\end{align}
in the limits $t_i\to -\infty$ and $t_f\to\infty$. Derivation of the 
asymptotic Landau-Zener populations requires much tedious calculation. 
Given the known general form of the exponent $\delta_\mathrm{LZ}$, 
however, a short calculation using Eq.~(\ref{resonancewidth}) reveals its 
simple dependence on the background scattering length $a_\mathrm{bg}$, 
the resonance width $(\Delta B)$ and the ramp speed 
$\left|\frac{d B}{d t}(t_\mathrm{res})\right|$:
\begin{equation}
  \delta_\mathrm{LZ}=
  \frac{\hbar|\Omega|^2}{4\left|\frac{d E_\mathrm{res}}{d B}
    (B_\mathrm{res})\right|\left|\frac{d B}{d t}(t_\mathrm{res})
    \right|}\\
  =\frac{(N-1)4\pi\hbar|a_\mathrm{bg}||\Delta B|}{\mathcal{V}m\left|
    \frac{d B}{d t}(t_\mathrm{res})\right|}.
  \label{deltaLZ}
\end{equation}
Although Eq.~(\ref{deltaLZ}) can be derived on the basis 
of the simple two level Rabi 
flopping model in Eqs.~(\ref{Rabibg}) and (\ref{Rabicl}), the two-body ($N=2$) 
Landau-Zener prediction is accurate even in applications to transition 
probabilities that include a continuum of energy levels above the 
dissociation threshold (see Fig.~\ref{fig:transprob}). In fact, the 
Landau-Zener coefficient for the asymptotic population of the resonance state
$\phi_\mathrm{res}(r)$ can be derived rigorously for an arbitrary number of 
linear curve crossings associated with a quasi continuum of two-body 
energy states \cite{Demkov68}. We note, however, that 
despite the universality of the asymptotic populations, a two level model 
is not suited to provide an adequate description of the intermediate states 
and their intermediate populations. From this viewpoint, the agreement between 
the asymptotic two- and multilevel descriptions is coincidental.

When applied to a gas of many atoms the linear two level Rabi flopping model 
in Eqs.~(\ref{Rabibg}) and (\ref{Rabicl}) yields analytic predictions on the 
efficiency of molecular production. The treatment of the Bose enhancement in 
the linear model, however, does not account for the depletion of the 
condensate mode in the course of the adiabatic association. This depletion 
can be accounted for in a straightforward way by replacing the initial number 
of condensate atoms $N$ by the actual number $N|C_0(t)|^2$ in the enhancement 
factor. This inclusion of the depletion modifies the linear Rabi flopping 
model in Eqs.~(\ref{Rabibg}) and (\ref{Rabicl}) to the nonlinear dynamic 
equations:
\begin{align}
  \label{NLRabibg}
  i\hbar\dot{C}_0(t)&=E_0C_0(t)+\frac{1}{2}
  \hbar\Omega^* |C_0(t)| C_\mathrm{res}(t),\\
  i\hbar\dot{C}_\mathrm{res}(t)&=E_\mathrm{res}(B(t))C_\mathrm{res}(t)+
  \frac{1}{2}\hbar\Omega |C_0(t)| C_0(t).
  \label{NLRabicl}
\end{align}
At final times $t_f\to\infty$ we shall interpret $N|C_0(t_f)|^2$ as the number 
of atoms in the remnant Bose-Einstein condensate and 
$N|C_\mathrm{res}(t_f)|^2$ as the number of atoms converted into molecules. 
The number of diatomic molecules produced in the adiabatic association is 
then given by $N|C_\mathrm{res}(t_f)|^2/2$. We show in Subsection 
\ref{subsec:comparison} that Eqs.~(\ref{NLRabibg}) and (\ref{NLRabicl}) 
significantly improve the predictions of the asymptotic Landau-Zener formulae 
in Eqs.~(\ref{LZbg}) and (\ref{LZcl}), in particular when the molecular 
production begins to saturate. 

\subsection{Dissociation of molecules}
\label{subsec:dissociation}
We note that the chemical bond of the diatomic molecules shifts the atomic 
spectral lines. This line shift can prevent the bound atoms from scattering 
light of probe lasers even in the case of the very weakly bound Feshbach 
molecules that are produced with the adiabatic association technique. 
As a consequence, many present day experimental molecular detection schemes 
rely upon the spatial separation of bound and free atoms in the cloud and 
the subsequent dissociation of the molecules (cf., e.g., Ref.~\cite{Duerr03}). 
The highest excited vibrational 
molecular bound state can be dissociated by crossing the Feshbach resonance 
from positive to negative scattering lengths which corresponds to an upward 
ramp in the case of the 100 mT Feshbach resonance of $^{87}$Rb. 
The crossing of the Feshbach resonance transfers the bound molecules into 
correlated pairs of atoms which can have a comparatively high 
relative velocity, depending on the ramp speed. In the following we shall 
characterise the dissociation energy spectra for the 100 mT Feshbach resonance 
of $^{87}$Rb, under the assumption that many-body phenomena can be neglected.
We expect this assumption to be accurate because the final continuum states
of the molecular fragments have a low occupancy. Consequently, phenomena 
related to Bose enhancement should be negligible. 

\subsubsection{General determination of dissociation energy spectra}
We shall consider linear upward ramps from the initial magnetic 
field strength $B_i$ at time $t_i$ across the Feshbach resonance to the final 
field strength $B_f$ at time $t_f$. Starting from the highest excited 
multichannel vibrational molecular bound state $\phi_\mathrm{b}(B_i)$ at the 
magnetic field strength $B_i$, the state of a pair of atoms at any time $t$ is 
determined in terms of the time evolution operator in Eq.~(\ref{SEU2B}) by:
\begin{equation}
  \Psi(t)=U_\mathrm{2B}(t,t_i)\phi_\mathrm{b}(B_i).
  \label{twobodystate}
\end{equation}
The dissociation energy spectrum is usually measured in a time of flight 
experiment that allows the fragments to evolve freely after the final time 
$t_f$ of the ramp. At any time $t$ after $t_f$ the two-body Hamiltonian is 
stationary, and the time evolution operator can be factorised into its 
contribution of the linear ramp of the magnetic field strength and a part 
that describes the subsequent relative motion of a pair of atoms:
\begin{equation}
U_\mathrm{2B}(t,t_i)=U_\mathrm{2B}(t-t_f)U_\mathrm{2B}(t_f,t_i).
\label{factorizationU2B}
\end{equation}
Here $U_\mathrm{2B}(t-t_f)$ is determined in terms of the stationary 
two-body Hamiltonian $H_\mathrm{2B}(B_f)$ at the final magnetic field strength 
$B_f$ by:
\begin{equation}
  U_\mathrm{2B}(t-t_f)=e^{-iH_\mathrm{2B}(B_f)(t-t_f)/\hbar}.
\end{equation}
We shall insert Eq.~(\ref{factorizationU2B}) into Eq.~(\ref{twobodystate})
and represent the time evolution operator of the relative motion of a 
pair of atoms after the time $t_f$ in terms of the complete set of 
multichannel energy states at the final magnetic field strength $B_f$. 
The state $\Psi(t)$ can then be decomposed as 
\begin{equation}
  \Psi(t)=\Psi_\mathrm{free}(t)+\Psi_\mathrm{bound}(t),
\end{equation}  
where $\Psi_\mathrm{free}(t)$ describes the dissociation into two
asymptotically free fragments, while $\Psi_\mathrm{bound}(t)$ describes those 
events in which the bound state $\phi_\mathrm{b}(B_i)$ is transferred into 
more tightly bound multichannel molecular vibrational states at the final 
magnetic field strength $B_f$. It is the continuum part 
$\Psi_\mathrm{free}(t)$ of the wave function 
$\Psi(t)$ that determines the measurable dissociation energy spectrum.  
In terms of the continuum multichannel energy states $\phi_\mathbf{p}(B_f)$ 
at the final magnetic field strength $B_f$ this is given by
%\begin{widetext}
\begin{align}
  \Psi_\mathrm{free}(t)= \int d\mathbf{p} \ \phi_\mathbf{p}(B_f)
  e^{-iE(t-t_f)/\hbar} \langle\phi_\mathbf{p}(B_f)|
  U_\mathrm{2B}(t_f,t_i)|\phi_\mathrm{b}(B_i)\rangle.
  \label{Psifree}
\end{align}
%\end{widetext}
Here $E=p^2/m$ is the energy of the relative motion of the fragments that 
corresponds to their relative momentum $\mathbf{p}$. From Eq.~(\ref{Psifree}) 
we deduce the probability of detecting a pair of atoms with a relative energy 
between $E$ and $E+dE$ to be:
\begin{equation}
  n(E)dE=p^2dp\int d\Omega_\mathbf{p} \ \left|\langle\phi_\mathbf{p}(B_f)|
  U_\mathrm{2B}(t_f,t_i)|\phi_\mathrm{b}(B_i)\rangle\right|^2.
  \label{dissociationspectrumgeneral}
\end{equation}
Here $d\Omega_\mathbf{p}$ denotes the angular component of $d\mathbf{p}$.

\subsubsection{Exact dissociation energy spectra}
\begin{figure}[htb]
  \includegraphics[width=\columnwidth,clip]{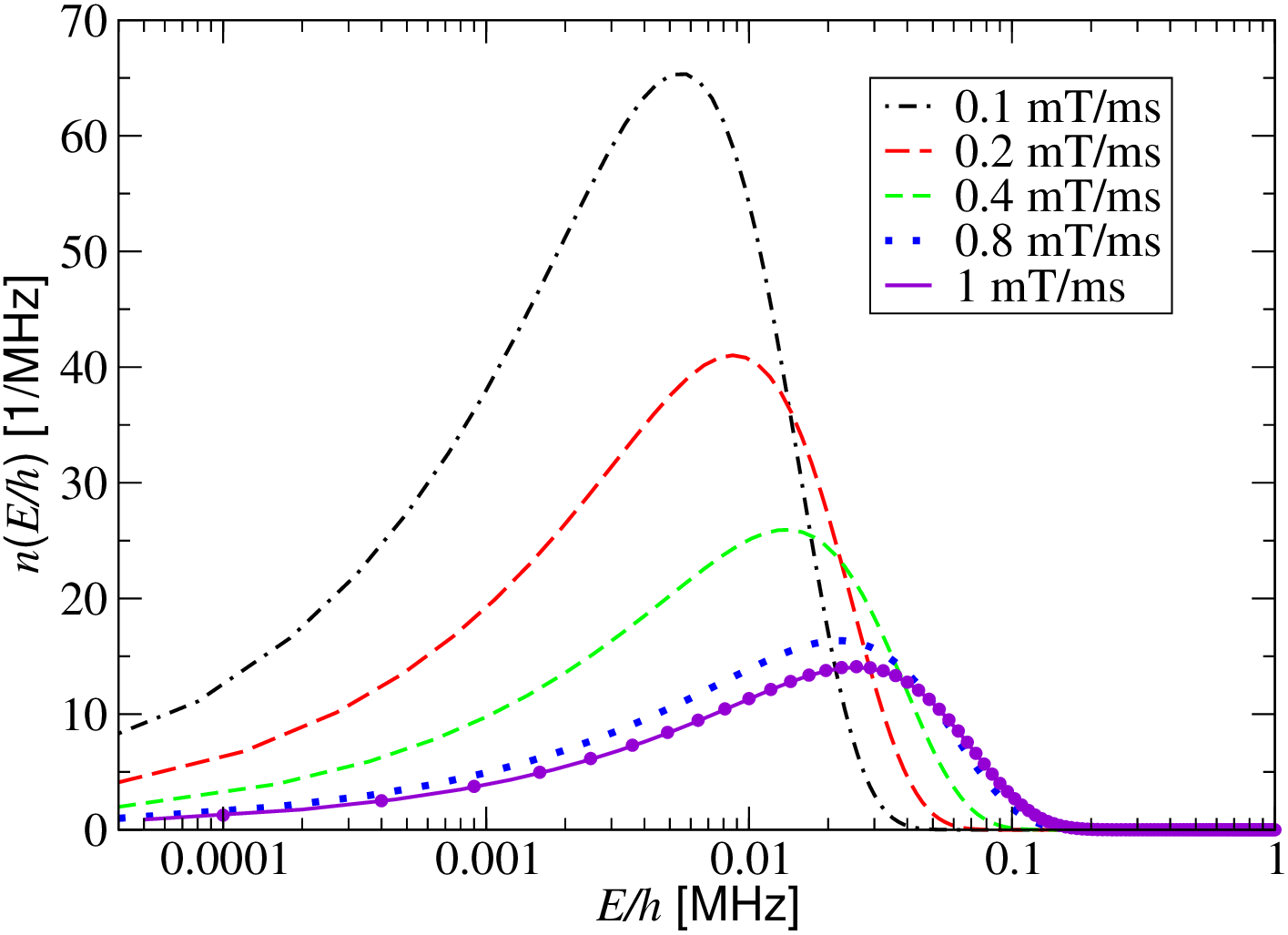}
  \caption{Dissociation spectra of the highest excited vibrational bound state
    of $^{87}$Rb$_2$ as a function of the relative energy of the fragments. The
    speeds of the upward ramps across the 100 mT Feshbach resonance were 
    varied between 0.1 mT/ms (dashed dotted curve) and 1 mT/ms (solid curve). 
    The spectra are rather insensitive to the range of magnetic field 
    strengths covered by the ramp; the solid curve indicates a 1 mT/ms ramp 
    with $B_i=100.7$ mT and $B_f=100.78$ mT, while the data indicated by the 
    dots on top of the solid curve correspond to a calculation for a 1 mT/ms 
    ramp with initial and final magnetic field strengths that are half way 
    closer to the resonance position of $B_0=100.74$ mT. We note that the 
    energies are given on a logarithmic scale.}
  \label{fig:dissociationspectra}
\end{figure}

Figure \ref{fig:dissociationspectra} shows the spectral density $n(E)$ for 
different ramp speeds as obtained from an exact solution of the Schr\"odinger
equation (\ref{SEU2B}) with the low energy two channel Hamiltonian in the
appendix. Although the low energy two-body Hamiltonian 
supports a comparatively tightly bound state at the high field side of the 
Feshbach resonance (see Fig.~\ref{fig:EbofBtwochannel}), the calculations do 
not indicate any transfer into this state for the realistic ramp speeds under 
consideration. Within this range of ramp speeds between 0.1 and 1 mT/ms, the 
spectra cover a broad range of energies of the relative motion of a pair of 
atoms that is much larger than the typical energy spread of a Bose-Einstein 
condensate. Consequently, the molecular fragments will be detected as a burst 
of correlated pairs of atoms moving radially outward from the original 
position of the molecular cloud.

\subsubsection{Dependence on the physical parameters of a Feshbach resonance}
In the following, we shall study in more detail the dependencies of the 
dissociation spectra in Fig.~\ref{fig:dissociationspectra} on the five 
physical parameters of a Feshbach resonance 
(cf.~Subsection \ref{subsec:energystates}), which completely characterise the 
resonance enhanced low energy collision physics. To this end, we shall neglect
the vibrational bound states of the background scattering potential and 
reformulate Eq.~(\ref{dissociationspectrumgeneral}) in terms of the energy 
states in the absence of interchannel coupling, in the formal asymptotic 
limits $t_i\to -\infty$ and $t_f\to \infty$. The idealising assumption 
that the background scattering potential does not support any vibrational 
bound states implies that the two channel Hamiltonian in 
Eq.~(\ref{H2B2channel}) supports just the initial vibrational bound state 
$\phi_\mathrm{b}(B_i)$ of Eq.~(\ref{dissociationspectrumgeneral}). In the 
formal limit $t_i\to-\infty$, i.e.~when the linear ramp of the magnetic field 
strength starts asymptotically far from the resonance position, 
$\phi_\mathrm{b}(B_i)$ is then identical to the closed channel resonance state 
$\phi_\mathrm{res}$. Furthermore, in accordance with Eqs.~(\ref{phipcl}),
(\ref{phipbg}) and (\ref{amplitude}), the final continuum state 
$\phi_\mathbf{p}(B_f)$ of Eq.~(\ref{dissociationspectrumgeneral}) is 
transferred into the energy state $\phi_\mathbf{p}^{(+)}$ of the background 
scattering [cf.~Eq.~(\ref{BCphipplus})], in the formal limit $t_f\to\infty$.
Under these assumptions, Eq.~(\ref{dissociationspectrumgeneral}) can be 
reformulated to be  
\begin{equation}
  n(E)dE=p^2dp\int d\Omega_\mathbf{p} \ 
  \left|
  \langle\phi_\mathbf{p}^{(+)},\mathrm{bg}|
  U_\mathrm{2B}(t_f,t_i)
  |\phi_\mathrm{res},\mathrm{cl}\rangle
  \right|^2.
  \label{dissociationspectrumasymptotic}
\end{equation}
Here $|\mathrm{bg}\rangle$ and $|\mathrm{cl}\rangle$ denote the internal 
states of an atom pair in the asymptotic open channel and the closed channel
strongly coupled to it, respectively. The asymptotic dissociation spectrum in 
Eq.~(\ref{dissociationspectrumasymptotic}) can then be determined from the 
results of Ref.~\cite{Demkov68} in analogy to the derivation of the  
Landau-Zener formulae of Subsection \ref{subsec:LandauZener}. This yields 
the analytic result for the spectral density \cite{Mukaiyama03}:
\begin{equation}
  n(E)=-\frac{\partial}{\partial E}
  \exp\left(-\frac{4}{3}\sqrt{\frac{mE}{\hbar^2}}\left|a_\mathrm{bg}\right|
  \frac{E|\Delta B|}{\hbar\left|\frac{dB}{dt}(t_\mathrm{res})\right|}\right).
  \label{finalspectraldensity}
\end{equation}

We note that the asymptotic energy density of the dissociation spectrum in 
Eq.~(\ref{finalspectraldensity}) depends, like the Landau-Zener coefficient 
in Eq.~(\ref{deltaLZ}), only on those physical parameters that determine the 
universal low energy scattering properties in the close vicinity of a Feshbach 
resonance (cf.~Subsection \ref{subsec:universal}), while the slope 
$\left[\frac{dE_\mathrm{res}}{dB}(B_\mathrm{res})\right]$ of the resonance
and the van der Waals dispersion coefficient $C_6$ do not contribute to
Eq.~(\ref{finalspectraldensity}). This corroborates the observation in 
Fig.~\ref{fig:transprob} that the relevant physics occurs in a small region 
of magnetic field strengths in which the modulus of the scattering length 
by far exceeds all the other length scales set by the binary interactions. 
In our applications to the 100 mT Feshbach resonance of $^{87}$Rb, the results 
obtained from Eq.~(\ref{finalspectraldensity}) are virtually indistinguishable 
from those of the exact calculations in Fig.~\ref{fig:dissociationspectra}. 
We note, however, that Eq.~(\ref{finalspectraldensity}) may not be applicable 
when the ramp of the magnetic field strength starts in the close vicinity of 
a Feshbach resonance. This situation may occur, for instance, in experimental 
applications involving the broad [$(\Delta B)=1.1$ mT] resonance of $^{85}$Rb 
at 15.5 mT \cite{Cornish00}. 
 
\subsubsection{Mean kinetic energy}
In the following, we shall study the mean kinetic energies 
\begin{equation} 
  \langle E_\mathrm{kin}\rangle 
  =\frac{1}{2}\int_0^\infty dE \ E n(E)
  \label{definitionmeankineticenergies}
\end{equation}
of the molecular fragments after the dissociation. These energies
characterise the speed of expansion of the gas of molecular fragments
before the detection in related experiments \cite{Duerr03}. 
We note that the kinetic energy of a single atom is 
$E_\mathrm{kin}=p^2/(2m)$, which is half the energy of the relative motion 
of a pair. Equations (\ref{finalspectraldensity}) and 
(\ref{definitionmeankineticenergies}) allow us to represent 
$\langle E_\mathrm{kin}\rangle$ in terms of physical parameters of a Feshbach 
resonance, of the ramp speed and of Euler's $\Gamma$ function:
\begin{equation} 
  \langle E_\mathrm{kin}\rangle =\frac{1}{3}
  \left[
    \frac{3}{4}\sqrt{\frac{\hbar^2}{m\left(a_\mathrm{bg}\right)^2}}
    \frac{\hbar\left|\frac{dB}{dt}(t_\mathrm{res})\right|}{|\Delta B|}
    \right]^{2/3}\Gamma(2/3).
  \label{asymptoticmeankineticenergies}
\end{equation}
We expect this prediction to be accurate provided that the initial and final 
magnetic field strengths of the linear ramp are asymptotically far from the 
resonance position.

\begin{figure}[htb]
  \includegraphics[width=\columnwidth,clip]{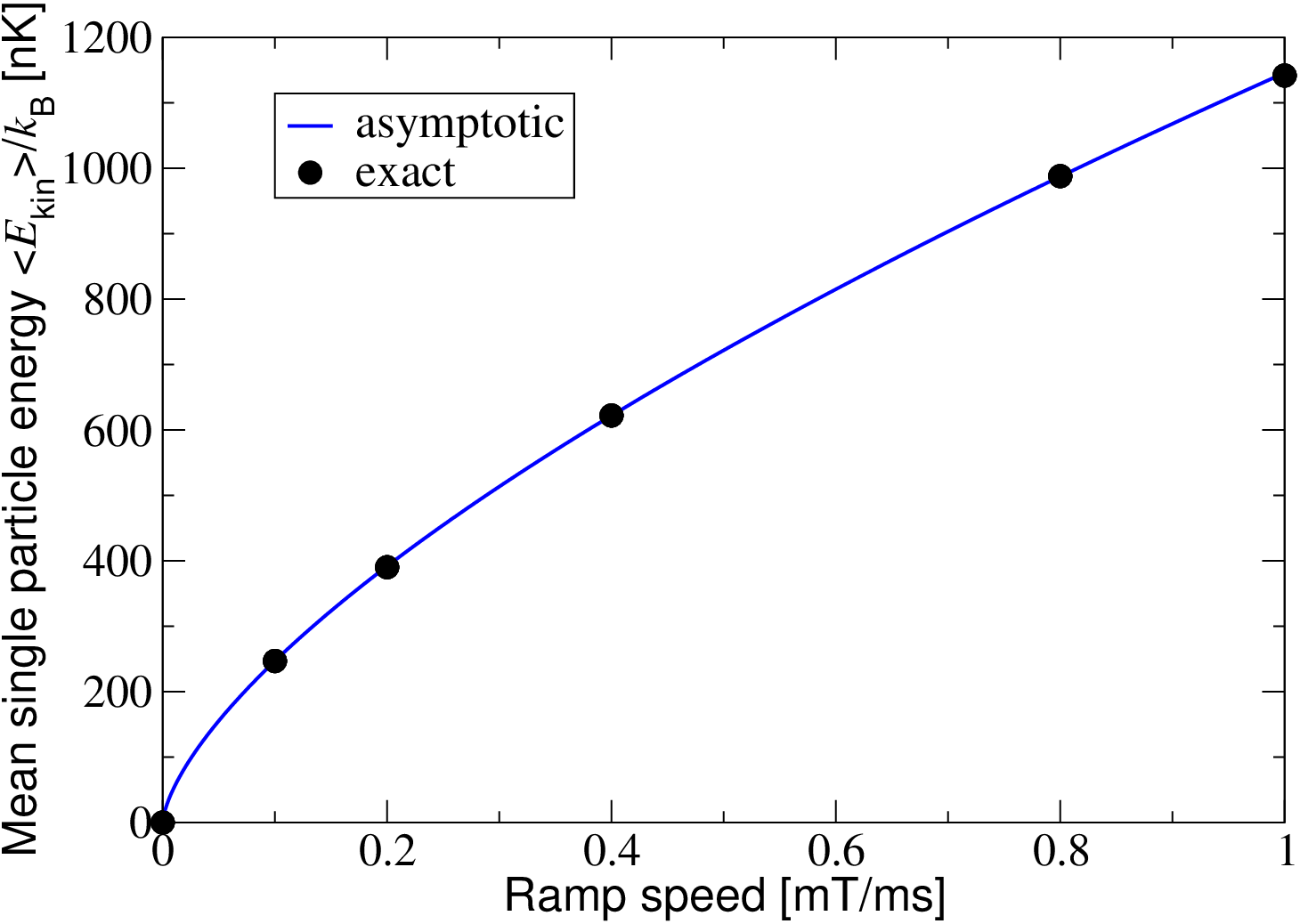}
  \caption{Mean single particle kinetic energies of the molecular fragments 
    associated with the exact dissociation spectra in 
    Fig.~\ref{fig:dissociationspectra} (circles) in dependence on the ramp 
    speed. The solid curve has been obtained from the asymptotic prediction
    in Eq.~(\ref{asymptoticmeankineticenergies}). The single particle kinetic 
    energy is half the energy of the relative motion of the fragments in 
    Fig.~\ref{fig:dissociationspectra}.}
  \label{fig:meandissociationenergies}
\end{figure}

Figure \ref{fig:meandissociationenergies} shows the strong dependence of the 
mean single particle kinetic energies on the ramp speed. The circles
indicate the mean energies of the exact dissociation energy spectra in
Fig.~\ref{fig:dissociationspectra}, while the solid curve has been obtained
from Eq.~(\ref{asymptoticmeankineticenergies}). Both approaches lead to
virtually the same predictions for the realistic ramp speeds and physical
parameters of the 100 mT Feshbach resonance of $^{87}$Rb.
Figure \ref{fig:meandissociationenergies} also reveals that ramp speeds
of less than about 0.03 mT/ms would be required to suppress the mean single 
particle energies of the fragments to a range below 
$\langle E_\mathrm{kin}\rangle/k_\mathrm{B}=100 \ \mathrm{nK}$
($k_\mathrm{B}=1.3806503\times 10^{-23}$ J/K), which would be 
close to the typical energy spread in a Bose-Einstein condensate
\cite{Dalfovo99}.

\section{Adiabatic association of molecules in a trapped Bose-Einstein 
condensate}
\label{sec:manybody}
The results in Subsection \ref{subsec:2Bapproach} allow us to rigorously treat 
the two-body physics of the adiabatic association. In this section we shall 
study the many-body physics of the molecular production in a trapped dilute 
Bose-Einstein condensate. Inhomogeneous Bose-Einstein condensates are subject 
to a rich spectrum of collective energy modes that depend sensitively on 
the number of condensate atoms and their binary interactions, as well as on 
the confining atom trap. The crossing of a singularity of the scattering 
length leads to a dramatic change of the intermediate collision physics 
that transfers a substantial fraction of the initially coherent atomic 
cloud into strongly correlated pairs of atoms in the highest excited 
vibrational bound state. This violent dynamics can be expected to couple 
all energy modes. A proper description of the complex interplay between the 
macroscopic collective behaviour and the microscopic binary collision 
physics requires a full many-body treatment. We shall provide an appropriate 
many-body description of the adiabatic association of molecules in a trapped 
dilute Bose-Einstein condensate and compare it to previous theoretical 
predictions at different levels of approximation. This will highlight in what
parameter regimes approximate descriptions are valid and to which 
physical observables they are applicable, as well as when we expect them to 
break down.
 
\subsection{Microscopic quantum dynamics approach}
\label{subsec:microscopicquantumdynamics}
The general many-body approach to the dynamics of atomic gases that we
shall apply has been derived in Refs.~\cite{KB02,KGB03}. This approach has 
been applied previously to several different physical situations that involve 
the production of correlated pairs of atoms in four wave mixing experiments 
with Bose-Einstein condensates 
\cite{KB02}, the determination of the mean field energy associated with 
three-body collisions \cite{TK02}, and the dynamics of atom molecule coherence 
\cite{KGB03,TKTGPSJKB03}
as well as Feshbach resonance crossing experiments with 
degenerate Bose gases of $^{85}$Rb and $^{23}$Na atoms
\cite{TKKG03}. The underlying method of cumulants \cite{KB02,Fricke96}
is based on the exact time dependent many-body Schr\"odinger equation. As the 
general technique has been derived before, we shall outline only those details 
specific to the adiabatic association of molecules in a Bose-Einstein 
condensate. 

\subsubsection{Multichannel approach}
We shall formulate the approach in terms of the full multichannel 
many-body Hamiltonian, which couples all the internal hyperfine states of 
the single atoms. We label the quantum numbers associated with the 
single atom energy states by Greek indices. In our applications to the low 
energy scattering in the vicinity of the $100$ mT Feshbach resonance of 
$^{87}$Rb, these indices are sufficiently characterised by the total angular 
momentum quantum number $F$ and its orientation quantum number $m_F$. In 
its second quantised form, the full many-body Hamiltonian then reads:   
\begin{widetext}
  \begin{align}
    H=\sum_\mu\int d\mathbf{x} \ \psi_\mu^\dagger(\mathbf{x})
    H^\mathrm{1B}_\mu(B)\psi_\mu(\mathbf{x})
    +\frac{1}{2}\sum_{\mu\nu\kappa\lambda}
    \int d\mathbf{x}d\mathbf{y} \ \psi^\dagger_\mu(\mathbf{x})
    \psi^\dagger_\nu(\mathbf{y})V_{\{\mu\nu\},\{\kappa\lambda\}}
    (\mathbf{x}-\mathbf{y})
    \psi_\kappa(\mathbf{x})\psi_\lambda(\mathbf{y}).
    \label{MBHamiltonian}
  \end{align}
\end{widetext}
Here the single particle annihilation and creation field operators satisfy 
the Bose commutation rules:
\begin{align}
  \left[\psi_\mu(\mathbf{x}),\psi_\nu^\dagger(\mathbf{y})\right]&=
  \delta_{\mu\nu}\delta(\mathbf{x}-\mathbf{y}),\\
  \left[\psi_\mu(\mathbf{x}),\psi_\nu(\mathbf{y})\right]&=0.
\end{align}
Furthermore, $H_\mu^\mathrm{1B}(B)$ is the one body Hamiltonian associated 
with the internal atomic state $\mu$ that contains the kinetic energy of the 
atom, the external potential of the optical atom trap, and the internal 
hyperfine energy $E_\mu^\mathrm{hf}(B)$:
\begin{equation}
  H^\mathrm{1B}_\mu(B)=-\frac{\hbar^2}{2m}\nabla^2+V_\mathrm{trap}+
  E_\mu^\mathrm{hf}(B).
\end{equation}
The hyperfine energy depends on the 
magnetic field strength through the Zeeman effect. The microscopic binary 
potential $V_{\{\mu\nu\},\{\kappa\lambda\}}(\mathbf{r})$ 
in Eq.~(\ref{MBHamiltonian}) is associated with the asymptotic incoming and 
outgoing binary scattering channels that can be labelled by the pairs of 
internal atomic quantum numbers $\{\kappa\lambda\}$ and $\{\mu\nu\}$, 
respectively. We note that in this general formulation of the many-body 
Hamiltonian all potentials associated with the asymptotic binary scattering
channels are chosen to vanish at infinite interatomic distances.

All physical properties of a gas of atoms are determined by correlation 
functions, i.e.~quantum expectation values 
(denoted by $\langle\cdot\rangle_t$) of normal ordered products of field 
operators for the quantum state at time $t$. The correlation functions of 
main interest in the adiabatic association of molecules involve the 
atomic mean field $\langle\psi_\mu(\mathbf{x})\rangle_t$, the anomalous 
average $\langle\psi_\nu(\mathbf{y})\psi_\mu(\mathbf{x})\rangle_t$, and the 
one-body density matrix 
$\langle\psi_\nu^\dagger(\mathbf{y})\psi_\mu(\mathbf{x})\rangle_t$. The 
dynamics of the correlation functions is determined by the many-body 
Schr\"odinger equation through an infinite hierarchy of coupled dynamic 
equations. The cumulant approach consists in transforming the coupled set 
of dynamic equations for the correlation functions into an equivalent but 
more practical infinite set of dynamic equations for noncommutative 
cumulants \cite{KB02,KGB03}. The transformed equations of motion for the 
cumulants can be truncated, at any desired degree of accuracy, in accordance 
with Wick's theorem of statistical mechanics
\cite{Fetter71}. The noncommutative cumulants that we 
shall consider in the following are the atomic mean field 
\begin{equation}
  \Psi_\mu(\mathbf{x},t)=\langle\psi_\mu(\mathbf{x})\rangle_t,
\end{equation} 
the pair function
\begin{equation}
  \Phi_{\mu\nu}(\mathbf{x},\mathbf{y},t)=
  \langle\psi_\nu(\mathbf{y})
  \psi_\mu(\mathbf{x})\rangle_t-\Psi_\mu(\mathbf{x},t)\Psi_\nu(\mathbf{y},t),
\end{equation}
and the density matrix of the noncondensed fraction 
\begin{equation}
  \Gamma_{\mu\nu}(\mathbf{x},\mathbf{y},t)=
  \langle\psi_\nu^\dagger(\mathbf{y})\psi_\mu(\mathbf{x})\rangle_t-
  \Psi_\mu(\mathbf{x},t)\Psi_\nu^*(\mathbf{y},t). 
\end{equation}
In the first order cumulant approach 
\cite{KB02,KGB03}
that we shall apply in this paper the 
atomic mean field and the pair function are determined by the coupled dynamic 
equations: 
\begin{widetext}
\begin{align}
  \label{meanfieldgeneral}
  i\hbar\frac{\partial}{\partial t}\Psi_\mu(\mathbf{x},t)=&H_\mu^\mathrm{1B}(B)
  \Psi_\mu(\mathbf{x},t)+
  \sum_{\nu\kappa\lambda}\int d\mathbf{y} \ \Psi_\nu^*(\mathbf{y},t)
  V_{\{\mu\nu\},\{\kappa\lambda\}}(\mathbf{x}-\mathbf{y})
  \left[
    \Phi_{\kappa\lambda}(\mathbf{x},\mathbf{y},t)+
    \Psi_\kappa(\mathbf{x},t)\Psi_\lambda(\mathbf{y},t)
    \right]\\
  i\hbar\frac{\partial}{\partial t}\Phi_{\mu\nu}(\mathbf{x},\mathbf{y},t)
  =&\sum_{\kappa\lambda}
  \left[
    H^\mathrm{2B}_{\{\mu\nu\},\{\kappa\lambda\}}(B)
    \Phi_{\kappa\lambda}(\mathbf{x},\mathbf{y},t)+
    V_{\{\mu\nu\},\{\kappa\lambda\}}(\mathbf{x}-\mathbf{y})
    \Psi_\kappa(\mathbf{x},t)\Psi_\lambda(\mathbf{y},t)
    \right].
  \label{pairfunctiongeneral}
\end{align}
\end{widetext}
Here $H_\mathrm{2B}(B)$ is the Hamiltonian matrix of two atoms whose matrix 
elements
\begin{equation}
  H_{\{\mu\nu\},\{\kappa\lambda\}}^\mathrm{2B}(B)=
  [H_\mu^\mathrm{1B}(B)+H_\nu^\mathrm{1B}(B)]
  \delta_{\kappa\mu}\delta_{\lambda\nu}+V_{\{\mu\nu\},\{\kappa\lambda\}} 
  \label{H2Bgeneral}
\end{equation}
involve all incoming and outgoing asymptotic binary scattering channels 
associated with the pairs of atomic indices $\{\kappa\lambda\}$ and 
$\{\mu\nu\}$, respectively. Given the solution of 
Eqs.~(\ref{meanfieldgeneral}) and (\ref{pairfunctiongeneral}), 
the density matrix of the noncondensed fraction is determined in terms of 
the pair function by \cite{KGB03}
\begin{equation}
\Gamma_{\mu\nu}(\mathbf{x},\mathbf{x}',t)=\sum_\kappa\int d\mathbf{y} \
\Phi_{\mu\kappa}(\mathbf{x},\mathbf{y},t)
\Phi^*_{\nu\kappa}(\mathbf{x}',\mathbf{y},t).
\label{Gammageneral}
\end{equation}
The general first order dynamic equations (\ref{meanfieldgeneral}), 
(\ref{pairfunctiongeneral}) and (\ref{Gammageneral}) not only strictly 
conserve the total number of atoms
\begin{equation}
  N=\sum_{\mu}\int d\mathbf{x} \ 
  \left[\Gamma_{\mu\mu}(\mathbf{x},\mathbf{x},t)+
    |\Psi_\mu(\mathbf{x},t)|^2
    \right]
  \label{numberconservationgeneral}
\end{equation}
at all times, but the explicit form of Eq.~(\ref{Gammageneral}) also ensures 
the crucial property of positivity of the one-body density matrix.

Equation (\ref{Gammageneral}) reveals that the density of the noncondensed 
fraction stems directly from the pair function. Consequently, the build up of 
pair correlations is the main source of atom loss from a Bose-Einstein 
condensate. The energy delivered by a time dependent homogeneous magnetic 
field can transfer a pair of condensate atoms either into the highest 
excited vibrational molecular bound state or into the quasi continuum 
of two-body energy states above the dissociation threshold. Not only the 
molecular association, but also the production of correlated pairs of atoms
in the scattering continuum
has been observed recently \cite{Donley02} as a burst of atoms  
ejected from the remnant condensate. As a time dependent homogeneous magnetic 
field delivers energy but no momentum to the gas, the centres of mass of all 
correlated bound and free pairs of atoms have the same momentum 
distribution as the initial Bose-Einstein condensate. In this sense, the 
molecules produced in the adiabatic association may be considered as a 
degenerate quantum gas. The number $N_\mathrm{b}$ of diatomic molecules in 
the state $\phi_\mathrm{b}$ is determined by counting the overlap of each pair 
of atoms with the multichannel molecular bound state \cite{Dollard73}. 
This relates the number of molecules to the two-body correlation function 
\begin{equation}
  G^{(2)}_{\mu\nu\kappa\lambda}
  (\mathbf{x},\mathbf{y};\mathbf{x}',\mathbf{y}')=
  \langle\psi^\dagger_\kappa(\mathbf{x}')\psi^\dagger_\lambda(\mathbf{y}')
  \psi_\nu(\mathbf{y})\psi_\mu(\mathbf{x})\rangle
\end{equation}
by \cite{KGB03}:
\begin{widetext}
  \begin{equation}
    N_\mathrm{b}=\frac{1}{2}\sum_{\mu\nu\kappa\lambda}
    \int d\mathbf{r} d\mathbf{r}' d\mathbf{R} 
    \left[\phi_\mathrm{b}^{\{\mu\nu\}}(\mathbf{r})\right]^* 
    G^{(2)}_{\mu\nu\kappa\lambda}
    \left(\mathbf{R}+\frac{\mathbf{r}}{2},\mathbf{R}-
    \frac{\mathbf{r}}{2};\mathbf{R}+\frac{\mathbf{r}'}{2},
    \mathbf{R}-\frac{\mathbf{r}'}{2}\right)
    \phi_\mathrm{b}^{\{\kappa\lambda\}}(\mathbf{r}').
    \label{Nbgeneral}
  \end{equation}
\end{widetext}
Here the spatial integration variables can be interpreted in terms of two-body
centre of mass and relative coordinates $\mathbf{R}=(\mathbf{x}+\mathbf{y})/2$ 
and $\mathbf{r}=\mathbf{x}-\mathbf{y}$, respectively. The number of correlated 
pairs of atoms in the scattering continuum can be deduced from the two-body 
correlation function in a similar way \cite{KGB03}
with the multichannel bound state wave function replaced by the 
continuum states. We note that Eq.~(\ref{Nbgeneral}) neither assumes any 
particular class of many-body states nor any approximation to the many-body 
Schr\"odinger equation. By expanding the two-body correlation function in 
Eq.~(\ref{Nbgeneral}) into cumulants and truncating the expansion in 
accordance with the first order cumulant approach, the density of bound
molecules can be represented in terms of a molecular mean field \cite{KGB03}:
\begin{widetext}
  \begin{equation}
    \Psi_\mathrm{b}(\mathbf{R},t)=\frac{1}{\sqrt{2}}\sum_{\mu\nu}\int 
    d\mathbf{r}
    \left[\phi_\mathrm{b}^{\{\mu\nu\}}(\mathbf{r})\right]^*
    \left[\Phi_{\mu\nu}(\mathbf{R},\mathbf{r},t)+
      \Psi_\mu\left(\mathbf{R}+\frac{\mathbf{r}}{2},t\right)
      \Psi_\nu\left(\mathbf{R}-\frac{\mathbf{r}}{2},t\right)\right].
    \label{psibgeneral}
  \end{equation}
\end{widetext}
Here we have introduced the centre of mass and relative coordinates 
$\mathbf{R}$ and $\mathbf{r}$
and represented the pair function in terms of these variables. The molecular 
mean field determines the density of diatomic molecules in the state 
$\phi_\mathrm{b}$ by 
$n_\mathrm{b}(\mathbf{R},t)=|\Psi_\mathrm{b}(\mathbf{R},t)|^2$. The molecular
mean field as well as the fraction of pairs of correlated atoms in the
scattering continuum are determined completely by the solution of the coupled 
equations (\ref{meanfieldgeneral}) and (\ref{pairfunctiongeneral}). 

\subsubsection{Two channel approach} 
The general form of the two-body Hamiltonian in Eq.~(\ref{H2Bgeneral}) 
with a realistic potential matrix allows us to describe the binary 
collision physics over a wide range of energies and magnetic field strengths, 
as indicated in Fig.~\ref{fig:Eboverview}. 
As the present applications involve only the adiabatic association of 
molecules in the vicinity of the 100 mT Feshbach resonance of $^{87}$Rb, 
we can restrict the asymptotic binary scattering channels to those that we 
have identified in Section \ref{sec:twobody}. We shall thus insert the 
two channel description of Section \ref{sec:twobody} into
Eqs.~(\ref{meanfieldgeneral}) and (\ref{pairfunctiongeneral}) and perform 
the pole approximation of Subsection \ref{subsec:energystates}. The relevant 
potentials then involve the background scattering potential $V_\mathrm{bg}(r)$
and the off diagonal matrix element $W(r)$ between the  open 
channel and the closed channel strongly coupled to it. In accordance with the 
pole approximation, the only configuration of the atoms in this 
closed channel is the diatomic resonance state $\phi_\mathrm{res}(r)$. 
Consequently, the atomic mean field is restricted to its component in the 
$(F=1,m_F=+1)$ state, which we shall denote simply by $\Psi(\mathbf{x},t)$.
According to Eq.~(\ref{meanfieldgeneral}),
the associated mean field dynamic equation is then given by
\begin{widetext}
\begin{align}
  %\nonumber
  i\hbar\frac{\partial}{\partial t}\Psi(\mathbf{x},t)=
  \left[
    -\frac{\hbar^2}{2m}\nabla^2+V_\mathrm{trap}(\mathbf{x})
    \right]\Psi(\mathbf{x},t)
  +\int d\mathbf{y} \ \Psi^*(\mathbf{y},t)
  \left(
  W(|\mathbf{x}-\mathbf{y}|)
  \Phi_\mathrm{cl}(\mathbf{x},\mathbf{y},t) 
  +V_\mathrm{bg}(|\mathbf{x}-\mathbf{y}|)
  \left[
    \Phi_\mathrm{bg}(\mathbf{x},\mathbf{y},t)+
    \Psi(\mathbf{x},t)\Psi(\mathbf{y},t)
    \right]
  \right)
  \label{Psi2ch}
\end{align}
\end{widetext}
The pair function has a component in the open channel and in the 
closed channel strongly coupled to it. In accordance with 
Eq.~(\ref{pairfunctiongeneral}), their coupled dynamic equations read:
\begin{widetext}
  \begin{align}
    i\hbar\frac{\partial}{\partial t}
    \left(
    \begin{array}{c}
      \Phi_\mathrm{bg}(\mathbf{x},\mathbf{y},t)\\
      \Phi_\mathrm{cl}(\mathbf{x},\mathbf{y},t)
    \end{array}
    \right)
    =H_\mathrm{trap}^\mathrm{2B}(t)
    \left(
    \begin{array}{c}
      \Phi_\mathrm{bg}(\mathbf{x},\mathbf{y},t)\\
      \Phi_\mathrm{cl}(\mathbf{x},\mathbf{y},t)
    \end{array}
    \right)+
    \left(
    \begin{array}{c}
      V_\mathrm{bg}(|\mathbf{x}-\mathbf{y}|)\\
      W(|\mathbf{x}-\mathbf{y}|)
    \end{array}
    \right)
    \Psi(\mathbf{x},t)
    \Psi(\mathbf{y},t).
    \label{Phi2ch}
  \end{align}
\end{widetext}
Here $H_\mathrm{trap}^\mathrm{2B}(t)$ is the general two channel two-body 
Hamiltonian [cf.~Eq.~(\ref{H2B2channel})] that includes the centre of mass 
kinetic energy of a pair of atoms as well as the confining harmonic potential 
of the atom trap. In the two channel formulation of the first order microscopic
quantum dynamics approach, Eqs.~(\ref{Psi2ch}) and (\ref{Phi2ch}) completely 
determine the dynamics of the coherent atomic condensate and the fraction of
correlated pairs of atoms.

For the purpose of numerical convenience, we shall solve the inhomogeneous 
linear Schr\"odinger equation (\ref{Phi2ch}) formally in terms of 
the complete two-body time evolution operator 
$U_\mathrm{trap}^\mathrm{2B}(t,\tau)$ 
[cf.~Eq.~(\ref{SEU2B})] that accounts for the centre of mass motion of the 
atoms and for the confining trap potential. We shall then insert the solution 
into the mean field equation (\ref{Psi2ch}) to eliminate the pair 
function. We shall assume that at the initial time $t_i$ at the start of the 
ramp the gas is well described by a dilute zero temperature 
Bose-Einstein condensate, so 
that initial two-body correlations are negligible. The resulting dynamic 
equation for the atomic mean field can then be expressed in terms of the 
time dependent two-body transition matrix  
\begin{widetext}
\begin{align}
  T_\mathrm{trap}^\mathrm{2B}(t,\tau)
  =\langle\mathrm{bg}|
  \left[
    V(t)\delta(t-\tau)
    +\frac{1}{i\hbar}\theta(t-\tau)
    V(t)U_\mathrm{trap}^\mathrm{2B}(t,\tau)V(\tau)
    \right]
  |\mathrm{bg}\rangle,
  \label{T2Btrap}
\end{align}
\end{widetext}
which involves the potential matrix
\begin{equation}
  V(t)=
  \left(
  \begin{array}{cc}
    V_\mathrm{bg} & W\\
    W             & V_\mathrm{cl}(B(t))
  \end{array}
  \right)
\end{equation}
and the two-body time evolution operator $U_\mathrm{trap}^\mathrm{2B}(t,\tau)$.
Here $|\mathrm{bg}\rangle$ denotes the internal state of a pair of atoms in
the asymptotic open scattering channel, and $\theta(t-\tau)$ is 
the step function, which yields unity at $t>\tau$ and zero elsewhere. A short 
calculation applying the methods derived in Refs.~\cite{KB02,KGB03} then 
shows that the elimination of the pair function in Eq.~(\ref{Psi2ch}) leads to 
a closed dynamic equation for the atomic mean field. Expressed in terms of 
the two-body transition matrix in Eq.~(\ref{T2Btrap}), this is given by:
\begin{widetext}
\begin{align}
  i\hbar\frac{\partial}{\partial t}\Psi(\mathbf{x},t)=
  \left[
    -\frac{\hbar^2}{2m}\nabla^2+V_\mathrm{trap}(\mathbf{x})
    \right]\Psi(\mathbf{x},t)
  +\int_{t_i}^\infty d\tau\int d\mathbf{y} d\mathbf{x}' d\mathbf{y}' \ 
  \Psi^*(\mathbf{y},t)
  T_\mathrm{trap}^\mathrm{2B}(\mathbf{x},\mathbf{y},t;
  \mathbf{x}',\mathbf{y}',\tau)
  \Psi(\mathbf{x}',\tau)
  \Psi(\mathbf{y}',\tau).
  \label{NLSgeneral}
\end{align}
\end{widetext}

In the case of a harmonic trap potential the two-body transition matrix in 
Eq.~(\ref{T2Btrap}) factorises into a centre of mass part and a contribution
from the relative motion of an atom pair. In the following, we shall assume 
the confinement of the atom trap to be sufficiently weak and the ramp speeds
to be sufficiently high that the time spent within the width of the resonance 
is much smaller than the trap periods in all spatial directions. 
The centre of mass motion of a pair of atoms then becomes negligible on this
time scale. The explicit form of Eq.~(\ref{T2Btrap}) also reveals that the 
time evolution operator of the relative motion of two atoms is evaluated 
only within the spatial range of the binary interaction potentials, where it 
is hardly modified by the presence of the atom trap. Consequently, the trap
potential can also be neglected in the two-body time evolution operator of 
the relative motion of an atom pair. Furthermore, as the atomic mean field 
is slowly varying on the length scales set by the binary interaction 
potentials, Eq.~(\ref{NLSgeneral}) becomes local in the spatial variables 
and acquires the form \cite{KB02,KGB03}:
%\begin{widetext}
\begin{align}
  \nonumber
  i\hbar\frac{\partial}{\partial t}\Psi(\mathbf{x},t)=& 
  \left[-\frac{\hbar^2}{2m}\nabla^2+V_\mathrm{trap}(\mathbf{x})\right]
  \Psi(\mathbf{x},t)\\
  &-\Psi^*(\mathbf{x},t)\int_{t_i}^\infty d\tau \ 
  \Psi^2(\mathbf{x},\tau)\frac{\partial}{\partial \tau}h(t,\tau).
  \label{NLSElocal}
\end{align}
%\end{widetext}
Here we have introduced the coupling function  
\begin{align}
  h(t,\tau)=\theta(t-\tau)(2\pi\hbar)^3\langle0,\mathrm{bg}|V(t)
  U_\mathrm{2B}(t,\tau)|0,\mathrm{bg}\rangle,
  \label{h2B}
\end{align} 
which depends on the time evolution operator $U_\mathrm{2B}(t,\tau)$ 
associated with the relative motion of an atom pair in free space, while  
$|0,\mathrm{bg}\rangle$ denotes the zero momentum plane wave state of the 
relative motion of two atoms in the asymptotic  open scattering 
channel.  

\subsubsection{Atomic condensate fraction}
\begin{figure}[htb]
\includegraphics[width=\columnwidth,clip]{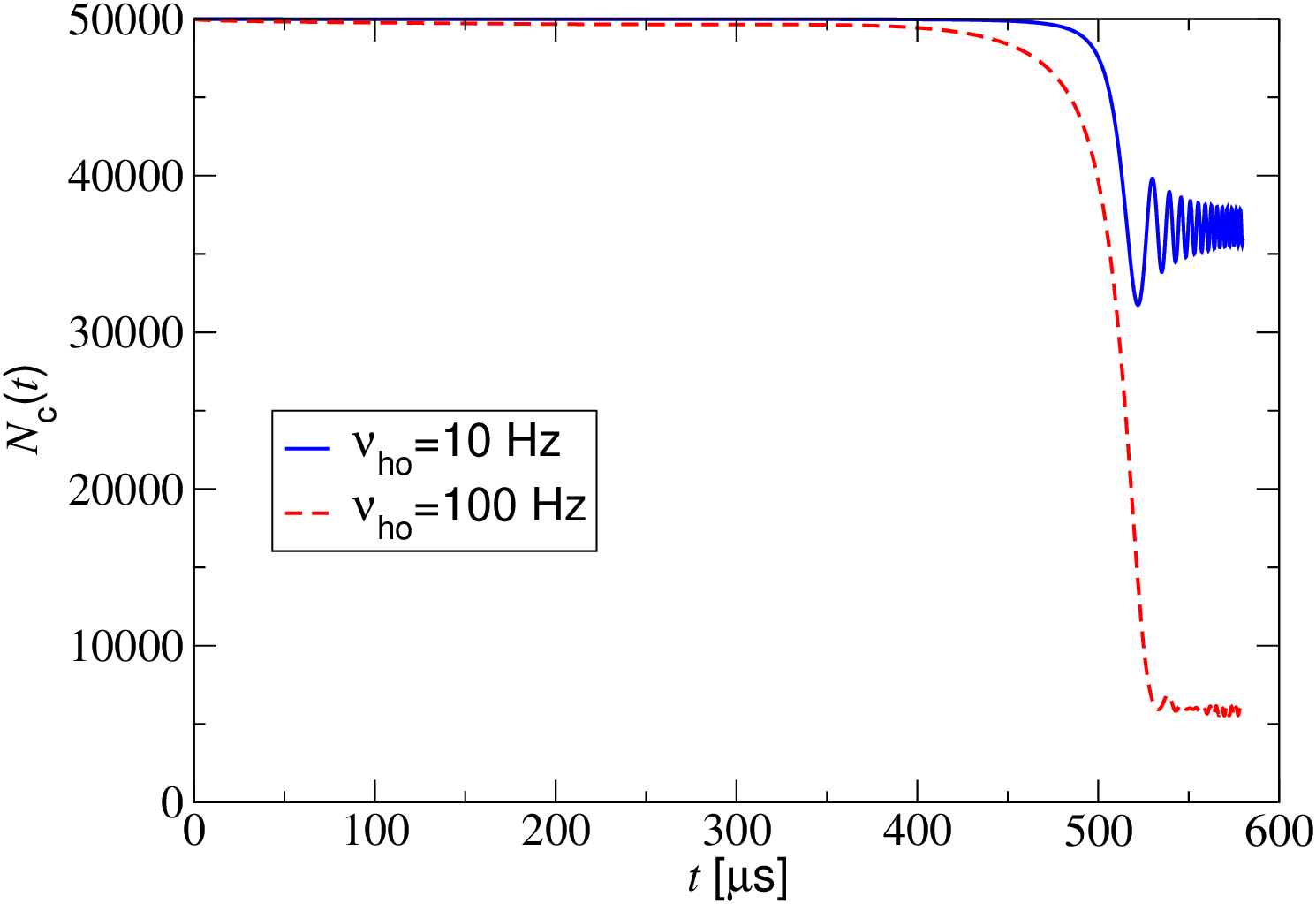}
\caption{The time evolution of the condensate fraction 
  ($N_\mathrm{c}(t)=\int d\mathbf{x} \ |\Psi(\mathbf{x},t)|^2$) for a 
  linear downward ramp of the magnetic field strength across the 100 mT 
  Feshbach resonance of $^{87}$Rb. The solid curve shows a calculation for a 
  weakly confining spherical atom trap with the frequency $\nu_\mathrm{ho}=10$ 
  Hz, while the dashed curve corresponds to a comparatively strongly confining 
  ($\nu_\mathrm{ho}=100$ Hz) spherical atom trap. The ramp speed in both 
  calculations is 0.1 mT/ms and the initial number of condensate atoms is 
  $N=50000$. The time at which the magnetic field strength 
  crosses the resonance position is $t=500 \ \mu$s.}
\label{fig:87RbNcoft}
\end{figure}

As we have shown in Subsections \ref{subsec:universal}, 
\ref{subsec:2Bapproach} and \ref{subsec:LandauZener}, the two-body time 
evolution of the main physical observables of interest in the adiabatic 
association of molecules is largely independent of the details of the 
implementation of the two-body Hamiltonian. The main requirement is that 
the two-body Hamiltonian properly accounts for the magnetic field dependence 
of the scattering length and the near resonant binding energy of the highest 
excited vibrational bound state. The differences between a single and a two 
channel treatment (cf.~Subsection \ref{subsec:2Bapproach}) in the dynamics of 
the atomic mean field in Eq.~(\ref{NLSElocal}) are marginal 
(cf., also, Ref.~\cite{TKKG03}). The following calculations have, therefore,
been performed with a single channel two-body Hamiltonian as introduced in 
Ref.~\cite{KGB03} and described in the appendix. 
In the course of our studies, we have solved Eq.~(\ref{NLSElocal}) for a 
variety of linear ramps of the magnetic field strength across the 100 mT 
Feshbach resonance of $^{87}$Rb in spherically as well as cylindrically 
symmetric harmonic atom traps. Figure \ref{fig:87RbNcoft} shows the time 
evolution of the number of condensate atoms 
$N_\mathrm{c}(t)=\int d\mathbf{x} \ |\Psi(\mathbf{x},t)|^2$ 
for linear downward ramps with a ramp speed of 0.1 mT/ms in 
spherically symmetric atom traps with the 
frequencies $\nu_\mathrm{ho}=100$ Hz and $\nu_\mathrm{ho}=10$ Hz. The 
atomic mean field at the initial time $t_i=0$ has been chosen in each 
calculation as the ground state of the Gross-Pitaevskii equation that 
corresponds to a dilute zero temperature Bose-Einstein condensate with 
$N=50000$ atoms and a scattering length of 
$a_\mathrm{bg}=100 \ a_\mathrm{Bohr}$. Figure \ref{fig:87RbNcoft} reveals 
that the loss of condensate atoms mainly occurs during the passage across the 
Feshbach resonance. In accordance with the higher local density, the losses 
of condensate atoms are more pronounced in the comparatively strongly 
confining 100 Hz atom trap. The pronounced oscillations immediately after the 
passage across the resonance ($t>500 \ \mu$s) indicate a rapid exchange 
between the condensed and the noncondensed phases of the gas. Exchange 
on these short time scales suggests that the crossing of a Feshbach resonance 
drives the gas into a highly excited nonequilibrium state.

\begin{figure}[htb]
\includegraphics[width=\columnwidth,clip]{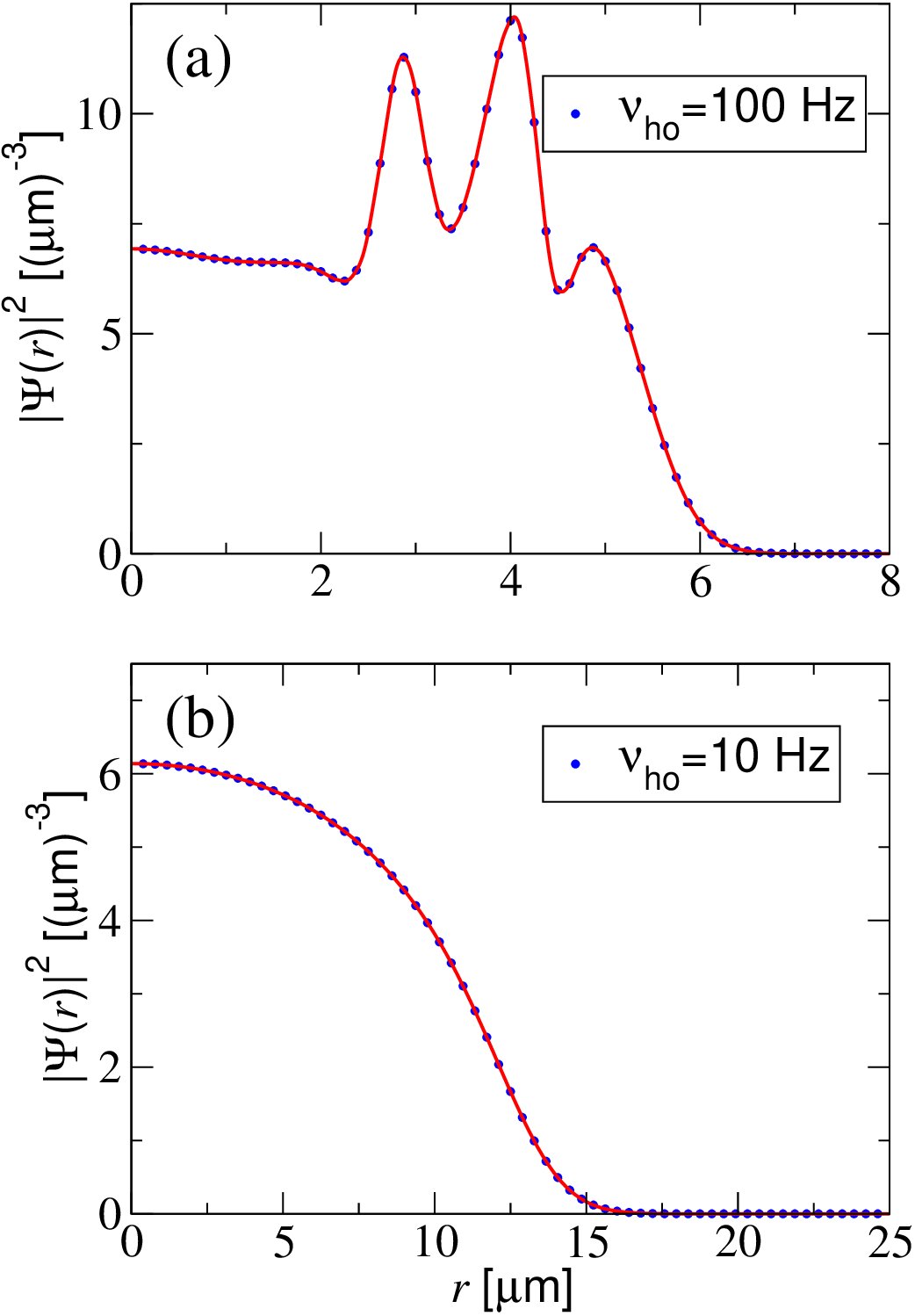}
\caption{Snapshots of the remnant condensate density after the crossing
  of the 100 mT Feshbach resonance of $^{87}$Rb in spherically symmetric 
  atom traps with
  the frequencies $\nu_\mathrm{ho}=100$ Hz (a) and $\nu_\mathrm{ho}=10$ Hz
  (b). The densities correspond to the calculations in 
  Fig.~\ref{fig:87RbNcoft} at the final time of the ramps.}
\label{fig:87Rbncfin1Gperms}
\end{figure}

Figure \ref{fig:87Rbncfin1Gperms} shows snapshots of the condensate density 
$n_\mathrm{c}(\mathbf{x},t)=|\Psi(\mathbf{x},t)|^2$ as a function of the 
radial coordinate 
$r=|\mathbf{x}|$ for the ramps and trap parameters in Fig.~\ref{fig:87RbNcoft} 
at the final time $t_f=580 \ \mu$s. The pronounced spatial variations of the
remnant condensate density indicate the simultaneous occupation of many 
collective energy modes. We note that these highly occupied excited modes are 
described by a coherent classical mean field and are therefore distinguished 
from the pairs of correlated atoms in the initially unoccupied scattering
continuum. In accordance with 
Eq.~(\ref{Gammageneral}), correlated pairs of atoms, as described by the pair 
function $\Phi(t)$, constitute the density of the noncondensed fraction.
 
\subsubsection{Noncondensed fraction}  
We shall study in the following how the final noncondensed fraction of the gas 
is distributed among the bound and free energy states of an atom pair. We 
shall focus on linear downward ramps of the magnetic field strength across the 
100 mT Feshbach resonance of $^{87}$Rb. Given
the solution of the dynamic equation (\ref{NLSElocal}) for the atomic 
mean field, 
the pair function $\Phi(t)$ is determined completely 
by the inhomogeneous linear 
two-body Schr\"odinger equation (\ref{Phi2ch}). We shall assume that the trap
is switched off immediately after the ramp so that the gas expands freely 
in space. Inserting the complete set of two-body energy states 
(cf.~Subsection \ref{subsec:energystates}) at the final magnetic field 
strength $B_f$ into Eq.~(\ref{Gammageneral}), and integrating with respect to 
the centre of mass coordinate 
$\mathbf{R}=(\mathbf{x}+\mathbf{y})/2$ yields the number of noncondensate
atoms \cite{KGB03}: 
\begin{align}
  \nonumber
  N_\mathrm{nc}=&\int d\mathbf{R}\int d\mathbf{p} \  
  \left|\langle\mathbf{R},\phi_\mathbf{p}(B_f)|\Phi(t_f)\rangle
  \right|^2\\
  &+\int d\mathbf{R} \ 
  \left|\langle\mathbf{R},\phi_\mathrm{b}(B_f)|\Phi(t_f)\rangle
  \right|^2.
  \label{Nnc}
\end{align}
Here we have presupposed that only the energetically accessible highest 
excited vibrational multichannel bound state $\phi_\mathrm{b}(B_f)$ on the low 
field side of the Feshbach resonance (cf.~Subsection \ref{subsec:universal}) 
will be populated in a downward ramp. We shall assume furthermore that the 
final magnetic field strength is sufficiently far from the Feshbach resonance 
that the gas is weakly interacting; i.e.~$n_\mathrm{peak}[a(B_f)]^3\ll 1$, 
where $n_\mathrm{peak}$ is the peak density of the gas, and $a(B_f)$ is the 
final scattering length. The second, factorised contribution to the molecular 
mean field on the right hand side of Eq.~(\ref{psibgeneral}) can then be 
neglected \cite{KGB03}. Consequently, the bound state contribution to the 
noncondensed density on the right hand side of Eq.~(\ref{Nnc}) determines 
the number of atoms associated to molecules. The contribution of the 
continuum part of the two-body energy spectrum yields the burst fraction, 
composed of correlated atoms with a comparatively high relative momentum 
$|\mathbf{p}|$ in initially unoccupied modes \cite{KGB03}. According to 
Eqs.~(\ref{numberconservationgeneral}) and (\ref{Nnc}), the number of 
condensate atoms, atoms associated to molecules and burst atoms then add up to 
the total number of atoms. We note that Eqs.~(\ref{numberconservationgeneral}) 
and (\ref{Nnc}) are also applicable in the small region of magnetic field 
strengths in the close vicinity of the Feshbach resonance, in which the gas 
becomes strongly interacting. The dilute gas parameter 
$n_\mathrm{peak}[a(B)]^3$ is then on the order of unity or larger. Under these 
conditions, however, a separation of the gas into bound and free atoms is 
physically meaningless \cite{TKTGPSJKB03,Molecularfraction}. 

\begin{figure}[htb]
  \includegraphics[width=\columnwidth,clip]{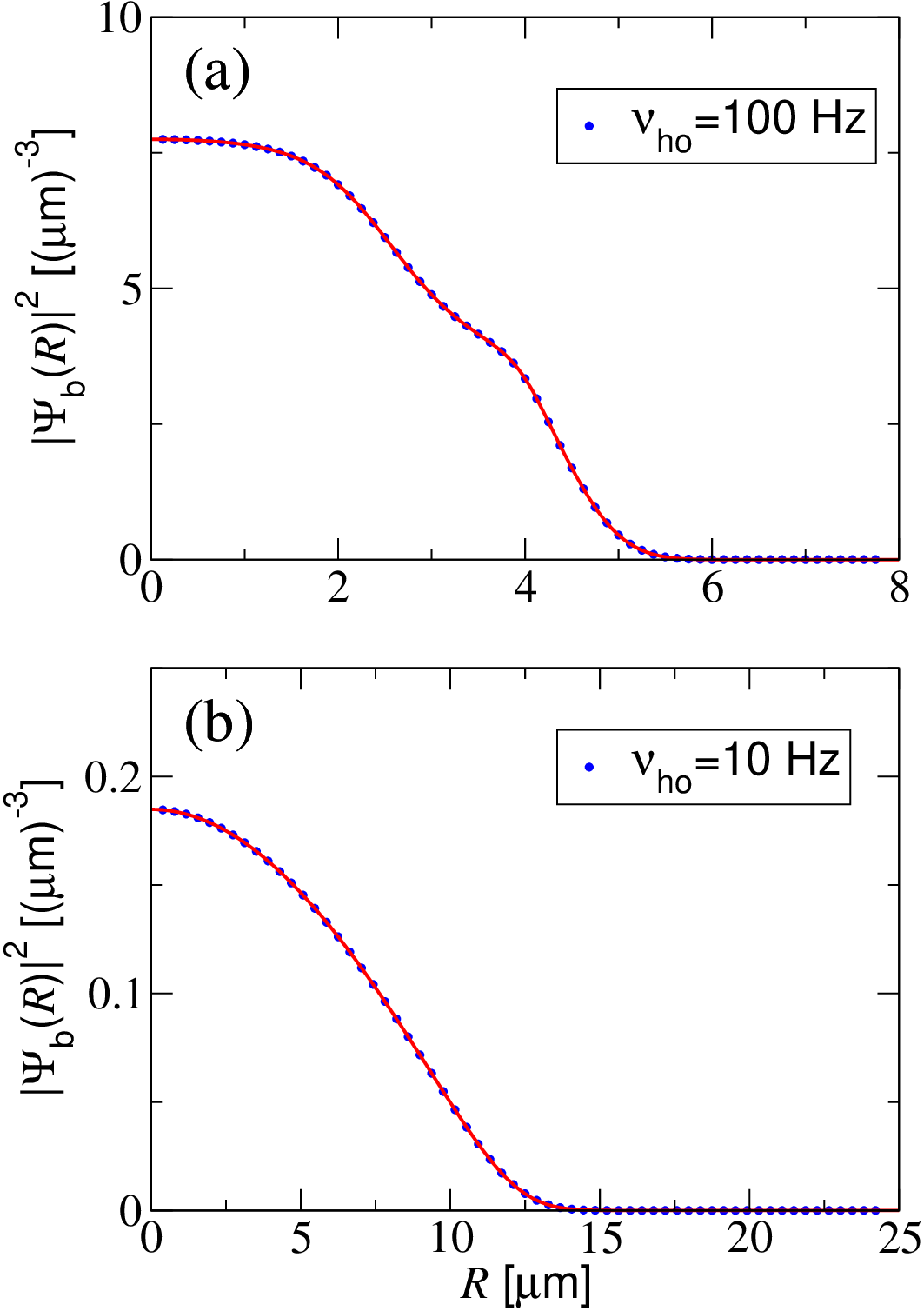}
  \caption{Density of $^{87}$Rb$_2$ molecules in the highest excited 
    vibrational bound state after the crossing of the 100 mT Feshbach 
    resonance in spherically symmetric atoms traps with the frequencies 
    $\nu_\mathrm{ho}=100$ Hz (a) and 
    $\nu_\mathrm{ho}=10$ Hz (b). The molecular densities 
    correspond to the calculations in Fig.~\ref{fig:87RbNcoft} at the final 
    time of the ramps.}
  \label{fig:87Rbnbfin1Gperms}
\end{figure}

Figure \ref{fig:87Rbnbfin1Gperms} shows the densities of $^{87}$Rb$_2$ 
molecules in the highest excited vibrational molecular bound state at the 
final magnetic field strength of the linear downward ramps across the
100 mT Feshbach resonance, which are described in Fig.~\ref{fig:87RbNcoft}.  
The densities have been obtained from the molecular mean field 
$\Psi_\mathrm{b}(\mathbf{R},t_f)$ in Eq.~(\ref{psibgeneral}) through 
$n_\mathrm{b}(\mathbf{R},t_f)=|\Psi_\mathrm{b}(\mathbf{R},t_f)|^2$ with the 
bound state wave function $\phi_\mathrm{b}(B_f)$ at the final magnetic field 
strength $B_f$ of the ramps. The spatial extent of the molecular clouds 
roughly corresponds to the size of the remnant condensate densities in 
Fig.~\ref{fig:87Rbncfin1Gperms}. In accordance with the higher local 
densities, the density of molecules in the tight 100 Hz atom 
trap is higher than in the 10 Hz trap. 

\begin{figure}[htb]
  \includegraphics[width=\columnwidth,clip]{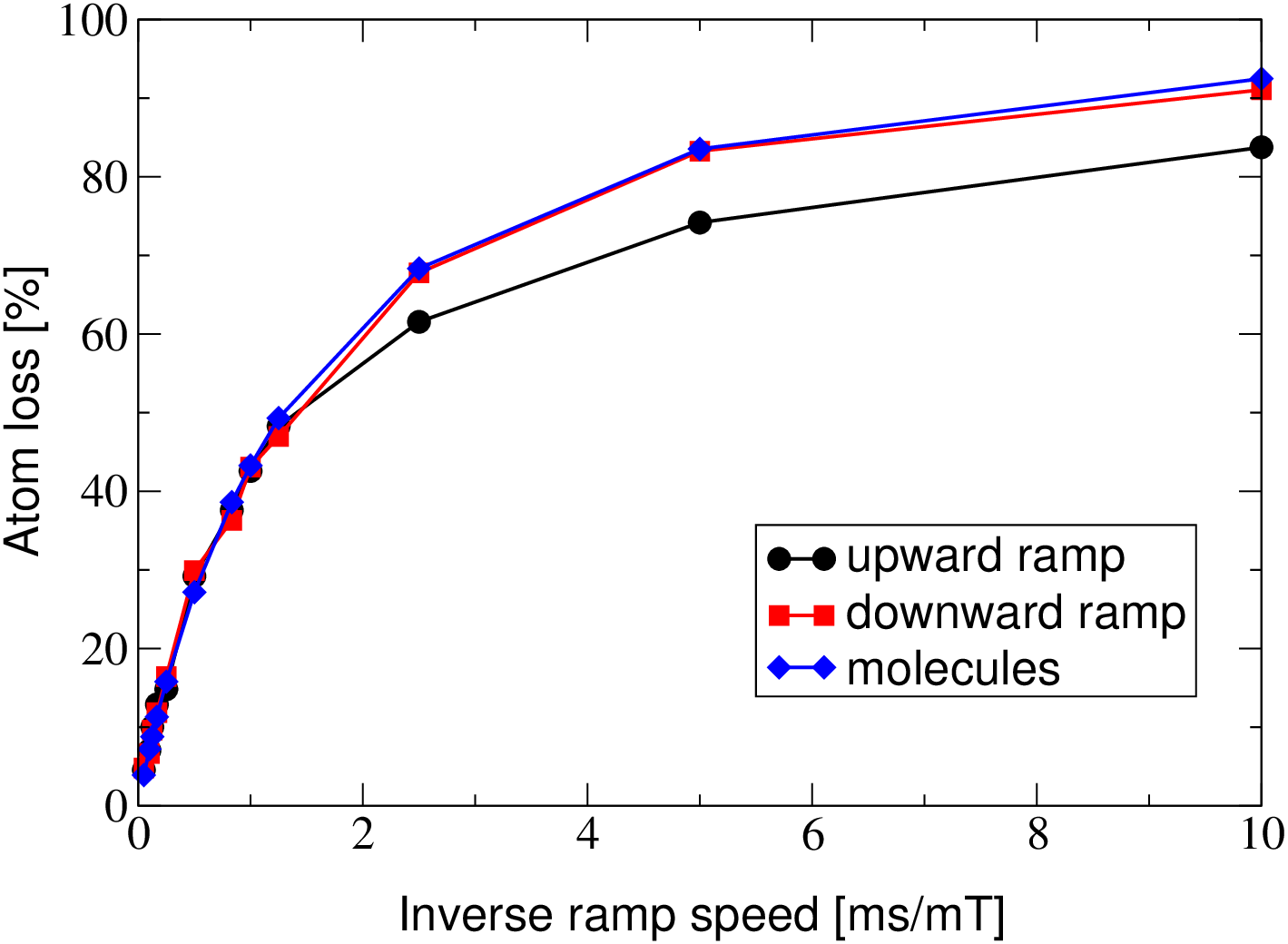}
  \caption{The relative loss of condensate atoms 
    $[N_\mathrm{c}(t_i)-N_\mathrm{c}(t_f)]/N_\mathrm{c}(t_i)$ in linear 
    upward ramps (circles) as well as downward ramps (squares) of the magnetic 
    field strength across the 100 mT Feshbach resonance of $^{87}$Rb as a 
    function of the inverse ramp speed. The upward ramps transfer the 
    condensate atoms into correlated pairs of atoms in initially unoccupied 
    excited states, while the downward ramps adiabatically associate a 
    substantial fraction of the condensate to bound molecules in the highest 
    excited vibrational state. The diamonds indicate the relative number 
    $2N_\mathrm{b}(t_f)/N_\mathrm{c}(t_i)$ of atoms associated to molecules in 
    the downward ramps. The calculations were performed for a cylindrical atom 
    trap with the axial (radial) frequencies of 
    $\nu_\mathrm{axial}=116$ Hz ($\nu_\mathrm{radial}=75$ Hz) and $N=140000$ 
    atoms. The small deviations in the calculated loss data of the remnant 
    atomic condensate (filled circles) from an entirely smooth curve are due 
    to rapid oscillations in the atomic condensate fraction after the 
    passage of the Feshbach resonance (see Fig.~\ref{fig:87RbNcoft}), which 
    have not entirely decayed at the final times that we have chosen for the 
    different inverse ramp speeds.}
  \label{fig:ox-ramps}
\end{figure}

Figure \ref{fig:ox-ramps} shows systematic studies of the dependence of the 
molecular production, after downward ramps of the magnetic field strength 
across the 100 mT Feshbach resonance of $^{87}$Rb, on the inverse ramp speed. 
The calculations have been performed on the basis of the dynamic equation
(\ref{NLSElocal}) for the atomic mean field. In these calculations we have 
chosen the potential of a comparatively tight cylindrically symmetric optical 
atom trap with axial and radial frequencies (see Fig.~\ref{fig:ox-ramps}) 
that resemble those in current experiments on the production of ultracold 
$^{87}$Rb$_2$ molecules in Oxford \cite{Boyer03}. The initial state has been 
chosen as the ground state of the Gross-Pitaevskii equation with 
$N_\mathrm{c}(t_i)=140000$ atoms and a scattering length of 
$a_\mathrm{bg}=100 \ a_\mathrm{Bohr}$. The squares in Fig.~\ref{fig:ox-ramps} 
indicate the relative loss of condensate atoms 
$[N_\mathrm{c}(t_i)-N_\mathrm{c}(t_f)]/N_\mathrm{c}(t_i)$ in the downward 
ramps, where $N_\mathrm{c}(t)=\int d\mathbf{x} \ |\Psi(\mathbf{x},t)|^2$ is
the number of condensate atoms at time $t$. As expected from the two-body 
considerations in Subsection \ref{subsec:LandauZener}, the loss curve is a  
monotonic function of the inverse ramp speed and saturates when the remnant 
condensate is completely depleted. The number of atoms associated to 
diatomic molecules in the highest excited vibrational bound state 
$\phi_\mathrm{b}(B_f)$ is given by 
$2N_\mathrm{b}(t_f)=2\int d\mathbf{R} \ |\Psi_\mathrm{b}(\mathbf{R},t_f)|^2$,
where $\Psi_\mathrm{b}(\mathbf{R},t_f)$ is the molecular mean field in 
Eq.~(\ref{psibgeneral}). The dependence of the relative molecular association
efficiency $2N_\mathrm{b}(t_f)/N_\mathrm{c}(t_i)$ on the inverse ramp speed in 
the downward ramps is indicated by the diamonds in Fig.~\ref{fig:ox-ramps} 
and follows the same monotonic trend as the loss curve of condensate atoms 
in the downward ramps (squares). The quantitative agreement between both curves
shows that the missing fraction of condensate atoms is transferred into 
diatomic molecules in the highest excited multichannel vibrational bound 
state. The excitation of atom pairs in continuum energy modes is suppressed.
The rapid loss of condensate atoms, however, leads to an overall heating of 
the atomic cloud due to the excitation of collective energy modes 
(see Fig.~\ref{fig:87Rbncfin1Gperms}). 

The circles in Fig.~\ref{fig:ox-ramps}
indicate predictions of the total loss of condensate atoms in upward ramps of 
the magnetic field strength across the 100 mT Feshbach resonance of $^{87}$Rb. 
The initial state of the gas as well as the confining atom trap in these 
calculations are the same as in the downward ramps. The final magnetic field 
strength $B_f$, after the passage across the resonance in an upward ramp, is 
on the high field side of the resonance position. 
Figure \ref{fig:EbofBtwochannel}
reveals that there is no energetically accessible bound state at the magnetic 
field strength $B_f$. Consequently, atoms lost from the condensate are 
entirely transferred into the burst fraction. Figure \ref{fig:ox-ramps} shows 
that despite an overall agreement in the monotonic behaviour of the condensate 
loss with increasing inverse ramp speeds, there is a small difference in the 
saturation region between the curves for the different ramp directions. This 
small deviation between the curves would be absent in the two-body description 
of Section \ref{sec:twobody} and may indicate phenomena related to the 
coherent nature of the gas. 

\subsection{Two level mean field approach}
\label{subsec:twolevel}
One of the most common approaches to the association of molecules in dilute 
Bose-Einstein condensates is based on a model many-body Hamiltonian that 
describes atoms and molecules in terms of separate quantum fields. The
coupling between these quantum fields leads to an exchange between the 
different species. This exchange then serves as a model for the molecular 
production. 

\subsubsection{Markov approximation to the first order microscopic quantum 
  dynamics approach}
In order to reveal the physical meaning of the molecular quantum 
field and its relation to the microscopic quantum dynamics approach of 
Subsection \ref{subsec:microscopicquantumdynamics}, we shall derive the mean
field equations associated with the two component model Hamiltonian on the 
basis of the first order coupled dynamic equations (\ref{Psi2ch}) and 
(\ref{Phi2ch}) that determine the atomic mean field and the pair function.
We shall represent the pair function in terms of the centre of mass and 
relative coordinates $\mathbf{R}$ and $\mathbf{r}$, respectively. In accordance
with the pole approximation of Subsection \ref{subsec:energystates}, the 
closed channel part of the pair function can be factorised into the resonance 
state $\phi_\mathrm{res}(r)$ and a centre of mass wave function:
\begin{equation}
  \Phi_\mathrm{cl}(\mathbf{R},\mathbf{r},t)=\sqrt{2}\phi_\mathrm{res}(r)
  \Psi_\mathrm{res}(\mathbf{R},t).
  \label{poleapproxPhicl}
\end{equation}
In the following derivation we shall show that the centre of mass wave 
function $\Psi_\mathrm{res}(\mathbf{R},t)$ is identical to the mean field 
associated with the molecular quantum field in the two component model
Hamiltonian. To this end we shall proceed in a way similar to the derivation 
of the dynamic equation (\ref{NLSElocal}) for the atomic mean field in 
Subsection \ref{subsec:microscopicquantumdynamics},
except that we shall eliminate only the component 
$\Phi_\mathrm{bg}(\mathbf{R},\mathbf{r},t)$ of the pair function in the
open channel from Eqs.~(\ref{Psi2ch}) and (\ref{Phi2ch}). 
The formal solution of the dynamic equation for 
$\Phi_\mathrm{bg}(\mathbf{R},\mathbf{r},t)$ 
[i.e.~the first component of Eq.~(\ref{Phi2ch})] can be expressed in terms 
of the two-body time evolution operator $U_\mathrm{trap}^\mathrm{bg}(t-\tau)$
that corresponds to the stationary diagonal element 
$\langle\mathrm{bg}|H_\mathrm{2B}|\mathrm{bg}\rangle$
of the full two-body Hamiltonian matrix $H_\mathrm{2B}(t)$ 
[cf.~Eq.~(\ref{Phi2ch})] in the open channel:
\begin{widetext}
  \begin{align}
    \nonumber
    \Phi_\mathrm{bg}(\mathbf{R},\mathbf{r},t)=
    &\int d\mathbf{R}' d\mathbf{r}' \
    \langle\mathbf{R},\mathbf{r}|
    U_\mathrm{trap}^\mathrm{bg}(t-t_i)|\mathbf{R}',\mathbf{r}'
    \rangle
    \Phi_\mathrm{bg}(\mathbf{R}',\mathbf{r}',t_i)\\
    \nonumber
    &+\frac{1}{i\hbar}\int_{t_i}^t d\tau \int d\mathbf{R}' d\mathbf{r}' \ 
    \langle\mathbf{R},\mathbf{r}|
    U_\mathrm{trap}^\mathrm{bg}(t-\tau)
    V_\mathrm{bg}|\mathbf{R}',\mathbf{r}'\rangle
    \Psi(\mathbf{R}'+\mathbf{r}'/2,\tau)\Psi(\mathbf{R}'-\mathbf{r}'/2,\tau)\\
    &+\frac{1}{i\hbar}\int_{t_i}^t d\tau\int d\mathbf{R}' \ 
    \langle\mathbf{R},\mathbf{r}|
    U_\mathrm{trap}^\mathrm{bg}(t-\tau)
    W|\mathbf{R}',\phi_\mathrm{res}\rangle
    \sqrt{2}\Psi_\mathrm{res}(\mathbf{R}',\tau).
    \label{formalsolutionPhibg}
  \end{align}
\end{widetext}
We note that $U_\mathrm{trap}^\mathrm{bg}(t-\tau)$ includes the centre of 
mass and the relative motion of two atoms as well as the confining potential 
of the atom trap. 

To obtain the mean field equations for $\Psi(\mathbf{x},t)$ 
and the centre of mass wave function $\Psi_\mathrm{res}(\mathbf{R},t)$ in 
Eq.~(\ref{poleapproxPhicl}), we shall insert Eq.~(\ref{formalsolutionPhibg}) 
into Eqs.~(\ref{Psi2ch}) and (\ref{Phi2ch}) and perform the Markov 
approximation. The Markov approximation relies upon the assumption that the 
main contribution to the time integrals in Eq.~(\ref{formalsolutionPhibg}) 
stems from a small region near $\tau=t$, in which the variation of the 
functions $\Psi(\mathbf{R}'+\mathbf{r}'/2,\tau)$, 
$\Psi(\mathbf{R}'-\mathbf{r}'/2,\tau)$ 
and $\Psi_\mathrm{res}(\mathbf{R}',\tau)$ is 
negligible. Under this assumption, the functions can be evaluated at $\tau=t$.
A similar argument applies to the spatial variation of 
$\Psi(\mathbf{R}'+\mathbf{r}'/2,\tau)$, 
$\Psi(\mathbf{R}'-\mathbf{r}'/2,\tau)$ 
and $\Psi_\mathrm{res}(\mathbf{R}',\tau)$
and leads to the replacements $\mathbf{R}'\to\mathbf{R}$ and 
$\mathbf{r}'\to 0$ in these functions. The subsequent formal procedure to 
derive the mean field equations is analogous to the derivation of the 
Gross-Pitaevskii equation in Ref.~\cite{KB02} and involves neglecting the 
initial pair function $\Phi_\mathrm{bg}(\mathbf{R}',\mathbf{r}',t_i)$ and 
the centre of mass motion as well as the confining potential of the atom trap 
in the time evolution operator $U_\mathrm{trap}^\mathrm{bg}(t-\tau)$. 
A short calculation then yields the coupled equations 
\begin{align}
  \nonumber
  i\hbar\frac{\partial}{\partial t}\Psi(\mathbf{x},t)=&
  \left[
    -\frac{\hbar^2}{2m}\nabla^2+\frac{m}{2}\omega_\mathrm{ho}^2|\mathbf{x}|^2
    \right]\Psi(\mathbf{x},t)\\
  \nonumber
  &+g_\mathrm{bg}|\Psi(\mathbf{x},t)|^2\Psi(\mathbf{x},t)\\
  &+g_\mathrm{res}\Psi^*(\mathbf{x},t)\sqrt{2}
  \Psi_\mathrm{res}(\mathbf{x},t)
  \label{2levelMFPsi}
\end{align}
for the atomic mean field, and
\begin{align}
  \nonumber
  i\hbar\frac{\partial}{\partial t}\Psi_\mathrm{res}(\mathbf{R},t)=&
  \left[
    -\frac{\hbar^2}{4m}\nabla^2+\frac{2m}{2}\omega_\mathrm{ho}^2|\mathbf{R}|^2
    \right]
  \Psi_\mathrm{res}(\mathbf{R},t)\\
  \nonumber
  &
  +\left[\frac{d E_\mathrm{res}}{d B}(B_\mathrm{res})\right]
  [B(t)-B_0]
  \Psi_\mathrm{res}(\mathbf{R},t)\\ 
  &+\frac{1}{\sqrt{2}}g_\mathrm{res}\Psi^2(\mathbf{R},t)
  \label{2levelMFPsires}
\end{align}
for the molecular mean field \cite{opticaltrap}. 
Here we have assumed the optical atom trap to 
be spherically symmetric. The coupling constants $g_\mathrm{bg}$ and 
$g_\mathrm{res}$ can be obtained by performing the spatial as well as the 
time integrations in Eq.~(\ref{formalsolutionPhibg}) in the limit 
$t-t_i\to\infty$. This yields
(cf., also, Ref.~\cite{KB02}):
\begin{align}
  \label{gbg}
  g_\mathrm{bg}=&\frac{4\pi\hbar^2}{m}a_\mathrm{bg}\\
  g_\mathrm{res}=&(2\pi\hbar)^{3/2}
  \langle\phi_\mathrm{res}|W|\phi_0^{(+)}
  \rangle.
  \label{gres}
\end{align}
Here $a_\mathrm{bg}$ is the background scattering length and 
$\phi_0^{(+)}(\mathbf{r})$ is the zero energy wave function 
associated with the background scattering 
(cf.~Eq.~(\ref{BCphipplus})). We note that the off diagonal coupling constant 
$g_\mathrm{res}$ is defined only up to a global phase and we shall, 
therefore, choose it to be real. In accordance with 
Eq.~(\ref{resonancewidth}), $g_\mathrm{res}$ is then determined by the 
resonance width $(\Delta B)$, the background scattering length $a_\mathrm{bg}$
and the slope of the resonance 
$\left[\frac{d E_\mathrm{res}}{dB}(B_\mathrm{res})\right]$.
The mean field equations (\ref{2levelMFPsi}) and (\ref{2levelMFPsires})
as well as the coupling constants in Eqs.~(\ref{gbg}) and (\ref{gres})
are analogous to those that have been applied, for instance, in 
Ref.~\cite{Abeelen99} to describe the atom loss in ramps of the magnetic 
field strength across Feshbach resonances in $^{23}$Na. In the applications 
of the mean field equations (\ref{2levelMFPsi}) and (\ref{2levelMFPsires}),
the densities $n_\mathrm{c}(\mathbf{x},t)=|\Psi(\mathbf{x},t)|^2$ and 
$n_\mathrm{res}(\mathbf{R},t)=|\Psi_\mathrm{res}(\mathbf{R},t)|^2$ 
are usually interpreted as atomic and molecular condensate densities, 
respectively.

\subsubsection{Two component model Hamiltonian}
Equations (\ref{2levelMFPsi}) and (\ref{2levelMFPsires}) can be derived 
formally as equations of motion for the mean fields 
\begin{align}
  \Psi(\mathbf{x},t)&=\langle\psi(\mathbf{x})\rangle_t \\
  \Psi_\mathrm{res}(\mathbf{R},t)&=\langle\psi_\mathrm{res}(\mathbf{R})
  \rangle_t
\end{align}
in the Hartree approximation. The corresponding two component model 
Hamiltonian is given by:
\begin{align}
  \nonumber
  H=&\int d\mathbf{x} \ 
  \psi^\dagger(\mathbf{x})\left[
    -\frac{\hbar^2}{2m}\nabla^2+\frac{m}{2}\omega_\mathrm{ho}^2|\mathbf{x}|^2
    \right]\psi(\mathbf{x})\\
  \nonumber
  &+\int d\mathbf{R} \
  \psi^\dagger_\mathrm{res}(\mathbf{R})\left[
    -\frac{\hbar^2}{4m}\nabla^2+\frac{2m}{2}\omega_\mathrm{ho}^2|\mathbf{R}|^2
    \right]\psi_\mathrm{res}(\mathbf{R})\\
  \nonumber
  &+\int d\mathbf{R} \
  \psi^\dagger_\mathrm{res}(\mathbf{R})
  \left[
    \frac{dE_\mathrm{res}}{dB}(B_\mathrm{res})
  \right]
       [B(t)-B_0]
       \psi_\mathrm{res}(\mathbf{R})\\
       \nonumber
       &+\frac{1}{2}g_\mathrm{bg}\int d\mathbf{x} \
       \psi^\dagger(\mathbf{x})
       \psi^\dagger(\mathbf{x})
       \psi(\mathbf{x})\psi(\mathbf{x})\\
       &+\frac{1}{\sqrt{2}}g_\mathrm{res}\int d\mathbf{x} \ 
       \left[
	 \psi_\mathrm{res}^\dagger(\mathbf{x})\psi(\mathbf{x})\psi(\mathbf{x})
	 +\mbox{h.c.}
	 \right].
       \label{modelHamiltonian}
\end{align}
Here the field operators $\psi(\mathbf{x})$ and 
$\psi_\mathrm{res}(\mathbf{R})$ fulfil the usual Bose commutation relations 
and commute among the different species:
\begin{align}
[\psi(\mathbf{x}),\psi_\mathrm{res}(\mathbf{R})]=&0\\
[\psi(\mathbf{x}),\psi_\mathrm{res}^\dagger(\mathbf{R})]=&0.
\label{commPsiPsires}
\end{align}

\subsubsection{Deficits of the two level mean field approach}
Despite its common use in studies of molecular association in dilute 
Bose-Einstein condensates, the two level mean field model in 
Eqs.~(\ref{2levelMFPsi}) and (\ref{2levelMFPsires}) and its common
interpretation are subject to two serious deficits. First, in the 
derivation of the mean field equations we have shown that the mean field 
$\Psi_\mathrm{res}(\mathbf{R},t)$ is associated with a pair of atoms 
in the closed channel resonance state $\phi_\mathrm{res}(r)$ 
[see Eq.~(\ref{poleapproxPhicl})].  
Figure \ref{fig:mixingcoefficient} clearly reveals that, in general, such an 
atom pair cannot be associated with a molecular bound state because in the 
close vicinity of the 100 mT Feshbach resonance of $^{87}$Rb, as well as 
asymptotically far from it, the wave function of the highest excited 
vibrational bound state is dominated by its component in the 
open channel \cite{85Rbadmixture}. 

The interpretation of 
$N_\mathrm{res}(t)=\int d\mathbf{R} \ |\Psi_\mathrm{res}(\mathbf{R},t)|^2$ in 
terms of the population in the resonance state (as opposed to the number 
of bound molecules) is a direct consequence of the commutation relation 
in Eq.~(\ref{commPsiPsires}) and of the form of the Hamiltonian in 
Eq.~(\ref{modelHamiltonian}), rather than an artifact 
of the derivation of the mean field equations (\ref{2levelMFPsi}) and 
(\ref{2levelMFPsires}). In fact, according to Eq.~(\ref{commPsiPsires}), a 
localised pair of atoms created from the vacuum state $|\mathrm{vac}\rangle$ 
by the field operator $\psi_\mathrm{res}^\dagger(\mathbf{R})$
is always orthogonal to the localised state of any atom pair created by 
$\psi^\dagger(\mathbf{y})\psi^\dagger(\mathbf{x})$ in the open 
channel; i.e.
\begin{equation}
\langle\mathrm{vac}|\psi(\mathbf{x})\psi(\mathbf{y})
\psi_\mathrm{res}^\dagger(\mathbf{R})|\mathrm{vac}\rangle=0.
\end{equation}     
The two states of the atom pairs, therefore, correspond to different 
asymptotic scattering channels. It can also be verified from the Hamiltonian 
in Eq.~(\ref{modelHamiltonian}) that any physical two-body state associated 
with $\psi_\mathrm{res}^\dagger(\mathbf{R})|\mathrm{vac}\rangle$ is not 
stationary with respect to the time evolution in the relative motion of the
atom pairs. The off diagonal coupling in Eq.~(\ref{modelHamiltonian}) leads 
to a decay into the open channel, which is determined by the 
coupling constant $g_\mathrm{res}$. The energy 
$E_\mathrm{res}(B)=\left[\frac{d E_\mathrm{res}}{dB}(B_\mathrm{res})\right]
(B-B_0)$, which can be associated 
with $\psi_\mathrm{res}^\dagger(\mathbf{R})|\mathrm{vac}\rangle$, is linear 
in the magnetic field strength and corresponds to the resonance state
rather than to the bound molecular states 
(cf.~Fig.~\ref{fig:EbofBtwochannel}). The off diagonal coupling constant 
$g_\mathrm{res}$ then determines the corresponding decay width. 

Although it seems conceivable to measure the closed channel resonance state 
population 
$N_\mathrm{res}(t)=\int d\mathbf{R} \ |\Psi_\mathrm{res}(\mathbf{R},t)|^2$,
several present day experiments on molecular 
association in Bose-Einstein condensates have clearly revealed the 
multichannel nature of the bound states. The experiments in 
Refs.~\cite{Donley02,Claussen03}, for instance, have determined the near 
resonant universal 
form $E_\mathrm{b}(B)=-\hbar^2/\left\{m[a(B)]^2\right\}$ of the binding energy 
in Eq.~(\ref{Ebuniversal}), corresponding to a bound state wave function 
of the universal form of Eq.~(\ref{universalwavefunction}), which is dominated 
by its component in the asymptotic open channel 
(cf., also, Fig.~\ref{fig:haloboundstates}). The experiment 
in Ref.~\cite{Duerr03} has determined the change of the 
magnetic moment of the molecular bound states in dependence on the magnetic 
field strength $B$, which is a consequence of the nonlinear dependence of the 
binding energy on $B$ (see Fig.~\ref{fig:EbofBtwochannel}). 
The magnetic moment of the resonance state, however, 
is independent of $B$. These experiments clearly reveal that the number of
bound molecules in Eq.~(\ref{Nbgeneral}) is the relevant physical quantity
rather than the resonance state population.

The second major deficit of the mean field equations (\ref{2levelMFPsi}) and 
(\ref{2levelMFPsires}) in the description of the adiabatic association 
technique is related to their two level nature; the only configurations of two 
atoms are a pair of condensate atoms and the resonance state. 
Neither the near resonant diatomic bound states nor the continuum states can 
be represented solely in terms of these two configurations. In particular, any 
coupling between the initial atomic condensate and the quasi continuum of 
states above the dissociation threshold energy of the  open channel 
is ruled out. The two level nature of Eqs.~(\ref{2levelMFPsi}) and 
(\ref{2levelMFPsires}) is a consequence of the application of the 
Markov approximation to Eqs.~(\ref{Psi2ch}) and (\ref{Phi2ch}). This indicates 
that the presuppositions leading to the Markov approximation are violated 
during the passage across a Feshbach resonance. Consequently,
Eqs.~(\ref{2levelMFPsi}) and (\ref{2levelMFPsires}) fail quantitatively and 
even qualitatively in the dynamic description of the near resonant 
atom-molecule coherence experiments in Refs.~\cite{Donley02,Claussen03}. The 
most striking consequence of their two level nature, however, consists in the 
insensitivity of Eqs.~(\ref{2levelMFPsi}) and (\ref{2levelMFPsires}) with 
respect to the ramp direction. The approach thus predicts molecular production 
even for an upward ramp across the 100 mT Feshbach resonance of $^{87}$Rb, 
although there is no energetically accessible diatomic vibrational bound state 
on the high field side of the resonance (cf.~Fig.~\ref{fig:EbofBtwochannel}). 
The reason for this failure is the absence of continuum states that the closed 
channel resonance state could decay into. The results of Subsection 
\ref{subsec:LandauZener} suggest, however, that the two level mean field 
approach may recover the far resonant asymptotic molecular 
population in a downward ramp of the magnetic field strength across the 
100 mT Feshbach resonance of $^{87}$Rb.

\subsubsection{Applications of the two component model Hamiltonian beyond
  mean field}
When applied beyond mean field
\cite{Kokkelmans02,Mackie02,Yurovsky03}, the model Hamiltonian in 
Eq.~(\ref{modelHamiltonian}) leads to ultraviolet singularities that are 
related to the energy independence of the coupling constants $g_\mathrm{bg}$ 
and $g_\mathrm{res}$, i.e.~the associated contact potentials crucially lack any
spatial extent. The approach in Ref.~\cite{Kokkelmans02} circumvents
this ultraviolet problem in terms of an energy cutoff, which is adjusted in 
such a way that the Hamiltonian, when applied to only two atoms, recovers the 
exact binding energy of the highest excited vibrational bound state. This 
beyond mean field approach has been applied recently to the experiments in 
Ref.~\cite{Donley02}. We expect this approach to give results similar to 
those of the first order microscopic quantum dynamics approach. It has been 
pointed out \cite{KGB03,TKTGPSJKB03,Borca03,Braaten03,Duine03}, however, that 
the interpretation in Ref.~\cite{Kokkelmans02} of the closed channel resonance 
state population $N_\mathrm{res}(t)$ in terms of the number of bound molecules 
(i.e.~the number of undetected atoms in Ref.~\cite{Donley02}) is not 
applicable. 

Some approaches \cite{Braaten03,Duine03} have suggested curing this deficit, 
by multiplying the closed channel resonance state population 
$N_\mathrm{res}(t)$, as determined in Ref.~\cite{Kokkelmans02}, with a 
magnetic field dependent factor, termed the wave function renormalisation 
factor, which accounts for the component of the bound state wave 
function in the asymptotic open scattering channel. A short calculation  
using Eq.~(\ref{twochannelnormalisation}) shows that the 
wave function renormalisation factor is equivalent to the squared  
normalisation factor
\begin{equation}
  \mathcal{N}_\mathrm{b}^2=1+\frac{1}{2}
  \left[
    \frac{dE_\mathrm{res}}{dB}(B_\mathrm{res})
    \right]
  (\Delta B)\frac{a_\mathrm{bg}}{a(B)}\frac{m [a(B)]^2}{\hbar^2}
  \label{wavefunctionrenormalisation}
\end{equation}
of the two channel bound state wave function in Eq.~(\ref{phib}), 
provided that the magnetic field strength is sufficiently close to the 
resonance position that the bound state wave function and its binding energy 
are universal (see Subsection \ref{subsec:universal}). 
A comparison between the closed channel resonance state population in 
Ref.~\cite{Kokkelmans02}, revised by multiplication with the right hand side 
of Eq.~(\ref{wavefunctionrenormalisation}) (as provided 
by Ref.~\cite{Braaten03}), indeed, largely recovers the magnitude
of the molecular fractions predicted by the first order microscopic quantum 
dynamics approach in Refs.~\cite{KGB03,TKTGPSJKB03}. The revised resonance
state population, however, shows unphysical oscillations between the 
populations of bound and free atoms (cf.~Fig.~3 in Ref.~\cite{Kokkelmans02}), 
for magnetic field strengths at which the spatial extent of the highest 
excited vibrational molecular bound state is by far 
smaller than the mean interatomic distance in the dilute gas. This unphysical
behaviour is due to the fact that not only the bound state wave function
in Eq.~(\ref{phib}), but also the continuum wave functions in 
Eqs.~(\ref{phipcl}) and (\ref{phipbg}), have a closed channel resonance state
component. As a consequence, there is no {\em general\/} relationship 
between the   
closed channel resonance state population of a gas and the number of 
bound molecules. 

The general relationship between the number of bound 
molecules and the two-body correlation function in Eq.~(\ref{Nbgeneral}), 
however, is applicable also to the approach in Ref.~\cite{Kokkelmans02}. 
Equation (\ref{Nbgeneral}) applies the appropriate quantum mechanical 
observable, and will therefore automatically determine the measurable fraction 
of bound molecules in a gas, provided that molecules identified as 
separate entities are at all a reasonable concept \cite{TKTGPSJKB03}.
  
\subsection{Comparison between different approaches}
\label{subsec:comparison}
We have shown in Subsection \ref{subsec:microscopicquantumdynamics} that, 
during a linear ramp of the magnetic field strength across a Feshbach 
resonance, the full microscopic quantum dynamics approach can predict a 
transfer of atoms from the atomic condensate not only to the bound, but also 
to the continuum part of the two-body energy spectrum. The transfer into the 
continuum energy states, however, occurred only in upward ramps of the magnetic
field strength across the 100 mT Feshbach resonance of $^{87}$Rb. The two 
level configuration interaction approach of Subsection \ref{subsec:LandauZener}
restricts the configuration of an atom pair in the open channel to the lowest 
energetic quasi continuum state of the background scattering in the atom trap, 
while in the two level mean field model of Subsection \ref{subsec:twolevel} 
all atoms in the $(F=1,m_F=+1)$ hyperfine state are described by a coherent 
mean field. Both two level approaches, therefore, rule out the production of 
correlated pairs of atoms in the quasi continuum of two-body energy levels 
from the outset. 

Equations (\ref{NLRabibg}) and (\ref{NLRabicl}) of the configuration 
interaction approach and the mean field equations (\ref{2levelMFPsi}) and 
(\ref{2levelMFPsires}) describe the coupling between the atomic condensate 
and the closed channel resonance state in essentially the same way, except 
that the phases of the off diagonal coupling terms are different. The spatial 
configuration of the atomic condensate in the configuration interaction 
approach, however, is static, while the two level mean field approach allows 
the trapped atomic condensate to access coherent collective excitations with 
a high occupancy of the energy modes. Consequently, the configuration 
interaction approach can, at least to some extent, be interpreted as a local 
density approximation to the two level mean field model. In accordance with 
the derivations in Subsection \ref{subsec:twolevel}, both two level approaches 
can, therefore, be considered as the Markov approximation to the first order 
microscopic quantum dynamics approach of Subsection 
\ref{subsec:microscopicquantumdynamics}. 

Our previous considerations have shown that there exists a qualitative 
agreement between the approaches, at different levels of approximation,
with respect to the prediction of molecular formation in linear downward ramps 
of the magnetic field strength across the 100 mT Feshbach resonance of 
$^{87}$Rb. The description of the microscopic binary collision physics
as well as their treatment of the coherent nature of the Bose-Einstein 
condensate, however, differ considerably among these approaches. In the 
following, we shall provide quantitative comparisons between their predictions 
with respect to the molecular production efficiency of the adiabatic 
association technique. We will study in detail the dependence of the 
magnitude of the molecular fraction on experimentally accessible parameters.

\subsubsection{Universal properties of the molecular production efficiency 
  of a linear ramp of the magnetic field strength}
In the adiabatic association technique with linear ramps of the magnetic field 
strength, the ramp speed $\frac{d B}{d t}$ controls the interatomic 
interactions. Furthermore, the initial number $N$ of atoms can be varied over 
a wide range in present experiments, and the trap frequencies determine the 
confining potential of a harmonic atom trap. The atom trap, the number of atoms
as well as their binary interactions control the spectrum of coherent 
collective excitations of a Bose-Einstein condensate. In the following, we 
shall consider spherically symmetric atom traps with a radial trap frequency 
that we denote by $\nu_\mathrm{ho}$. 

\begin{figure}[htb]
  \includegraphics[width=\columnwidth,clip]{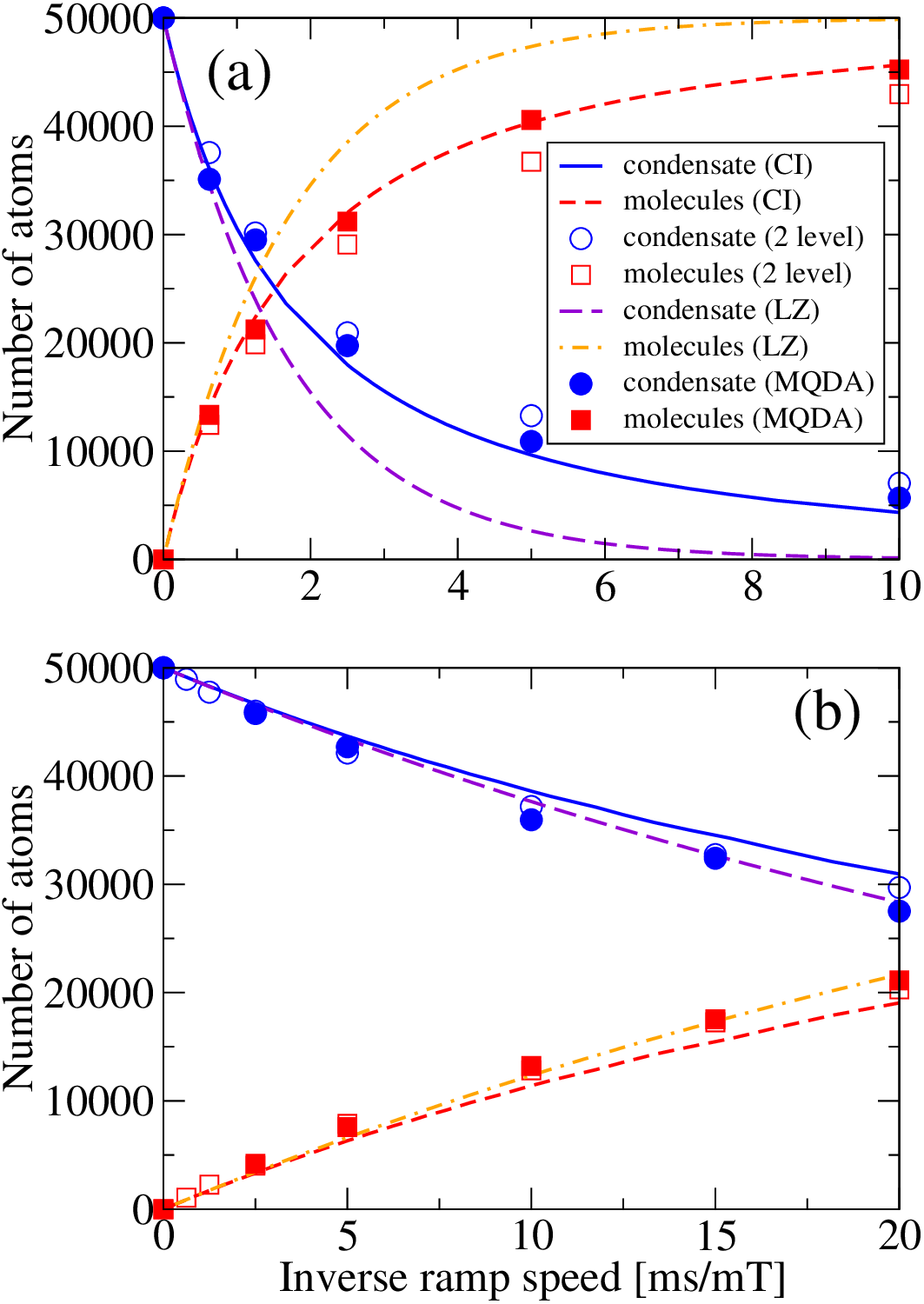}
  \caption{Predicted remnant condensate fraction and number of atoms 
    associated to diatomic molecules in the highest excited vibrational 
    bound state after a linear downward ramp of the 
    magnetic field strength across the 100 mT Feshbach resonance 
    of $^{87}$Rb in dependence on the inverse ramp speed. 
    The initial state of the gas corresponds to a dilute zero 
    temperature Bose-Einstein condensate with a scattering length of 
    $a_\mathrm{bg}=100 \ a_\mathrm{Bohr}$ and $N=50000$ atoms in spherically 
    symmetric atom traps with the frequencies 100 Hz (a) and 10 Hz (b).
    The abbreviations in the legend indicate the different approaches 
    applied in the calculations. These approaches contain the first order
    microscopic quantum dynamics approach (MQDA), the nonlinear configuration
    interaction equations (\ref{NLRabibg}) and (\ref{NLRabicl}) (CI),
    the two level mean field equations (\ref{2levelMFPsi}) and 
    (\ref{2levelMFPsires}) (2 level), and the Landau-Zener asymptotic 
    populations in Eqs.~(\ref{LZbg}) and (\ref{LZcl}) (LZ). The small 
    deviations in the calculated MQDA data of the remnant atomic condensate 
    (filled circles) from an entirely smooth curve are discussed in 
    Fig.~\ref{fig:ox-ramps}.}
  \label{fig:87RbNatomswithLZ}
\end{figure}

Figure \ref{fig:87RbNatomswithLZ} shows the predicted dependence of the 
remnant atomic condensate fraction and of the fraction of atoms associated to 
diatomic molecules in the highest excited vibrational bound state on the 
inverse speed of linear downward ramps of the magnetic field strength across 
the 100 mT Feshbach resonance of $^{87}$Rb. We have chosen trap frequencies 
of 100 Hz (a) and of 10 Hz (b) that correspond to a comparatively strongly 
and a weakly confining harmonic atom trap, respectively. The initial number 
of $N_\mathrm{c}(t_i)=50000$ condensate atoms is fixed in all calculations. 
The predictions correspond to the full first order microscopic quantum 
dynamics approach (MQDA), the nonlinear configuration interaction equations 
(\ref{NLRabibg}) and (\ref{NLRabicl}) (CI), the two level mean field equations 
(\ref{2levelMFPsi}) and (\ref{2levelMFPsires}) (2 level), and the 
Landau-Zener asymptotic populations in Eqs.~(\ref{LZbg}) and (\ref{LZcl}) 
(LZ). The comparison reveals that the approaches predict similar conversion
efficiencies with respect to the association of molecules as well as similar
remnant condensate fractions. Only the saturation behaviour of the exponential 
Landau-Zener (LZ) curves [cf.~Eqs.~(\ref{LZbg}) and (\ref{LZcl})] differs 
significantly from all other approaches because it corresponds to the 
asymptotic populations of the linear equations (\ref{Rabibg}) and 
(\ref{Rabicl}). These results suggest a remarkable insensitivity of the 
molecular production in linear ramps of the magnetic field strength with 
respect to both the details of the microscopic binary collision physics and 
the coherent nature of the Bose-Einstein condensate.

\begin{figure}[htb]
  \includegraphics[width=\columnwidth,clip]{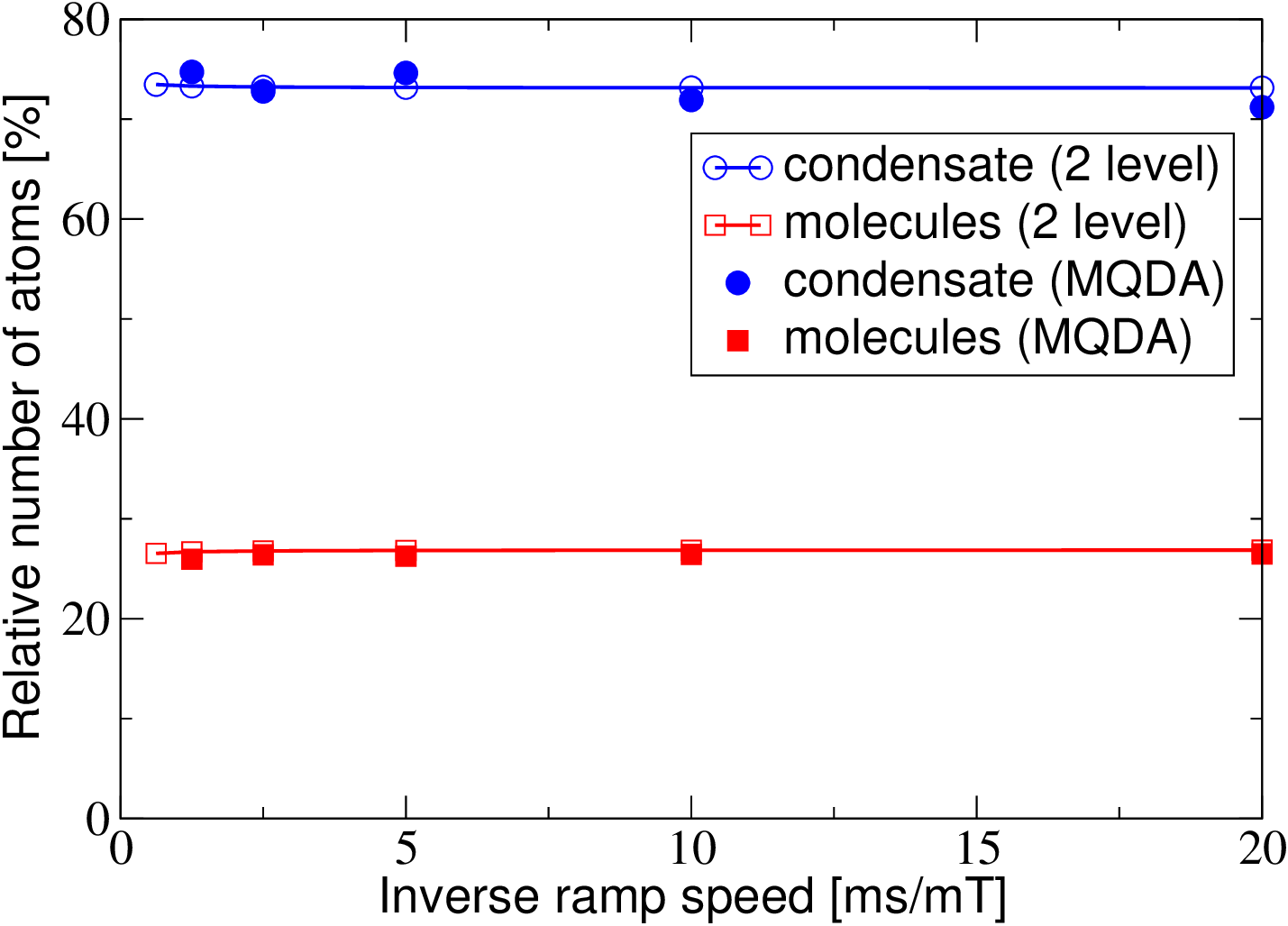}
  \caption{Predicted proportion of remnant condensate atoms and proportion of 
    atoms associated to diatomic molecules in the highest excited vibrational 
    bound state after a linear downward ramp of the magnetic field strength 
    across the 100 mT Feshbach resonance of $^{87}$Rb in dependence on the 
    inverse ramp speed. The atom number and the angular frequency 
    $\omega_{\rm ho}$ of the atom trap are adjusted to keep the nonlinearity 
    parameter $k_{\rm bg}$ and the quantity $\omega_{\rm ho}t_\mathrm{nat}$ 
    constant, which also fixes the Landau-Zener parameter 
    $\delta_{\mathrm{LZ}}$. The initial states of the gas correspond to a 
    dilute zero temperature Bose-Einstein condensate with a scattering length 
    of $a_\mathrm{bg}=100 \ a_\mathrm{Bohr}$ in spherically symmetric 
    atom traps. In each case the shape of the initial condensate mode is, in 
    units of $l_{\rm ho}$, identical to that of a dilute Bose-Einstein 
    condensate with 50000 atoms in a spherically symmetric atom trap with a 
    frequency of $\nu_\mathrm{ho}=10$ Hz. The abbreviations in the legend 
    indicate the different approaches applied in the calculations. These 
    approaches contain the first order microscopic quantum dynamics approach 
    (MQDA), and the two level mean field equations (\ref{2levelMFPsi}) and 
    (\ref{2levelMFPsires}) (2 level).}
  \label{fig:87RbLZconst}
\end{figure}

\subsubsection{Dependence of the molecular production efficiency on 
  experimentally accessible parameters}
We shall now identify the universal physical parameter that controls the 
molecular production efficiency of the adiabatic association technique. 
As shown in Subsection \ref{subsec:LandauZener}, the linear Rabi flopping 
model in Eqs.~(\ref{Rabibg}) and (\ref{Rabicl}) leads to an asymptotic 
fraction of diatomic molecules that can be described by a single parameter, 
the Landau-Zener coefficient $\delta_\mathrm{LZ}$. We shall show in the 
following that also in the nonlinear two level approaches the asymptotic 
molecular population is characterised by $\delta_\mathrm{LZ}$. To this end, 
we take Eqs.~(\ref{2levelMFPsi}) and (\ref{2levelMFPsires}) of the two level 
mean field approach, where it is revealing to introduce a natural time scale 
$t_\mathrm{nat}=(\Delta B)/\left(\frac{d B}{d t}\right)$, as well as the 
harmonic oscillator length scale 
$l_{\mathrm{ho}}=\sqrt{\hbar/(m 2\pi\nu_\mathrm{ho})}$. Applying these natural 
time and length scales to the two level mean field equations reveals that 
they can be characterised purely in terms of the nonlinearity parameter 
$k_{\mathrm{bg}}=Na_\mathrm{bg}/l_\mathrm{ho}$ of the 
Gross-Pitaevskii equation \cite{Dalfovo99}, in addition to the 
quantities $\omega_{\mathrm{ho}}t_\mathrm{nat}$ and 
\begin{equation}
  k_\mathrm{eff}=
  \frac{1}{\hbar}
  \left[\frac{d E_{\mathrm{res}}}{d B}(B_{\mathrm{res}})\right]
  (\Delta B)^{2}/\left(d B/d t\right). 
\end{equation}
Here $N$ is the total number of 
atoms and $\omega_\mathrm{ho}=2\pi\nu_\mathrm{ho}$ is the angular frequency of 
the spherically symmetric atom trap. We note that all three of these 
parameters can in principle be independently varied by manipulating the trap 
frequency, the ramp speed, and the total number of atoms. Of these three 
dimensionless quantities, the Landau-Zener parameter $\delta_\mathrm{LZ}$ in 
Eq.~(\ref{deltaLZ}) can be written as a function of just the nonlinearity 
parameter $k_{\mathrm{bg}}$ and $\omega_{\mathrm{ho}}t_\mathrm{nat}$. 
Indeed, as we observe in Fig.\ \ref{fig:87RbLZconst}, keeping 
$k_{\mathrm{bg}}$ and $\omega_{\mathrm{ho}}t_\mathrm{nat}$ constant while 
varying $k_\mathrm{eff}$ reveals a very weak dependence of the molecular 
conversion efficiency on this variation. Using identical input parameters for 
the dynamic equation (\ref{NLSElocal}) of the first order microscopic quantum 
dynamics approach reveals that this remarkable universality of the molecular 
production during a linear passage across a Feshbach resonance is preserved 
even when the complete quasi continuum of excited two-body energy modes is 
explicitly accounted for by the theory. 

\section{Conclusions}
\label{sec:conclusions}
We have presented a comprehensive theoretical analysis of the  
adiabatic association to diatomic molecules of initially Bose-Einstein 
condensed $^{87}$Rb atoms via magnetic field tunable interactions. In 
particular, we have considered the situation in which a gas of Bose-Einstein 
condensed atoms in the $(F=1,m_{F}=+1)$ state is exposed to a homogeneous
magnetic field whose strength is varied linearly to cross the broadest Feshbach
resonance at 100 mT, in order to produce strongly correlated pairs 
of atoms in the highest excited vibrational bound state. We have compared the 
predictions of Landau-Zener, nonlinear configuration interaction, and two 
level mean field approaches with the full first order microscopic quantum 
dynamics approach, which explicitly includes all energy states of two atoms.

We found that despite decisive differences between these many-body approaches 
with respect to the description of the underlying microscopic binary collision 
dynamics, the predicted molecular production efficiencies obtained from linear
ramps of the magnetic field strength are virtually independent of the level 
of approximation. We have shown that the Landau-Zener coefficient of 
Eq.~(\ref{deltaLZ}) is the main parameter that controls the molecular 
production in all theoretical approaches. Consequently, the efficiency of 
molecular production in linear ramps of the magnetic field strength is 
remarkably insensitive with respect to the details of the binary collision 
dynamics and to the coherent nature of the gas. The adiabatic association of 
molecules in dilute Bose-Einstein condensed gases is thus subject to 
universal physical properties similar to those that we have identified in the 
associated two-body problem.
This indicates that related experiments on the formation of molecules as well
as their subsequent dissociation via linear ramps of the magnetic field 
strength (see, e.g., \cite{Mukaiyama03}) are largely inconclusive with respect
to the details of the intermediate microscopic binary collision dynamics
(cf., also, Ref.~\cite{TKKG03}). 
 
The universal properties of the molecular production efficiencies and 
dissociation spectra reported in this paper are restricted to {\em linear}
ramps of the magnetic field strength. The interferometric studies of 
Refs.~\cite{Donley02,Claussen03} and their subsequent theoretical analysis 
\cite{Kokkelmans02,Mackie02,KGB03}, for instance, have clearly revealed that 
sequences of linear variations of the magnetic field strength lead to 
molecular production efficiencies beyond the predictions of Landau-Zener or
two level mean field models.

\acknowledgments %{Acknowledgements}
We thank Vincent Boyer, Donatella Cassetari, Rachel Godun, Eleanor Hodby, 
Giuseppe Smirne, C.M. Chandrashekar, Christopher Foot, Stephan D\"{u}rr, 
Thomas Gasenzer, and Keith Burnett for stimulating discussions.
This research has been supported by the European Community Marie Curie
Fellowship under Contract no.\ HPMF-CT-2002-02000 (K.G.), the United Kingdom 
Engineering and Physical Sciences Research Council as well as NASA (S.A.G.),
the US Office of Naval Research (E.T. and P.S.J.), and a University Research 
Fellowship of the Royal Society (T.K.).

\begin{appendix}
  \section{Low energy two channel Hamiltonian}
  \label{app:separablepotential} 
  In this appendix we shall provide the universal and practical model of the 
  low energy two-channel potential matrix that we have used to determine the 
  static as well as the dynamic properties of the resonance enhanced 
  scattering in Sections \ref{sec:twobody} and \ref{sec:manybody}. Based on 
  the general form of the energy states in Subsection 
  \ref{subsec:energystates}, we shall determine the explicit parameters of the 
  low energy potentials that correspond to the 100 mT Feshbach resonance of 
  $^{87}$Rb. 
  
  \subsection{Low energy background scattering potential}
  \label{app:background}
  The two-channel bound and continuum energy states in Subsection 
  \ref{subsec:energystates} depend on the complete Green's function 
  $G_\mathrm{bg}(z)$ of the background scattering in Eq.~(\ref{Gbg}).
  Here $z$ is the complex parameter in Eqs.~(\ref{phipbg}) and 
  Eq.~(\ref{phib}), respectively, which determines the asymptotic form of
  the continuum and bound state wave functions at large interatomic 
  distances. The Green's function $G_\mathrm{bg}(z)$ is determined by all
  bound and continuum energy states associated with the background 
  scattering. The detailed form of the binary interaction potential is not 
  resolved in ultracold collisions
  (cf.~discussion in \ref{subsubsec:background}). At magnetic field strengths
  asymptotically far from the position of a Feshbach resonance, the low energy 
  scattering observables can be determined in terms of just two 
  parameters, the background scattering length $a_\mathrm{bg}$ and the 
  binding energy $E_{-1}$ (see Fig.~\ref{fig:EbofBtwochannel}) of the highest 
  excited vibrational bound state of the background scattering potential 
  $V_\mathrm{bg}$ \cite{Gao98,secondchoice}. This universality allows us to 
  choose a potential model of $V_\mathrm{bg}$ that recovers the exact 
  scattering length $a_\mathrm{bg}$ as well as the exact binding energy 
  $E_{-1}$. In the following, we thus use a most convenient separable 
  potential energy operator, of the form \cite{Lovelace64}
  \begin{equation}
    V_\mathrm{bg}=|\chi_\mathrm{bg}\rangle \xi_\mathrm{bg}
    \langle\chi_\mathrm{bg}|,
    \label{separablepotentialgeneral}
  \end{equation}
  to determine $G_\mathrm{bg}(z)$. We choose the arbitrary but convenient 
  Gaussian form of the function 
  \begin{equation}
    \langle\mathbf{p}|\chi_\mathrm{bg}\rangle=\chi_\mathrm{bg}(p)=
    \frac{\exp\left(-\frac{p^2\sigma_\mathrm{bg}^2}{2\hbar^2}\right)}
	 {(2\pi\hbar)^{3/2}}
	 \label{separablepotentialGaussian}
  \end{equation}
  in momentum space. We shall then adjust the amplitude $\xi_\mathrm{bg}$ and
  the range parameter $\sigma_\mathrm{bg}$ in such a way that they exactly
  recover $a_\mathrm{bg}$ as well as $E_{-1}$ 
  (see Fig.~\ref{fig:EbofBtwochannel}). 
  
  The choice of a separable potential energy operator in 
  Eq.~(\ref{separablepotentialgeneral}) allows us to determine the 
  bound state $\phi_{-1}$ and the continuum energy states of the background 
  scattering in an analytic form. The bound state $\phi_{-1}$ and its
  binding energy $E_{-1}$ are determined by the integral form of the 
  Schr\"odinger equation, which for the separable potential in 
  Eq.~(\ref{separablepotentialgeneral}) is given by:
  \begin{align}
    |\phi_{-1}\rangle &=G_0(E_{-1})V_\mathrm{bg}|\phi_{-1}\rangle
    =G_0(E_{-1})|\chi_\mathrm{bg}\rangle\xi_\mathrm{bg}
    \langle\chi_\mathrm{bg}|\phi_{-1}\rangle.
    \label{SEboundseparable}
  \end{align}
  Here $G_0(z)=\left[z-\left(-\hbar^2\nabla^2/m\right)\right]^{-1}$ is the 
  free Green's function, i.e.~the Green's function of the relative motion 
  of an atom pair in the absence of interactions. The factor 
  $\langle\chi_\mathrm{bg}|\phi_{-1}\rangle$ in Eq.~(\ref{SEboundseparable})
  is determined by the unit normalisation of $\phi_{-1}$. The as yet 
  undetermined binding energy $E_{-1}$ can be obtained by multiplying 
  Eq.~(\ref{SEboundseparable}) by $\langle\chi_\mathrm{bg}|$ from the left. 
  This yields:
  \begin{equation}
    1-\xi_\mathrm{bg}\langle\chi_\mathrm{bg}|G_0(E_{-1})
    |\chi_\mathrm{bg}\rangle=0.
    \label{Eminus1determination}
  \end{equation}
  A short calculation using the Gaussian form of $\chi_\mathrm{bg}$ in
  Eq.~(\ref{separablepotentialGaussian}) determines the matrix element of the
  free Green's function in Eq.~(\ref{Eminus1determination}) by:
  \begin{equation}
    \langle\chi_\mathrm{bg}|G_0(E_{-1})
    |\chi_\mathrm{bg}\rangle=
    \frac{m}{4\pi^{3/2}\hbar^2\sigma_\mathrm{bg}}
    \left\{
    \sqrt{\pi}xe^{x^2}\left[1-\mathrm{erf}(x)\right]-1
    \right\}.
    \label{matrixelementG0}
  \end{equation}
  Here $\mathrm{erf}(x)=\frac{2}{\sqrt{\pi}}\int_0^x e^{-u^2}du$ is the
  error function with the argument 
  $x=\sqrt{m|E_{-1}|}\sigma_\mathrm{bg}/\hbar$.
  
  In addition to Eq.~(\ref{Eminus1determination}),
  the second condition that is needed to relate the two parameters 
  $\xi_\mathrm{bg}$ and $\sigma_\mathrm{bg}$ of the separable potential
  operator to the background scattering length and the binding energy 
  $E_{-1}$ is provided by the continuum energy wave functions, or 
  equivalently, by the $T$ matrix. The $T$ matrix of the background scattering 
  is determined by the Lippmann-Schwinger equation \cite{Newton82}:
  \begin{equation}
    T_\mathrm{bg}(z)=V_\mathrm{bg}+V_\mathrm{bg}G_0(z)T_\mathrm{bg}(z).
    \label{LSTmatrix}
  \end{equation}
  For a separable potential energy operator Eq.~(\ref{LSTmatrix}) can be 
  solved by iteration, which yields the sum of the Born series:
  \begin{align}
    T_\mathrm{bg}(z)=V_\mathrm{bg}\sum_{j=0}^\infty 
    \left[G_0(z)V_\mathrm{bg}\right]^j
    =\frac{|\chi_\mathrm{bg}\rangle\xi_\mathrm{bg}\langle\chi_\mathrm{bg}|}
    {1-\xi_\mathrm{bg}\langle\chi_\mathrm{bg}|G_0(z)
      |\chi_\mathrm{bg}\rangle}.
    \label{Bornseries}
  \end{align}
  The background scattering length is then determined in terms of the 
  $T$ matrix in Eq.~(\ref{Bornseries}) by
  \begin{align}
    a_\mathrm{bg}=\frac{m}{4\pi\hbar^2}(2\pi\hbar)^3
    \langle 0|T_\mathrm{bg}(0)|0\rangle
    =\frac{\frac{m}{4\pi\hbar^2}\xi_\mathrm{bg}}
    {1-\xi_\mathrm{bg}\langle\chi_\mathrm{bg}|G_0(0)|\chi_\mathrm{bg}\rangle}.
    \label{abgofTbg}
  \end{align}
  Here $|0\rangle$ is the zero momentum plane wave of the relative motion
  of an atom pair. The denominator on the right hand side of 
  Eq.~(\ref{abgofTbg}) can be obtained directly from 
  Eq.~(\ref{matrixelementG0}) by replacing the energy argument 
  $E_{-1}$ by 0. This yields:
  \begin{equation}
    a_\mathrm{bg}=\frac{\frac{m}{4\pi\hbar^2}\xi_\mathrm{bg}}
    {1+\frac{m}{4\pi\hbar^2}\xi_\mathrm{bg}/
      \left(\sqrt{\pi}\sigma_\mathrm{bg}\right)}.
    \label{abgofxibgsigmabg}
  \end{equation}
  Equation (\ref{abgofxibgsigmabg}) can be used to eliminate 
  $\xi_\mathrm{bg}$ from Eq.~(\ref{Eminus1determination}), which, in turn, 
  determines the range parameter $\sigma_\mathrm{bg}$ in terms of the
  background scattering length $a_\mathrm{bg}$ and of the binding energy 
  $E_{-1}$. Given the range parameter $\sigma_\mathrm{bg}$ 
  the remaining amplitude $\xi_\mathrm{bg}$ can then be obtained from 
  Eq.~(\ref{abgofxibgsigmabg}). For the $^{87}$Rb parameters 
  $a_\mathrm{bg}=100 \ a_\mathrm{Bohr}$ and $E_{-1}=-h\times 23$ MHz,
  as obtained from Fig.~\ref{fig:Eboverview},
  this yields $\sigma_\mathrm{bg}=42.90753599 \ a_\mathrm{Bohr}$ and 
  $m\xi_\mathrm{bg}/\left(4\pi\hbar^2\right)=-317.5649079 \ a_\mathrm{Bohr}$.
  
  \subsection{Off diagonal coupling}
  In order to recover the universal properties of the near resonant bound 
  states in Subsection \ref{subsec:universal} 
  (cf.~Fig.~\ref{fig:haloboundstates}), the transition probabilities
  in Subsection \ref{subsec:2Bapproach} (cf.~Fig.~\ref{fig:transprob}), 
  as well as the time evolution operator in Eq.~(\ref{NLSElocal}), it is 
  sufficient to use a single channel Hamiltonian, as explained in 
  Subsection \ref{subsec:universal}. To this end, we have extended the 
  potential model in Eq.~(\ref{separablepotentialgeneral}) to recover not 
  only the background scattering, but also the scattering length $a(B)$. 
  We have thus adjusted the separable potential energy operator in 
  Eq.~(\ref{separablepotentialgeneral}), at each magnetic field strength $B$,
  to the magnetic field dependent scattering length 
  $a(B)=a_\mathrm{bg}\left[1-(\Delta B)/(B-B_0)\right]$ 
  and to the energy $E_{-1}$ (cf., also, Ref.~\cite{KGB03}).  

  The dissociation spectra in Subsection \ref{subsec:dissociation} have been 
  obtained from a two channel description. A two channel approach recovers
  not only the exact magnetic field dependence of the scattering length, but 
  it also accurately describes the binding energies of the multichannel bound 
  states over a much wider range of magnetic field strengths 
  (see Fig.~\ref{fig:EbofBtwochannel}) than a single channel treatment. Apart 
  from the background scattering, a two channel description of resonance 
  enhanced collisions requires us to specify the coupling between the 
  open and the closed channels. Equations (\ref{resonancewidth}) and 
  (\ref{resonanceshift}) relate the off diagonal coupling element $W(r)$ of 
  the general two channel Hamiltonian in Eq.~(\ref{H2B2channel}) to the 
  resonance width $(\Delta B)$ as well as to the shift $B_0-B_\mathrm{res}$. 
  Although the resonance shift is not directly measurable, 
  Eq.~(\ref{magicformula}) relates it to the van der Waals dispersion 
  coefficient $C_6$ and to the width $(\Delta B)$, which can usually be 
  determined from experimental data. In the pole approximation to the 
  closed channel Green's function [cf.~Eq.~(\ref{poleapproximation})], the
  closed channel part of the two channel Hamiltonian in Eq.~(\ref{H2B2channel})
  reduces to Eq.~(\ref{replacementHcl}), which restricts the state of an atom 
  pair in the closed channels to the resonance state $\phi_\mathrm{res}$. 
  Given the linear dependence of the energy $E_\mathrm{res}(B)$, associated 
  with $\phi_\mathrm{res}$, on the magnetic field strength $B$, the closed 
  channel part of the two channel Hamiltonian is determined completely by the 
  slope $\left[\frac{dE_\mathrm{res}}{dB}(B_\mathrm{res})\right]$ of the
  Feshbach resonance (cf.~Subsection \ref{subsec:energystates}). 

  The only remaining quantity that needs to be determined is the 
  product $W(r)\phi_\mathrm{res}(r)$, which provides the interchannel 
  coupling in the pole approximation. Similar to the choice of the separable 
  potential in Eq.~(\ref{separablepotentialgeneral}), we shall use a two 
  parameter description of the interchannel coupling, in terms of a real 
  amplitude $\zeta$ and a range parameter $\sigma$, to recover the width and 
  the shift of the Feshbach resonance. This leads to the {\em ansatz}
  \begin{equation}
    W|\phi_\mathrm{res}\rangle=|\chi\rangle\zeta,
  \end{equation}
  where we have chosen the arbitrary but convenient Gaussian form of the
  function
  \begin{equation}
    \langle\mathbf{p}|\chi\rangle=\chi(p)=
    \frac{\exp\left(-\frac{p^2\sigma^2}{2\hbar^2}\right)}
	 {(2\pi\hbar)^{3/2}}
	 \label{couplingGaussian}
  \end{equation}
  in momentum space. In the following, we shall adjust the parameters 
  $\zeta$ and $\sigma$ in such a way that the two channel Hamiltonian 
  recovers the resonance width $(\Delta B)$ and the shift 
  $B_0-B_\mathrm{res}$ via Eqs.~(\ref{resonancewidth}) and 
  (\ref{resonanceshift}), respectively. To this end, we shall first 
  determine the zero energy continuum state $\phi_0^{(+)}$  
  in Eq.~(\ref{resonancewidth}) and the zero energy Green's function 
  $G_\mathrm{bg}(0)$ in Eq.~(\ref{resonanceshift}) in terms of the $T$ matrix 
  associated with the background scattering, which is given by 
  Eq.~(\ref{LSTmatrix}). The continuum energy states associated with 
  the relative momentum $\mathbf{p}$ fulfil the Lippmann-Schwinger equation
  \cite{Newton82} ($E=p^2/m$):
  \begin{equation}
    |\phi_\mathbf{p}^{(+)}\rangle=|\mathbf{p}\rangle+
    G_0(E+i0)T_\mathrm{bg}(E+i0)|\mathbf{p}\rangle.
  \end{equation}
  Here the energy argument ``$z=E+i0$'' ensures that the wave function
  $\phi_\mathbf{p}^{(+)}(\mathbf{r})$
  has the long range asymptotic form of Eq.~(\ref{BCphipplus}). Furthermore, 
  the Green's function $G_\mathrm{bg}(z)$ is completely determined by the
  $T$ matrix in Eq.~(\ref{LSTmatrix}) in terms of the general formula
  \cite{Newton82}:
  \begin{equation}
    G_\mathrm{bg}(z)=G_0(z)+G_0(z)T_\mathrm{bg}(z)G_0(z).
  \end{equation}
  In the separable potential approach the exact $T$ matrix of the 
  background scattering is given by Eq.~(\ref{Bornseries}), so that the 
  subsequent determination of the right hand sides of 
  Eqs.~(\ref{resonancewidth}) and (\ref{resonanceshift}) involves just
  the evaluation of matrix elements of the zero energy free Green's 
  function $G_0(0)$ between the wave functions $\chi_\mathrm{bg}$
  and $\chi$. Given the Gaussian form of these wave functions in 
  Eqs.~(\ref{separablepotentialGaussian}) and (\ref{couplingGaussian}), 
  the matrix elements can be readily determined analytically in complete 
  analogy to Eq.~(\ref{matrixelementG0}). This yields
  \begin{equation}
    (\Delta B)=\frac{\zeta^2}{
      \left[
	\frac{dE_\mathrm{res}}{dB}(B_\mathrm{res})
	\right]}
    \frac{m}{4\pi\hbar^2a_\mathrm{bg}}
    \left(
      1-\frac{a_\mathrm{bg}}{\sqrt{\pi}\overline{\sigma}}
	\right)^2
    \label{deltaBofzetasigma}
  \end{equation}
  for the resonance width and
  \begin{equation}
    B_0-B_\mathrm{res}=(\Delta B)\frac{a_\mathrm{bg}}{\sqrt{\pi}\sigma}
    \frac{1-\frac{a_\mathrm{bg}}{\sqrt{\pi}\sigma}
      \left(\frac{\sigma}{\overline{\sigma}}\right)^2}
	 {\left(
	   1-\frac{a_\mathrm{bg}}{\sqrt{\pi}\sigma}
	   \frac{\sigma}{\overline{\sigma}}
	   \right)^2}
	 \label{shiftozetasigma}
  \end{equation}
  for the shift. Here we have introduced the mean range parameter
  \begin{equation}
    \overline{\sigma}=
    \sqrt{\frac{1}{2}
      \left(
      \sigma^2+\sigma_\mathrm{bg}^2
      \right)}.
  \end{equation}
  Inserting the experimentally known width of the resonance and 
  Eq.~(\ref{magicformula}) for the shift into the left hand side of 
  Eqs.~(\ref{deltaBofzetasigma}) and (\ref{shiftozetasigma}), respectively,
  then determines $\zeta$ and $\sigma$ in terms of $(\Delta B)$, $C_6$ and
  the slope of the resonance.
  Using $(\Delta B)=0.02$ mT \cite{Volz03}, $C_6=4660$ a.u.~\cite{Roberts01} 
  and $\left[\frac{dE_\mathrm{res}}{dB}(B_\mathrm{res})\right]=h\times 38$ 
  MHz/mT for the 100 mT Feshbach resonance of $^{87}$Rb, the parameters of the
  interchannel coupling can be summarised as
  $\sigma=21.50035463 \ a_\mathrm{Bohr}$ and 
  $m\zeta^2/(4\pi\hbar^2\sigma)=h\times 8.0536126$ MHz.

\end{appendix}


\begin{thebibliography}{99}
\bibitem{opticaltrap}
  We note that the $^{87}$Rb atoms in the $(F=1,m_F=+1)$ electronic ground 
  state cannot be magnetically trapped. The harmonic potential is, 
  therefore, provided by crossed laser beams. These optical atom traps can
  confine atoms and molecules at the same time.
\bibitem{Molecularfraction}
  The second, factorised contribution to the molecular mean field on the 
  right hand side of Eq.~(\ref{psibgeneral}) can be interpreted as an overlap 
  between the atomic mean field and the molecular bound state wave function. 
  When the gas is strongly interacting the molecular bound state has a spatial 
  extent comparable to the mean distance of the atoms in the gas 
  (cf.~Eq.~(\ref{bondlength})), and the overlap becomes significant. As a 
  consequence, one and the same atom could contribute to several bound 
  molecules, so that a separation of the gas into bound and free atoms becomes 
  physically meaningless.
\bibitem{burstspectra}
  An explicit determination of the energy spectra of the burst atoms and their 
  typical single particle energy scale of $k_\mathrm{B}\times$150 nK in the
  atom-molecule coherence experiments of Ref.~\cite{Donley02} has been 
  performed in Ref.~\cite{KGB03}.
\bibitem{85Rbadmixture}
  The admixture of the resonance state to the highest excited vibrational
  bound state depends on the Feshbach resonance. Over the whole range of 
  magnetic field strengths between 15.5 mT and 16.22 mT in the atom-molecule 
  coherence experiments \cite{Donley02} with $^{85}$Rb Bose-Einstein 
  condensates, the admixture of the resonance state is even less than 30 \%
  (cf., also, Ref.~\cite{TKTGPSJKB03}).
\bibitem{secondchoice}
  An equivalent choice of parameters would be $a_\mathrm{bg}$ and the van der 
  Waals dispersion coefficient $C_6$ \cite{Gao98}.
\end{thebibliography}
\end{document}